\documentclass[11pt]{article}

\usepackage[reqno]{amsmath}
\usepackage{epsfig}
\usepackage{array}
\usepackage{float}

\begin{document}


\def\ltap{\ \raisebox{-.4ex}{\rlap{$\sim$}} \raisebox{.4ex}{$<$}\ }
\def\gtap{\ \raisebox{-.4ex}{\rlap{$\sim$}} \raisebox{.4ex}{$>$}\ }
\newcommand{\deltaatm}{\mbox{$\Delta  m^2_{\mathrm{atm}} \ $}}
\newcommand{\deltasol}{\mbox{$ \Delta  m^2_{\odot} \ $}}
\newcommand{\deltalsnd}{\mbox{$ \Delta  m^2_{\mathrm{SBL}} \ $}}
\newcommand{\thetasbl}{\mbox{$ \theta_{\mathrm{SBL}}  \ $}}
\newcommand{\utre}{\mbox{$|U_{\mathrm{e} 3}|$}}
\newcommand{\betabeta}{\mbox{$(\beta \beta)_{0 \nu}  $}}
\newcommand{\mefff}{\mbox{$ < \! m  \! > $}}
\newcommand{\meff}{\mbox{$\left|  < \! m  \! > \right| \ $}}
\newcommand{\hbeta}{$\mbox{}^3 {\rm H}$ $\beta$-decay \ }
\newcommand{\eV}{\mbox{$ \  \mathrm{eV} \ $}}
\newcommand{\deltatre}{\mbox{$ \ \Delta m^2_{32} \ $}}
\newcommand{\deltadue}{\mbox{$ \ \Delta m^2_{21} \ $}}
\newcommand{\ueuno}{\mbox{$ \ |U_{\mathrm{e} 1}|^2 \ $}}
\newcommand{\uedue}{\mbox{$ \ |U_{\mathrm{e} 2}|^2 \ $}}
\newcommand{\uetre}{\mbox{$ \ |U_{\mathrm{e} 3}|^2  \ $}}


\hyphenation{par-ti-cu-lar}
\hyphenation{ex-pe-ri-men-tal}
\hyphenation{dif-fe-rent}
\hyphenation{bet-we-en}
\hyphenation{mo-du-lus}
\hyphenation{}
%
\rightline{Ref. SISSA 13/2001/EP}
\rightline{TUM-HEP-410/01}
\rightline{April 2001}
\rightline{hep -- ph/0104218}
\vskip 0.6cm
\begin{center}
{\bf
Majorana Neutrinos, Neutrino Mass Spectrum, CP-Violation and \\ 
Neutrinoless Double $\beta$-Decay: II. Mixing of Four Neutrinos\\
} 

\vspace{0.3cm} 
S. M. Bilenky$^{(a,b)}$
\footnote{Also at: Joint Institute for Nuclear Research, Dubna, Russia}
,~S. Pascoli$^{(b,c)}$~and~
S. T. Petcov$^{(b,c)}$ 
\footnote{Also at: Institute of Nuclear Research and
Nuclear Energy, Bulgarian Academy of Sciences, 1784 Sofia, Bulgaria}

\vspace{0.3cm}
{\em $^{(a)}$ Physik Department, Technische Universitat Munchen,
D-85748, Garching, Germany \\
}
\vspace{0.2cm}   
{\em $^{(b)}$ Scuola Internazionale Superiore di Studi Avanzati, 
I-34014 Trieste, Italy\\
}
\vspace{0.2cm}   
{\em $^{(c)}$ Istituto Nazionale di Fisica Nucleare, 
Sezione di Trieste, I-34014 Trieste, Italy\\
}
\end{center}

\vskip 0.4cm
\begin{abstract}
Assuming four-neutrino mixing 
and massive Majorana neutrinos,
we study the implications of the neutrino 
oscillation solutions of the solar 
and atmospheric neutrino problems,
of the results of the LSND experiment
and of the constraints on neutrino oscillations,
obtained in reactor and accelerator experiments,
for the predictions of the
effective Majorana mass in
neutrinoless double beta
(\betabeta-)decay, \meff. 
All 4-neutrino mass spectra compatible 
with the existing neutrino mass 
and oscillation data are considered:
2+2A,B and 3+1A,B,C. 
The general case of CP-nonconservation 
is investigated.
The predicted values of  \meff
depend strongly on the value of the 
lightest neutrino mass $m_1$, on the 
type of the neutrino mass spectrum, 
on the LSND neutrino mass-squared difference
\deltalsnd,
on the solution of the solar neutrino problem, 
as well as on the values of the three
Majorana CP-violating phases,
present in the lepton mixing matrix.
If CP-invariance holds, 
\meff is very sensitive
to the values of the relative 
CP-parities of the massive 
Majorana neutrinos.
We have also analyzed in detail 
the question of whether a measurement of 
$\meff \gtap 0.01$ eV in the next generation of
\betabeta-decay experiments (NEMO3, 
CUORE, EXO, GENIUS), 
combined with the data from the solar, atmospheric,
reactor and accelerator neutrino 
oscillation experiments and from the 
future neutrino mass $^3$H $\beta-$decay 
experiment KATRIN 
would allow, and under what conditions,
i) to determine the absolute values of the 
neutrino masses and thus the 
neutrino mass spectrum, and
ii) to establish the existence of 
CP-violation in the lepton sector. 
We have pointed out, in particular, that 
the 2+2A and 3+1A spectra can be 
critically tested by the
KATRIN experiment. The latter, in particular,
can provide information on 
the value of the lightest neutrino
mass, $m_1$, in the cases of the
spectra 2+2A, 3+1A, 3+1B and 3+1C.
For these 
neutrino mass spectra  
there exists a direct 
relation between 
\meff or $m_1$ and the neutrino mass 
measured in $^3$H $\beta-$decay,
$m_{\nu_e}$, 
and the measurement of 
$\meff \gtap 0.01$ eV 
and of $m_{\nu_e} \gtap 0.4$ eV 
will give 
the unique possibility to determine the 
absolute values of all four 
neutrino masses and 
to obtain information on
CP-violation in the lepton sector.

\end{abstract}
\vfill
\newpage

\hyphenation{par-ti-cu-lar}
\hyphenation{ex-pe-ri-men-tal}
\hyphenation{dif-fe-rent}
\hyphenation{bet-we-en}
\hyphenation{mo-du-lus}
\hyphenation{}
\section{Introduction} \indent 

 With the accumulation of more and stronger 
evidences for oscillations of the
atmospheric ($\nu_{\mu}$, $\bar{\nu}_{\mu}$)
\cite{SKatm98,SKatm00}
and solar ($\nu_e$)  
\cite{Cl98,Fu96,Ab99,Ha99,SK99,
SKYSuz00,GNO,SNO,BP00}
neutrinos,  
caused by  neutrino mixing 
and nonzero neutrino masses in vacuum 
(see, e.g., \cite{BiPont78,BiPe87,SP97,BGG99}), 
the problem of the nature of massive neutrinos 
emerges as one of the fundamental problems 
in the studies of neutrino mixing. 
Massive neutrinos, as is well known,
can be Dirac or Majorana particles.
In the former case they possess 
a conserved lepton charge and
distinctive antiparticles, 
while in the latter there is no conserved 
lepton charge 
and massive neutrinos  are truly neutral 
particles identical with their antiparticles
(see, e.g., \cite{BiPe87}).
Massive Dirac neutrinos 
and neutrino mixing arise in gauge theories
in which the individual lepton charges, 
$L_e$, $L_{\mu}$ and $L_{\tau}$, are 
not conserved, but a specific combination
of the latter, which could be the total 
lepton charge $L = L_e + L_{\mu} + L_{\tau}$,
or, e.g., the charge \cite{SPPD82}
$L' = L_e - L_{\mu} + L_{\tau}$, is conserved. 
Massive Majorana neutrinos arise 
if no lepton charge is conserved by the 
interactions responsible 
for the neutrino mass generation.
Thus, the question of the nature of 
massive neutrinos is directly related 
to the question of the 
basic symmetries of the 
fundamental particle interactions.

  The Majorana nature of massive neutrinos can 
be revealed by investigation 
of processes in which the total lepton charge $L$
is not conserved and changes by two units, $\Delta L = 2$. 
The process most sensitive to the 
existence of massive Majorana neutrinos 
(coupled to the electron)
is the neutrinoless double $\beta$ 
($\betabeta-$) decay of
certain even-even nuclei 
(see, e.g., \cite{bb0nu1,bb0nu2,BiPe87}):
\begin{equation}
(A,Z) \rightarrow (A,Z+2) + e^{-} + e^{-}.
\end{equation}
\noindent If the $\betabeta-$ decay is generated 
{\it only by the (V-A) charged current weak interaction
through the exchange of virtual 
massive Majorana neutrinos}, 
the probability amplitude of this process 
is proportional in the case of 
neutrinos having masses not exceeding few MeV 
to the so-called ``effective Majorana mass parameter''
\begin{equation}
\mefff \equiv  \sum_{j=1} U_{\mathrm{e} j}^2~ m_j,
\label{meff}
\end{equation}

\noindent where $m_j$ is the mass of the 
Majorana neutrino $\nu_j$  
and $U_{ej}$ is the element of neutrino (lepton)
mixing matrix.

  Assuming mixing of three 
massive Majorana neutrinos,
we have studied in \cite{BPP1} 
the implications of the 
solar and atmospheric 
neutrino oscillation data
and of the data of the 
long baseline reactor experiments
CHOOZ and Palo Verde 
for the predictions of the
effective Majorana mass 
\meff, which determines
\footnote{There exists an extensive literature on 
the predictions for \meff
in the case of three-neutrino  
mixing and massive Majorana neutrinos:
see, e.g., refs. \cite{SPAS94,BBGK9596,
BGKP96,Vissani-bb, 
BGGKP99,SPWIN99,
Pols00,KPS00,Rodej00}.
}
the rate of the \betabeta-decay. 
Our study was stimulated, in part, 
by the expected future 
progress in the experimental searches for
$\betabeta-$decay. 
The presently existing most stringent constraint 
on the value of the effective Majorana mass parameter
was obtained in the $^{76}$Ge Heidelberg-Moscow 
experiment \cite{76Ge00}: 
\begin{equation}
\meff < 0.35 \ \mathrm{eV},~~~90\%~{\rm C.L.}.
\label{76Ge0}
\end{equation}
\noindent Taking into account a factor of 3 
uncertainty associated with the calculation 
of the relevant nuclear matrix element 
(see, e.g., \cite{bb0nu1,bb0nu2}) one finds
\begin{equation}
\meff < (0.35 \div 1.05) \ \mathrm{eV},~~~90\%~{\rm C.L.}
\label{76Ge00}
\end{equation}
\noindent The IGEX collaboration  has obtained \cite{IGEX00}:
\begin{equation}
\meff < (0.33 \div 1.35) \ \mathrm{eV},~~~90\%~{\rm C.L.}
\label{IGEX00}
\end{equation}

\noindent A sensitivity to $\meff \sim 0.10$ eV
is foreseen to be reached in the currently operating 
NEMO3 experiment \cite{NEMO3},
while the next generation 
of $\betabeta-$decay experiments CUORE,
EXO, GENIUS \cite{CUORE,GENIUS,EXO},
is planned to reach a sensitivity 
to values of $\meff \gtap 0.01$ eV,
which are considerably smaller than the 
presently existing most stringent upper
bounds (\ref{76Ge00}) and (\ref{IGEX00}).

  We have found in \cite{BPP1}, in particular, 
that  the observation of the
\betabeta-decay with a rate corresponding to 
$\meff \gtap 0.02~$eV,
which is in the range of sensitivity of the 
future \betabeta-decay experiments, 
will provide unique information 
not only on the nature of massive neutrinos,
but also on the neutrino mass spectrum.
Combined with information on the value of the 
lightest neutrino mass
or on the type of the neutrino mass spectrum,
it can provide also information 
on the CP-violation in the lepton sector,
and if CP-invariance holds - on the 
relative CP-parities
of the massive Majorana neutrinos.
A measured value of 
$\meff \gtap (2 - 3)\times 10^{-2}~$eV,
for instance, would strongly disfavor 
(if not rule out), 
under the general assumptions of our study
(3-neutrino mixing, \betabeta-decay generated 
only by the charged (V-A) current weak interaction 
via the exchange of the three Majorana neutrinos,
neutrino oscillation solutions of the solar 
neutrino problem and atmospheric
neutrino anomaly) the possibility of a 
hierarchical neutrino mass spectrum,
while a value of 
$\meff \gtap (2 - 3)\times 10^{-1}~$eV
would rule out the 
hierarchical neutrino mass spectrum,
strongly disfavor the spectrum with
inverted mass hierarchy and favor
the quasi-degenerate spectrum.

 In the present article 
we present predictions  
for \meff in the case of  
mixing of four massive Majorana neutrinos. 
The assumption of mixing of four massive 
neutrinos allows 
to explain by neutrino oscillations
both the results of the 
solar and atmospheric
neutrino experiments, which requires two 
different neutrino mass-squared differences,
and the indications for 
$\bar{\nu}_{\mu}\leftrightarrow \bar{\nu}_{e}$
oscillations reported by
the LSND collaboration \cite{LSND},
which requires an additional much larger 
neutrino mass-squared difference.
The fourth weak-eigenstate 
neutrino involved in the mixing 
must be (see, e.g., \cite{BGG99})
a sterile neutrino, $\nu_s$ \cite{Pont67}.
The latter can be accommodated, e.g., 
in extensions of the Standard Theory, 
which include $SU(2)_L$ singlet 
right-handed (RH) neutrino field(s) 
(see, e.g. \cite{4numodels}).
More specifically, we study in detail the
implications of the neutrino 
oscillation solutions of the solar 
and atmospheric neutrino problems,
of the results of the analysis of the LSND 
data and of the constraints on neutrino oscillations,
obtained in reactor and accelerator experiments,
for the predictions of the
effective Majorana mass, 
\meff. The effective 
Majorana mass \meff 
in schemes
with 4-neutrino mixing 
was discussed earlier in, e.g.,
\cite{BGGKP99,Giunti99,
KPS00,Pols002,Fins00}.
We consider five different 
types of neutrino mass spectrum,
compatible with the 
data on neutrino oscillations: 
using 
standard notations we denote them as
2+2A, 2+2B, 3+1A, 3+1B and 3+1C.
These are essentially all 
possible types of 4-neutrino
mass spectra compatible with the existing 
data. The general case of 
CP-nonconservation in the lepton sector
is investigated, although we study
the possibility of CP-invariance as well.
In the latter case 
we obtain predictions for \meff for 
all different sets of values of
the relative CP-parities 
of the massive Majorana neutrinos.
In both cases we pay special 
attention to the possibility
of cancellations between the 
contributions to \meff
of the different massive Majorana neutrinos.
We give, in particular,
detailed predictions for 
the value of \meff for the five different 
types of neutrino mass spectrum
indicated above.
We analyze also in detail 
the question of whether a measurement of 
$\meff \gtap 0.01$ eV in the next generation of
\betabeta-decay experiments, 
combined with the data from the solar, atmospheric,
reactor and accelerator neutrino 
oscillation experiments and from the 
future neutrino mass $^3$H $\beta-$decay 
experiments (KATRIN, etc.) would allow, and 
under what conditions,
i) to determine the absolute values of the 
neutrino masses and thus the 
neutrino mass spectrum, and
ii) to establish the existence of 
CP-violation in the lepton sector.


\section{\textbf{Experimental Data and Constraints on Neutrino 
Masses and Mixing Angles}}
\label{data}
We present in this Section
the constraints on 
the neutrino mass squared differences, neutrino (lepton) 
mixing angles and on the neutrino masses,  
which follow from the existing 
experimental data and
will be used in our further analyzes.

 Strong evidences for neutrino 
oscillations have been obtained
in the experiments  with atmospheric \cite{SKatm98,SKatm00} 
and solar \cite{Cl98,Fu96,Ab99,Ha99,SK99,
SKYSuz00,GNO,SNO,BP00} neutrinos.
Indications for $\bar{\nu}_{\mu}\leftrightarrow \bar{\nu}_{e}$
oscillations were reported by
the accelerator LSND collaboration \cite{LSND}.
The interpretation of these results 
in terms of neutrino oscillations
requires three independent neutrino 
mass-squared differences,
$\deltaatm$, $\deltasol$, $\deltalsnd$, respectively,
and thus (at least) 4-neutrino mixing. 
The fourth (weak-eigenstate)
neutrino participating in the mixing 
together with the three flavour neutrinos,
$\nu_e$, $\nu_{\mu}$ and $\nu_{\tau}$,
must be a sterile neutrino,
$\nu_s$ (see, e.g., \cite{BGG99,4numodels}). 
The combined neutrino oscillation
analyzes of the atmospheric, solar and 
LSND data have been performed so far 
assuming that
$\deltalsnd$, $\deltaatm$ and 
$\deltasol$ obey the relation
(see, e.g., \cite{Gonza4nu,Fogli4nu}):
\begin{equation}
|\deltasol | \ll |\deltaatm | \ll |\deltalsnd|.
\label{3deltamh}
\end{equation}

\noindent This hierarchical relation is suggested by the 
results of the two-neutrino oscillation analyzes of the
atmospheric, solar and LSND data. The predictions for the 
effective Majorana mass, \meff, we are interested in,
depend, in general, on the values of $|\deltalsnd|$, $|\deltaatm|$ and 
$|\deltasol|$, on the mixing angles which 
determine the solar neutrino
oscillations/transitions, as well as 
on the limits on the neutrino oscillation parameters
obtained in the reactor $\bar{\nu}_e$ disappearance experiments 
BUGEY, CHOOZ  and Palo Verde.

  There are two general types of 4-neutrino mass
spectra compatible with the existing data 
on neutrino masses and neutrino oscillations:
i) the 2+2 one which is characterized
by two light neutrinos and 
two quasi-degenerate heavier ones, the two 
corresponding pairs of 
neutrino mass-squared differences being separated 
by the LSND ``gap''
$\sim \deltalsnd$, and 
ii) the 3+1 type consisting 
of three neutrinos which are quasi-degenerate
in mass and of one much lighter or much heavier neutrino
separated (in mass-squared) from the other three 
by the LSND ``gap'' $\sim \deltalsnd$.  

  In the case of 3+1 type of neutrino mass spectrum,
the negative results
of reactor and accelerator short baseline 
disappearance neutrino oscillation
experiments 
imply a strong upper bound on the
amplitude of the $\nu_{\mu} \to \nu_{e} $
transition, $A_{\mu;e}$. Comparing
this bound  with the allowed values of 
$A_{\mu;e}$, obtained 
in the analysis of  
the LSND data, lead to the conclusion 
\cite{BGG4nu98,Gosw4nu97,Barger4nu98}
that the 3+1 schemes are disfavored by the data. 
At the same time
it was shown \cite{BGG4nu98,Gosw4nu97,Barger4nu98}
that the 2+2 schemes are fully compatible 
with all existing neutrino oscillation data.
Recently, as a result of a new LSND data analysis, the
region of allowed values of $A_{\mu;e}$
changed and it was demonstrated
\cite{Barger31,PeresS31,Giunti31} that
the 3+1 schemes are compatible with all 
existing data for
some specific values of $\deltalsnd$.
This problem was further investigated in \cite{Grimus31}.
It was shown by a comprehensive statistical analysis 
that in the case of 
3+1 schemes there is no overlapping of the 
region limited by the 
95\% C.L. upper bound on $A_{\mu;e}$ and of 
the 99\% C.L. LSND allowed region.
All existing neutrino oscillation data
can be described at
99\% C.L. if  \deltalsnd has values 
in limited regions around the points 
$\sim {\rm 6~eV^{2},~1.7~eV^{2},~0.9~eV^{2},~0.3~eV^{2}}$. 
Thus, the 3+1 schemes cannot be completely excluded by
the analysis of the existing data. 
Taking this into account we will
consider in this paper both the 2+2 and the 
3+1 types of 4-neutrino mass spectra.

   The results of the combined analysis of 
the solar and atmospheric neutrino data depend on 
the spectrum considered.
At the same time, for the 
indicated two types of 4-neutrino mass spectra,
the results of the 2-neutrino mixing analysis of the LSND
data, which are given below 
(see eqs. (\ref{mlsnd}) and (\ref{anglsnd})),
can be used to constrain the elements
of the neutrino mixing matrix in the
4-neutrino mixing case.

  Let us discuss next briefly the LSND result.
Searching for $\bar{\nu}_{\mu}\leftrightarrow \bar{\nu}_{e}$
oscillations with a high intensity
source of $\nu_{\mu}$, $\bar{\nu}_{\mu}$ and $\nu_{e}$ 
in the range of energies $E \sim (30 - 60)$ MeV
and detector located at a distance $L \sim 30$ m 
from the neutrino source,
the LSND collaboration observed an excess of
$e^{+}-$like events over the background \cite{LSND}.
This excess was interpreted as being due to the
$\bar{\nu}_{\mu}\leftrightarrow \bar{\nu}_{e}$
transitions. The search for
the same transitions 
under somewhat less favorable experimental 
conditions (neutrino beam of smaller intensity,
$L \sim 18$ m) by the KARMEN experiment
gave negative results \cite{KARMEN}.
Performing a joint analysis of the LSND and KARMEN
data and taking into account the 
limits from the BUGEY reactor antineutrino experiment
\cite{BUGEY}, one finds (at 95\% C.L.) the following
allowed region of values of the 
corresponding two-neutrino oscillation parameters, 
$\deltalsnd$ and $\sin^2 \thetasbl$, for 
which the LSND excess of events 
can be explained by 
$\bar{\nu}_{\mu}\leftrightarrow \bar{\nu}_{e}$
oscillations:
\begin{align}
1.6 \times 10^{-1} \eV^2  \leq & \ \ \deltalsnd   \leq  \  2.0 \eV^2
~~~(95\%~\mathrm{ C.L.}),
\label{mlsnd} \\
8.0 \times 10^{-4} \leq & \ \ \sin^2 \thetasbl   \leq  4 \times 10^{-2}
~~~(95\%~\mathrm{ C.L.}).  
\label{anglsnd}
\end{align}
%
\noindent The upper and lower  
limits in (\ref{mlsnd}) and (\ref{anglsnd})
are strongly correlated; in particular, 
both $\deltalsnd|_{\mathrm{MAX}}$
and $\deltalsnd|_{\mathrm{MIN}}$ 
are decreasing functions of $\sin^2 \thetasbl$. 
This inter-dependence of the allowed values of
$\deltalsnd$ and $\sin^2 \thetasbl$
has been taken into account in our analysis.
Note also that the sign of $\deltalsnd$ 
cannot be determined from the analysis of the data.
We have assumed above for concreteness 
that $\deltalsnd > 0$. Let us note that the LSND results
on $\bar{\nu}_{\mu}\leftrightarrow \bar{\nu}_{e}$
oscillations will be tested in the 
accelerator experiment
MiniBooNE \cite{MiniB} which is under preparation.

 Consider next the results of the analysis of 
the solar and atmospheric 
neutrino oscillation data.
We will discuss first the case of neutrino mass 
spectrum of the 2+2 type.
The Super-Kamiokande atmospheric neutrino data 
\cite{SKatm98,SKatm00}
is best interpreted in terms of dominant
$\nu_\mu \rightarrow \nu_\tau$
and $\bar{\nu}_\mu \rightarrow \bar{\nu}_\tau$
oscillations \cite{SKatm00};
some non-dominant fraction of the atmospheric 
$\nu_\mu$ ($\bar{\nu}_\mu$)
can oscillate into $\nu_s$ and $\nu_e$ 
($\bar{\nu}_e$).
The $\nu_\mu \rightarrow \nu_{\tau,s}$
and $\bar{\nu}_\mu \rightarrow \bar{\nu}_{\tau,s}$
oscillations of the atmospheric neutrinos of interest
are characterized by three parameters 
(see, e.g., \cite{Gonza4nu,Fogli4nu}):
$\deltaatm$, $\theta_{\mathrm{atm}}$
and $\cos^2\beta$, where $\beta$ is a
neutrino mixing angle.
Of these three parameters only $\deltaatm$
enters into the expression for \meff.
Under the condition
(\ref{3deltamh}), the 
oscillations/transitions of solar
neutrinos are characterized by 
three parameters \cite{Gonza4nu} as well:
$\deltasol > 0$, $\theta_{\odot}$ 
(or $\tan^2 \theta_{\odot}$) and
$\cos^2 \beta$.
As we have indicated, the probability of the 
atmospheric $\nu_{\mu} \rightarrow \nu_{\tau,s}$
($\bar{\nu}_{\mu} \rightarrow \bar{\nu}_{\tau,s}$)
oscillations depends also on
$\cos^2 \beta$. 
The parameter $\cos^2 \beta$
determines, in particular,
what fraction of the solar 
$\nu_e$ and of the atmospheric $\nu_{\mu}$ 
($\bar{\nu}_{\mu}$)
can transform into a 
sterile neutrino $\nu_s$ ($\bar{\nu}_s$).
For, e.g., $\cos^2 \beta = 0$,
the solar $\nu_e \rightarrow \nu_s$
transitions do not take place at all,
while the atmospheric 
$\nu_{\mu}$ ($\bar{\nu}_{\mu}$)
can oscillate {\it only}
into $\nu_s$ ($\bar{\nu}_s$)~
\footnote{This case is 
strongly disfavored 
(if not excluded) by the atmospheric
neutrino data \cite{SKatm00,Fogli4nu}.}.
If $\cos^2 \beta = 1.0$, the solar $\nu_e$ 
can undergo transitions 
{\it only} into $\nu_s$,
while the atmospheric $\nu_{\mu}$ 
($\bar{\nu}_{\mu}$)
{\it cannot undergo transitions into}
$\nu_s$ ($\bar{\nu}_s$).

 For the 2+2 type neutrino mass spectrum 
and  the hierarchy (\ref{3deltamh})
of neutrino mass-squared differences,
a 4-neutrino oscillation analysis 
of the most recent solar neutrino data,
including the new high precision Super-Kamiokande 
results \cite{SKYSuz00} on
the spectrum of the recoil electrons 
and on the day-night (D-N) effect,
was performed in ref. \cite{Gonza4nu}.  
The regions of neutrino oscillation 
parameters, corresponding to 
the large mixing angle (LMA) MSW, 
small mixing angle (SMA)
MSW (see, e.g., \cite{SP97,BGG99}), 
of the LOW and the 
quasi-vacuum oscillation (QVO)
solutions of the solar neutrino problem
(see, e.g., \cite{Fogli00,PK00,Gonza3nu,KPVO96}), allowed by the  
data at a given C.L. were determined.
The existence of a given solution
and the corresponding 
solution regions were shown
to depend on the value of 
$\cos^2 \beta$. For $\cos^2 \beta = 0$,
one recovers the
results derived in the case 
of the two-neutrino 
$\nu_e \rightarrow \nu_{\mu (\tau)}$
oscillations analyzes of the solar neutrino data
(see, e.g., \cite{Fogli00,PK00,KPVO96}).
If, however, $\cos^2 \beta = 1$,
only the $\nu_e \rightarrow \nu_{s}$
transitions of solar neutrinos are possible 
and one finds that only the 
MSW SMA $\nu_e \rightarrow \nu_{s}$ transition
solution is
allowed by the data \cite{Gonza4nu}.
As $\cos^2 \beta$ changes from 0 to 1,
the LMA and LOW-QVO solution regions
diminish and disappear, while 
the SMA MSW solution region 
essentially changes only its position
with respect to the \deltasol-axis,
moving as a whole 
to smaller by a factor of $\sim 1.2$
values of \deltasol \cite{KPQ96}.
  For, e.g., $\cos^2 \beta = 0.30$,
the allowed regions of 
the LMA and LOW-QVO solutions
are somewhat smaller than in the case of
$\cos^2 \beta =0$ \cite{Gonza4nu}.
Both the LMA and LOW-QVO solution
regions include $\cos2\theta_{\odot} = 0$
at 99\% C.L., as like the analogous 
$\cos^2 \beta =0$ solution regions. 
The maximal values of 
$\cos^2 \beta$, for which one still has
LMA, LOW and QVO solutions 
(at 99\% C.L.) are \cite{Gonza4nu}
$\cos^2 \beta \cong 0.72;~0.77;~0.80$.
The $\chi^2-$ analysis performed in 
\cite{Gonza4nu} showed also that as 
$\cos^2 \beta$ increases from 0 to 1,
the quality of the fit of the data 
for the LMA and LOW-QVO solutions
decreases, while that provided by the
SMA MSW solution 
remains essentially unchanged. 
It should be added that 
the SMA MSW solution is excluded at 
68\% C.L., but becomes allowed 
approximately at 90\% C.L. \cite{Gonza4nu}.

   A neutrino oscillation analysis 
of the Super-Kamiokande atmospheric neutrino
data in the case of four-neutrino
mixing and 2+2 type of neutrino mass spectrum 
was performed in \cite{Fogli4nu}.
The analysis showed, in particular, that the 
atmospheric neutrino data 
excludes the possibility of 
$\cos^2 \beta = 0$: one finds that
$\cos^2 \beta \geq 0.33$ 
at $90\%$ C.L. \cite{Fogli4nu}. 
In the case of  $\cos^2 \beta = 0.4$, 
for instance,   
we have $\nu_\mu \rightarrow \nu_{\tau,s}$
and $\nu_e \rightarrow \nu_{s,\tau}$
transitions respectively of 
the atmospheric and solar neutrinos.
For $\cos^2 \beta = 0.30;~0.50$,
the allowed values of \deltaatm 
at $90\%$~C.L. ($99\%$~C.L.) 
are the following \cite{Fogli4nu}:
\begin{align}
\text{for} \ \cos^2 \beta = 0.3:~~~~~~~~~~~~~~         & 
                     2.0 \ (1.5) \times 10^{-3} \eV^2 \leq 
                     \deltaatm \leq 5.0 \ (6.5) \times 10^{-3} \eV^2, \\
\text{for} \ \cos^2 \beta = 0.5:~~~~~~~~~~~~~~         & 
                     2.5 \ (2.0) \times 10^{-3} \eV^2 \leq 
                     \deltaatm \leq 4.5 \ (6.5) \times 10^{-3} \eV^2.
\label{deltamatm05}
\end{align}    
\noindent \noindent The sign of $\deltaatm$ is 
undetermined by the data. The best fit value of 
$|\deltaatm |$ found in \cite{Fogli4nu} 
is: $|\deltaatm |_{\mathrm{BF}} = 
3.2 \times 10^{-3} \mathrm{eV}^2$. 
 
 It should be obvious from the preceding 
discussion that there exists 
a correlation between the allowed values of
\deltaatm, \deltasol, $\tan^2 \theta_\odot$
and of $\cos^2 \beta$. Thus, although 
$\cos^2 \beta$ does not enter explicitly into
the expression for \meff, the latter 
depends implicitly on $\cos^2 \beta$
through the allowed values of \deltaatm, 
\deltasol and $\tan^2 \theta_\odot$, 
which depend on the value of $\cos^2 \beta$.
In what follows (Sections 4 and 5) 
we will present results for \meff
in the case of the 2+2 type of neutrino mass spectrum
for two values of $\cos^2 \beta = 0.3;~ 0.5$.
The corresponding values of \deltaatm, and of
\deltasol and $\tan^2\theta_{\odot}$, are given in
eqs. (9)-(\ref{deltamatm05}) and in Table 1, 
respectively. The predictions for \meff for 
$\cos^2 \beta \gtap 0.8$ practically 
coincide with those
obtained for $\cos^2 \beta = 0.3~(0.5)$
in the case of the SMA MSW 
solution of the solar neutrino problem.

\vspace{5mm}
\setlength{\extrarowheight}{4pt}
\hspace{-2cm}
\begin{table}
\caption{Values of \deltasol and 
$\tan^2 \theta_\odot$  obtained in the analysis of
the solar neutrino data in ref. \cite{Gonza4nu}
at 90\% (99\%) C.L. and used in the present 
study (see text for details).
The best fit values (B.F.V.)  are also given.
The results shown correspond to 
$\cos^2 \beta = 0$ (upper table), 
$\cos^2 \beta = 0.3$ (middle table) and 
$\cos^2 \beta = 0.5$ (lower table).
}
\setlength{\extrarowheight}{4pt}
\begin{center}
\vspace{3mm}
\begin{tabular}{|c|c|c|}  \hline
\setlength{\tabcolsep}{1pt} 
\label{tabella1}
           & $\deltasol [\mathrm{eV}^2]$  [B.F.V.]          & $\tan^2 \theta_\odot$  [B.F.V.]   \\ \hline \hline
LMA       & $ 1.0 \  (1.0) \times 10^{-5}  \div
  2.0 \  (2.0) \times 10^{-4} $   [$3 \times 10^{-5}$] &
 $ 0.2 \ (0.18) \div 0.8  \  (2.0)$ [0.35] \\
SMA       & $  5.0 \ (4.0) \times 10^{-6}  \div
 6.0 \ (10) \times 10^{-6}$  & 
$  6.0 \ (2.0) \times 10^{-4}  \div 
1.0 \  (2.0) \times 10^{-3}$   \\
LOW-QVO & $ 35 \ (0.5) \times 10^{-9} \div  
2.0 \ (3.0) \times 10^{-7}$ & 
$ 0.47 \ (0.4) \div 1.0 \ (3.0)$  \\ \hline
\end{tabular}

\vspace{3mm}

\begin{tabular}{|c|c|c|}  \hline
\setlength{\tabcolsep}{1pt} 
           & $\deltasol [\mathrm{eV}^2]$          & $\tan^2 \theta_\odot$   \\ \hline \hline
LMA       & $ 1.6 \  (1.2) \times 10^{-5}  \div
  0.8 \  (2.0) \times 10^{-4} $   &
 $ 0.33 \ (0.25) \div 0.62  \  (1.1)$  \\
SMA       & $  4.0 \ (4.0) \times 10^{-6}  \div
 7.5 \ (10) \times 10^{-6}$  & 
$  3.5 \ (2.0) \times 10^{-4}  \div 
1.0 \  (1.5) \times 10^{-3}$   \\
LOW-QVO & $ 60 \ (0.6) \times 10^{-9} \div  
6.0 \ (23) \times 10^{-8}$ & 
$ 0.7 \ (0.45) \div 0.7 \ (2.8)$  \\ \hline
\end{tabular}

\vspace{3mm}

\begin{tabular}{|c|c|c|}  \hline
\setlength{\tabcolsep}{1pt} 
           & $\deltasol [\mathrm{eV}^2]$    & $\tan^2 \theta_\odot$   \\ \hline \hline
LMA       & $ 2.2 \  (1.6) \times 10^{-5}  \div
  2.2 \  (20) \times 10^{-5} $   &
 $ 0.54 \ (0.35) \div 0.54  \  (0.9)$ \\
SMA       & $  4.0 \ (4.0) \times 10^{-6}  \div
 6.5 \ (10) \times 10^{-6}$  & 
$  4.0 \ (2.0) \times 10^{-4}  \div 
0.9 \  (1.5) \times 10^{-3}$   \\
LOW-QVO & $  \ (6.0 \times 10^{-10} \div  
 \ 1.5) \times 10^{-7}$ & 
$  \ (0.5 \div  \ 2.2)$  \\ \hline
\end{tabular}
\end{center}
\end{table}

  In the case of 3+1 type of neutrino mass spectrum,
the results of the combined analysis of solar and
atmospheric neutrino data reduce effectively 
to the results of the analysis 
performed under the assumption 
of 3-flavour neutrino mixing \cite{Gonza3nu}.
A detailed description of these 3-neutrino mixing 
results we will use in the corresponding analysis
in Sections 6, 7 and 8 is given in \cite{BPP1} (see Section 2)
and we are not going to reproduce them here.

  It should be noted that given the 
results of the analysis of
the atmospheric neutrino data, the inequality
$\deltasol \ll \deltaatm$ holds for 
\begin{equation}
\deltasol \ltap 2.0 \times 10^{-4} \eV^2.
\label{condition1}
\end{equation}
\noindent This condition 
is fulfilled 
in the SMA MSW and LOW-QVO solutions regions 
and in part of the LMA MSW one 
\cite{SKYSuz00,Gonza4nu,Fogli00,PK00,Gonza3nu}.
According to ref.~\cite{Gonza4nu,Gonza3nu}, values of 
$\deltasol \sim (7.0 - 8.0) \times 10^{-3} \eV^2$ are allowed 
in the LMA MSW solution region. In this case the 
condition $\deltasol \ll \deltaatm$ is not satisfied 
and the reliability (accuracy) of the results derived 
is somewhat questionable. Thus, we will 
use in our further analysis
the upper bound on \deltasol given 
in eq. (\ref{condition1}), adding comments 
about how the results change if
$\deltasol \sim 7.0 - 8.0 \times 10^{-3} \eV^2$.
This bound is also reported in Table 1. 

  Very important constraints on the oscillations of 
electron (anti-)neutrinos were obtained 
in the CHOOZ and Palo Verde disappearance 
experiments with reactor $\bar{\nu}_e$ \cite{CHOOZ,PaloV}.
For the 4-neutrino mixing under
discussion these constraints are 
relevant primarily in the case of 
neutrino mass spectrum of the
3+1 type. The two experiments 
were sensitive to values of
$\Delta m^2 \gtap 10^{-3}~{\rm eV^2}$,
which includes the region of the solution 
of the atmospheric neutrino problem
(see, e.g., eqs. (9) and (\ref{deltamatm05})). 
No disappearance of the reactor
$\bar{\nu}_e$ was observed. Performing a 
two-neutrino oscillation
analysis, the following rather stringent
upper bound on the value of the
corresponding mixing angle, $\theta$,
was obtained
\footnote{The possibility of large 
$\sin^2  \theta > 0.9$
which is admitted by the CHOOZ data alone
is incompatible with the neutrino oscillation
interpretation of the solar neutrino deficit
(see, e.g., \cite{BGG99,SPWIN99}).}
 at 95\% C.L. for 
$\Delta m^2 \geq 1.5 \times 10^{-3} \mathrm{eV}^2$ \cite{CHOOZ}:
\begin{equation}
\sin^2  \theta < 0.09.
\label{chooz}
\end{equation}

\noindent This limit holds in the case 
of the neutrino mass spectrum
of the 2+2 type. For the 3+1 type of spectrum 
we will use the limit on 
$\sin^2  \theta$, obtained in the combined 
3-neutrino mixing analysis 
of the solar, atmospheric and CHOOZ neutrino oscillation 
data in ref. \cite{Gonza3nu}: $\sin^2  \theta < 0.05 
~(0.08)$ (90\% (99\%) C.L.). 
The precise upper limit on $\sin^2 \theta$ 
following from the CHOOZ data 
is $\Delta m^2$-dependent \cite{CHOOZ}. 
This dependence is accounted for, whenever necessary,
in our analysis.
The sensitivity to the value of the  parameter
$\sin^2  \theta$ is expected
to be considerably improved by the MINOS 
experiment \cite{MINOS} in which 
the following upper limit can be reached:
$\sin^2 \theta < 5 \times 10^{-3}$.
%

  In the next few years 
the constraints on the
values of $\deltalsnd$, $\sin^2 \thetasbl$,
$\cos^2\beta$,
\deltasol, $\theta_\odot$,
\deltaatm, $\theta_{\mathrm{atm}}$
and $\theta$ will be 
improved due to the increase of the 
statistics of the currently running experiments
(e.g., SAGE \cite{Ab99}, GNO \cite{GNO}, 
Super-Kamiokande \cite{SKatm00,SKYSuz00}, 
SNO \cite{SNO}, K2K \cite{K2K})
and the upgrade of some of them,
as well as due to the data 
from the new experiments MiniBooNE \cite{MiniB}, 
BOREXINO \cite{BOREX}, KamLand \cite{KamL}, 
MINOS \cite{MINOS} and CNGS \cite{CERNGS}.
As a result, the values of the neutrino 
oscillation parameters will be know
with much better accuracy.
For the 4-neutrino mixing schemes under study
the results of the
MiniBooNE experiment which is scheduled to 
begin data-taking in December of 2001,
will be crucial: this experiment
will test the LSND indications for
$\bar{\nu}_{\mu} \rightarrow \bar{\nu}_e$
oscillations.

   The Troitzk~\cite{MoscowH3} and Mainz~\cite{Mainz} 
\hbeta experiments, studying the electron spectrum, 
provide information on the electron (anti-)neutrino mass  
$m_{\nu_e}$. The data contain features which are not well
understood (a peak in the end-point region
which varies with time \cite{MoscowH3}). 
The upper bounds given 
by the authors (at 95\%~C.L.) read:
\begin{eqnarray}
m_{\nu_e}  <  2.5 \eV~~\cite{MoscowH3},\ \ 
m_{\nu_e} <  2.9  \eV~~~\cite{Mainz}.
\label{H3beta}
\end{eqnarray}

\noindent There are prospects to 
increase the sensitivity  
of the \hbeta experiments
and probe the region of 
values of $m_{\nu_e} \sim (0.4 - 1.0)$ eV: 
the next generation of 
\hbeta experiment KATRIN \cite{KATRIN}
is planned to have a sensitivity 
to values of $m_{\nu_e} \sim 0.35$ eV.

 Cosmological and astrophysical data provide
information on the sum of the neutrino masses.
The current upper bound reads (see, e.g., \cite{Cosmo}
and the references quoted therein):
\begin{eqnarray}
\sum_{j} m_{j}  \ltap  5.5 \eV.
\end{eqnarray}

\noindent The future experiments MAP and PLANCK 
can be sensitive to \cite{MAP}
\begin{eqnarray}
\sum_{j} m_{j}  \cong  0.4 \eV.
\end{eqnarray}

  In the next Sections we show that the data
from the new generation of $\betabeta-$ decay 
experiments, which will be sensitive to values of 
$\meff \sim (0.01 - 0.10)~$eV, can provide 
information on the neutrino mass spectrum 
as well as on the leptonic CP-violation 
generated by Majorana CP-violating phases.

\section{\textbf{Four-Neutrino Mixing and the \betabeta-Decay 
Effective Majorana Mass}}

   The explanation of the data of atmospheric neutrino,
solar neutrino and LSND experiments
in terms of neutrino oscillations 
requires the existence of mixing
of (at least) four 
massive neutrinos.
This means that in addition to the three
active left-handed 
flavour neutrino fields 
$\nu_{l\mathrm{L}}$, $l= e,\mu,\tau$, 
which enter into the expression
of the weak charged lepton current,
one has to assume that there exists 
a sterile neutrino field, $\nu_{s L}$ -
a singlet of $SU(2)_{L}\times U(1)_{Y_{W}}$  
\footnote{The field $\nu_{s L}$ is 
related to the ``standard'' RH 
$SU(2)_{L}\times U(1)_{Y_{W}}$-singlet neutrino field 
$\nu_{R}$ through the charge conjugation matrix $C$:
$\nu_{s L} = C (\bar{\nu}_{R})^{T}$ 
(see, e.g., \cite{BiPe87}).}, and that
\begin{align}
\nu_{l \mathrm{L}} = & \sum_{j=1}^{4} U^{\ast}_{l j} \nu_{j \mathrm{L}},~~
l=e,\mu,\tau, \nonumber \\
\nu_{s \mathrm{L}} = &  \sum_{j=1}^{4} U^{\ast}_{s j} \nu_{j \mathrm{L}}.
\label{4numix}
\end{align}

\noindent Here $\nu_{j \mathrm{L}}$ is the 
left-handed field of the 
neutrino $\nu_j$ having a mass $m_j$
and $U$ is a $4 \times 4$ unitary mixing matrix.
We will assume that 
the neutrinos $\nu_j$ are Majorana particles
whose fields satisfy the Majorana condition:
\begin{equation}
C(\bar{\nu}_{j})^{T} = \nu_{j},~j=1,2,3,4,
\label{Majcond}
\end{equation}
\noindent where $C$ is the charge conjugation matrix.
We will numerate 
the neutrino masses in such way 
that $m_1 < m_2 < m_3 < m_4$.

   The effective Majorana mass parameter can be expressed as:
\begin{equation}
\meff \equiv \left| m_1 U_{\mathrm{e} 1}^2 
+ m_2 U_{\mathrm{e} 2}^2 + m_3 U_{\mathrm{e} 3}^2  
+ m_4 U_{\mathrm{e} 4}^2 \right|,
\label{effectivemass}
\end{equation}

\noindent
where $U_{\mathrm{e} i}$ are the entries 
of the $U$ mixing matrix, 
$m_i$ is the mass of the $\nu_i$ massive neutrino.
We have 
\begin{equation}
U_{\mathrm{e} j} = |U_{\mathrm{e} j}|~e^{i{\alpha_j}\over 2},
\label{Uejphase}
\end{equation}
\noindent where $\alpha_j$, $j=1,2,3,4$, are 
real phases. Only the phase differences
$(\alpha_j - \alpha_k) \equiv \alpha_{jk}$ ($j > k$)
can play a physical role. 
 
\indent The Majorana condition can also have the form 
\begin{equation}
C(\bar{\nu}_{j})^{T} = (\xi_j^{\ast})^2~\nu_{j},~j=1,2,3,4,
\label{Majcond2}
\end{equation}
\noindent where 
$\xi_j$, $j=1,2,3,4$, are 
arbitrary phase factors.
The effective Majorana mass parameter
is given now by:
\begin{equation}
\meff \equiv \left|~ m_1 U_{\mathrm{e} 1}^2 \xi_1^2
+ m_2 U_{\mathrm{e} 2}^2 \xi_2^2 + 
m_3 U_{\mathrm{e} 3}^2 \xi_3^2 +
m_4 U_{\mathrm{e} 4}^2 \xi_4^2~\right|.
\label{meffectivexi}
\end{equation}
\noindent Obviously, the  physical Majorana
CP-violating phases on which \meff depends 
are $\alpha_{21} \equiv  
{\rm arg}(U_{\mathrm{e} 2}^2 \xi_2^2)
-{\rm arg}(U_{\mathrm{e} 1}^2 \xi_1^2)$, 
$\alpha_{31} \equiv
{\rm arg}(U_{\mathrm{e} 3}^2 \xi_3^2)
-{\rm arg}(U_{\mathrm{e} 1}^2 \xi_1^2)$ and 
$\alpha_{41} \equiv
{\rm arg}(U_{\mathrm{e} 4}^2 \xi_4^2)
-{\rm arg}(U_{\mathrm{e} 1}^2 \xi_1^2)$.
All CP-violation effects
associated with the Majorana nature of
the massive neutrinos are generated 
by $\alpha_{21} \neq k\pi$,
$\alpha_{31} \neq k'\pi$ and 
$\alpha_{41} \neq k''\pi$, $k,k',k''=0,1,2,...$. 

  The $4\times 4$ mixing matrix $U$ can be parametrized
by 6 Euler-like angles and 10 phases. 
We can eliminate the unphysical phases
in $U$ by rephasing, e.g., the 
charged lepton, l(x), 
and the neutrino, $\nu_j(x)$, fields in 
the weak charged lepton
current, $l(x) \rightarrow e^{i \eta_l}l(x)$ \ and \ 
$\nu_j(x) \rightarrow e^{i \beta_j} \nu_j(x)$.
The phases which re-appear in the Majorana 
condition for the fields of 
massive neutrinos when the latter are
Majorana particles are physical \cite{BHP80,Kobz80}: 
these phases will be present also in \meff
\footnote{In the case of massive Dirac neutrinos the 
indicated phases will just be eliminated
from the neutrino mixing matrix $U$
and from the charged current weak interaction 
Lagrangian by the indicated rephasing of the charged 
lepton and neutrino fields; thus, they are unphysical.}.
The elements of the lepton mixing matrix and 
the phase factors in the Majorana condition
for massive neutrino fields change as follows:
\begin{align}
U_{\mathrm{l} j} & \rightarrow 
U_{\mathrm{l} j} e^{ -i ( \eta_l - \beta_j)},~l=e,\mu,\tau,~j=1,2,3,4, \\
\xi_j  & \rightarrow \xi_j  e^{ -i \beta_j}.
\end{align}
%
Thus, in the case of Dirac neutrinos
there are four CP-violating phases in the
neutrino (lepton) mixing matrix $U$,
while in the case of massive Majorana neutrinos
$U$ contains seven CP-violating phases.
However, since the sterile neutrino 
$\nu_{s}$ does not participate in 
electroweak interactions
\footnote{We suppose that the sterile neutrino 
field $\nu_{s L}$ is a singlet with respect to the 
electroweak gauge symmetry group, which is
assumed to be that of the Standard Theory, i.e.,
$SU(2)_{L}\times U(1)_{Y_{W}}$.}
(see, e.g., refs. \cite{BiPe87,BGG99,4numodels}), 
in both cases of massive Dirac 
and massive Majorana
neutrinos one CP-violating phase
does not appear in physical processes 
taking place at energy scales 
of the order of, or smaller than, the 
electroweak symmetry breaking scale.
This is the phase 
common for all elements of the
fourth row of $U$, i.e., for the elements
$U_{s j}$, $j=1,2,3,4$, associated with
the sterile neutrino field $\nu_{s L}$. 
Since the field $\nu_{s L}$
does not enter into  
the charged current and neutral
current weak interaction 
Lagrangian ${\cal L}^{CC+NC}$,
the indicated phase 
will not appear in processes
generated by  ${\cal L}^{CC+NC}$.
Thus, there exist at most six 
{\it relevant} physical CP-violating phases  
in the lepton sector of 
the theory with massive Majorana neutrinos
of interest, and consequently - 
at most six relevant rephasing invariants.
Three are the standard Dirac ones, 
$J_{1,2,3}$, present 
in the case of mixing of four massive
Dirac neutrinos 
\cite{CeciliaJ,NievPal87,Branco00} 
(see also \cite{KP88Earth}):
\begin{align}
J_1 & = \mathit{Im} \ \ ( U_{\mu 3} U_{e 4} U_{\mu 4}^{\ast} U_{e3}^{\ast}), \\
J_2 & = \mathit{Im} \ \ ( U_{\tau 3} U_{e 4} U_{\tau 4}^{\ast} U_{e3}^{\ast}), \\
J_3 & = \mathit{Im} \ \ ( U_{\mu 2} U_{e 4} U_{\mu 4}^{\ast} U_{e2}^{\ast}).
\end{align}
\noindent The existence of the other three, 
$S_1$, $S_2$ and $S_3$, 
is related to the Majorana nature of 
the massive neutrinos $\nu_j$ (see \cite{NievPal87,Branco00}): 
\begin{align}
S_1  & \equiv  \mathit{Im} \ \ (  U_{e1} U_{e4}^{\ast} 
\xi_4^{\ast} \xi_1), \\
S_2  & \equiv  \mathit{Im} \ \ (  U_{e2} U_{e4}^{\ast} 
\xi_4^{\ast} \xi_2), \\
S_3  & \equiv  \mathit{Im} \ \ (  U_{e3} U_{e4}^{\ast} \xi_4^{\ast} \xi_3). 
\label{Majrephinv}
\end{align}

The  Majorana CP-violating phases 
$\alpha_{21}$, $\alpha_{31}$ and $\alpha_{41}$
are determined by the three independent 
rephasing invariants, $S_{1,2,3}$. We have, for instance:
\begin{align}
\cos \alpha_{41} & =  1 - 2 \frac{S_1^2}{ |U_{\mathrm{e} 1}|^2 
|U_{\mathrm{e} 4}|^2}, \\
 \cos(\alpha_{41} -  \alpha_{21}) & =
\cos(\alpha_{4} - \alpha_{2}) =    
1 - 2 \frac{S_2^2}{ |U_{\mathrm{e} 2}|^2 |U_{\mathrm{e} 4}|^2}, \\
\intertext{and}
 \cos(\alpha_{41} -  \alpha_{31}) & =
\cos(\alpha_{4} - \alpha_{3}) =    
1 - 2 \frac{S_3^2}{ |U_{\mathrm{e} 3}|^2 |U_{\mathrm{e} 4}|^2}.
\end{align}
%
\noindent One can express the Majorana CP-violating 
phases entering into the expression for \meff 
in terms of the rephasing invariants 
\footnote{Let us note that the Majorana CP-violating
phases of interest do not enter into the expressions
for the $\nu_l \rightarrow \nu_l'$
($\bar{\nu}_l \rightarrow \bar{\nu}_l'$) and 
$\nu_{l (s)} \rightarrow \nu_{s(l)}$
($\bar{\nu}_{l (s)} \rightarrow \bar{\nu}_{s(l)}$),
$l,l'=e,\mu,\tau$, oscillation/transition 
probabilities \cite{BHP80,LPST87}.}
$S_{1,2,3}$: the former are independent 
of the Dirac rephasing 
invariants $J_{1,2,3}$.
This is a consequence of the specific choice of 
$S_{1,2,3}$, eqs. (27) - (\ref{Majrephinv}),  
which is not unique \cite{NievPal87}. 
With this choice
the amplitude of the
$K^{+} \rightarrow \ \pi^{-} + \mu^{+} + \mu^{+}$ decay,
for instance, which, as like the \betabeta-decay, is generated 
by the exchange of the four virtual massive Majorana neutrinos
in the scheme under discussion,
depends, as can be shown, on all six rephasing 
invariants $S_{1,2,3}$ and $J_{1,2,3}$.

  If CP-invariance holds in the lepton sector 
we have, in particular, $S_1= 0$
or $\mathit{Re} \ (U_{e1} U_{e4}^{\ast} 
\xi_4^{\ast} \xi_1) = 0$, 
$S_2= 0$ or
$\mathit{Re} 
(U_{e2} U_{e4}^{\ast} \xi_4^{\ast} \xi_2) = 0$, 
and $S_3 = 0$ or
$\mathit{Re} 
(U_{e3} U_{e4}^{\ast} \xi_4^{\ast} \xi_3) = 0$.
In terms of constraints on the phases
$\alpha_{21}$, $\alpha_{31}$ and $\alpha_{41}$ this implies 
$\alpha_{21} = k \pi$,
$\alpha_{31} = k^\prime \pi$,
$\alpha_{41} = k^{\prime \prime} \pi$, $k, 
k^\prime, k^{\prime \prime} = 0,1,2...$.

  In all our subsequent analyzes we will
set for convenience (and without loss of generality) 
$\xi_j = 1$, j=1,2,3,4, i.e., we will assume that 
the fields of the Majorana neutrinos $\nu_j$ satisfy
the Majorana conditions (\ref{Majcond}).
The CP-invariance constraint on the elements
of the lepton mixing matrix of interest reads 
\footnote{This constraint is obtained 
from the requirement of CP-invariance
of the charged current weak interaction 
Lagrangian ${\cal L}^{CC}$, 
by choosing the arbitrary phase factors 
in the CP-transformation laws of the electron 
and the $W$-boson fields equal to 1.}  
\cite{LW81,BNP84,Kayser84} (see also \cite{BiPe87}):
\begin{equation}
U^{\ast}_{\mathrm{e} j} = \eta^{CP}_j~U_{\mathrm{e} j},
\label{UejCPinv}
\end{equation}
\noindent where $\eta^{CP}_j = i \phi_j = \pm i$ is 
the CP-parity of the Majorana neutrino $\nu_j$
with mass $m_j >0$.
In this case \meff is given by:
\begin{equation}
\meff \equiv 
\left| \sum_{j=1}^{4}  \eta^{CP}_j |U_{\mathrm{e} j}|^2 m_j \right| =
      \left| \sum_{j=1}^{4}  \phi_j |U_{\mathrm{e} j}|^2 m_j \right|.
\label{meffCP}
\end{equation}

 The neutrino oscillation 
experiments provide information
on $\Delta m^2_{jk} = m^2_{j} - m^2_{k}~ (j > k)$.
In the case of 4-neutrino mixing (\ref{4numix})
there are three independent $\Delta m^2$ parameters.
The four neutrino masses $m_j$, $j=1,2,3,4$, can be 
expressed in terms of these three parameters and, e.g., of 
$m_1$. We have:
\begin{equation} 
m_{2} = \sqrt{m^2_{1} +  \Delta m^2_{21}}~,
\label{m2}
\end{equation}
\begin{equation} 
m_{3} = \sqrt{m^2_{1} + \Delta m^2_{21} +  \Delta m^2_{32}}~.
\label{m3}
\end{equation}
\begin{equation} 
m_{4} = \sqrt{m^2_{1} + \Delta m^2_{21} +  \Delta m^2_{32}+ \Delta m^2_{43} }~.
\label{m4}
\end{equation}
 The mass-squared difference 
inferred from the neutrino 
oscillation interpretation of the LSND data,
\deltalsnd, is equal to $\Delta m^2_{41}$,
\begin{equation} 
\deltalsnd = \Delta m^2_{41} = \Delta m^2_{21} +  \Delta m^2_{32}+ \Delta m^2_{43}~,
\label{dmlsnd0}
\end{equation}

\noindent while for  the ones deduced from 
the solar and atmospheric neutrino data, 
\deltasol and \deltaatm, we 
have several possibilities:
\begin{itemize}
\item
in the case of the 2+2 type of 
neutrino mass spectrum we can identify two sub-cases: 
i) 2+2A - with $\deltasol = \Delta m^2_{43}$
and $\deltaatm  = \Delta m^2_{21}$, 
and ii) 2+2B - with
$\deltasol = \Delta m^2_{21}$
and $\deltaatm  = \Delta m^2_{43}$;
\item
for the 3+1A type of neutrino mass spectrum 
characterized by one neutrino being much lighter 
than the other three which are nearly degenerate,
we have $\deltasol = \Delta m^2_{43}$,
$\deltaatm  = \Delta m^2_{32}$, 
or $\deltaatm  = \Delta m^2_{32}$,
$\deltasol =\Delta m^2_{43} $;
\item
for the 3+1B and in the 3+1C  neutrino mass spectra
with one neutrino being much heavier than the other three,
one finds, respectively, 
$\deltasol = \Delta m^2_{21}$, $\deltaatm  = \Delta m^2_{32}$, 
and $\deltasol = \Delta m^2_{32}$, 
$\deltaatm  = \Delta m^2_{31}$.
\end{itemize}
%

Evidently, getting information on the lightest neutrino mass
$m_1$ would be crucial for determining the values of 
$m_{2,3,4}$. The neutrino oscillation data and cosmological
arguments (see, e.g., eq. (14)) suggest that 
$m_1 \ltap 1$ eV.

 For each of the possible patterns of neutrino masses 
indicated above, we will study in detail in what follows 
the implications of the data on neutrino
oscillations for the searches for
\betabeta-decay.  


\section{\textbf{ The  2+2A  Neutrino Mass Spectrum}}
\label{2+2A}

The 2+2A neutrino mass spectrum 
is characterized by the following pattern 
of the neutrino masses $m_j$:

\unitlength=1mm
\begin{center}
\begin{picture}(100,40)
\put(20,5){\line(1,0){20}}
\put(20,10){\line(1,0){20}}
\put(20,33){\line(1,0){20}}
\put(20,35){\line(1,0){20}}
\put(30,5){\line(0,1){30}}
\put(45,5){\vector(0,1){5}}
\put(45,10){\vector(0,-1){5}}
\put(45,32.5){\vector(0,1){3}}
\put(45,35.5){\vector(0,-1){3}}
\put(72,5){\vector(0,1){30}}
\put(72,35){\vector(0,-1){30}}
\put(48,7){\deltaatm}
\put(48,33){\deltasol}
\put(75,18){\deltalsnd}
\put(17,4){1}
\put(17,9){2}
\put(17,31.5){3}
\put(17,34.2){4}
\end{picture}
\end{center}
%
This pattern corresponds to 
the inequalities
$m_1 < m_2 <~(\ll)~ m_3 \simeq m_4$, 
or equivalently to:
\begin{gather}
m_1 <~(\ll)~ \sqrt{\Delta m^2_{41}}, \nonumber \\
\sqrt{\Delta m^2_{43}} \ll \sqrt{\Delta m^2_{21}} 
\ll \sqrt{\Delta m^2_{41}}. 
\label{22Aspectrum}
\end{gather}
%
%
Taking into account these inequalities
one can make the identification:


\begin{equation}
\Delta m^2_{21}    \equiv    \Delta  m^2_{\mathrm{atm}}, \ \ 
\Delta m^2_{43}    \equiv    \Delta  m^2_{\odot}, \ \ 
\Delta m^2_{41}    \equiv    \Delta  m^2_{\mathrm{SBL}}.
\label{22apattern} 
\end{equation}
%
\noindent
We will suppose that $\deltasol$ takes values 
in the regions reported in Table 1 for 
$\cos^2\beta = 0.3;~0.5$,
$\deltaatm$ lies in the interval 
(9) or (10), and
\deltalsnd can have values in the interval (\ref{mlsnd}).
From eqs. (\ref{m2}) - (\ref{m4}), (\ref{dmlsnd0}) 
and (\ref{22Aspectrum}) it follows that:
\begin{align}
m_2 & = \sqrt{m_1^2 + \deltaatm}, \nonumber  \\
m_3 & \simeq m_4 = \sqrt{m_1^2 + \deltalsnd}.
\end{align}
 
 The condition $m_1,m_2 \ll m_3, m_4$,
which could be valid for the neutrino mass spectrum 
under discussion, is satisfied for 
\begin{equation}
m_1 < 0.25 \sqrt{\deltalsnd} \ \ \Rightarrow \ \ m_1 < 0.1 \eV.
\label{m1}
\end{equation}

 For the elements $U_{\mathrm{e} j}$ of the 
neutrino mixing matrix $U$ we have:
\begin{itemize}
\item[i)]
$|U_{\mathrm{e} \, 1}|^2$ and 
$|U_{\mathrm{e} \, 2}|^2$ are constrained 
by the LSND results \cite{LSND} and BUGEY
neutrino oscillation limits \cite{BUGEY}
to lie in the interval:
\begin{equation}
2 \times 10^{-4} \leq |U_{\mathrm{e} \, 1}|^2 +
 |U_{\mathrm{e} \, 2}|^2 \leq 1 \times 10^{-2};
\end{equation}
%
\item[ii)]
$|U_{\mathrm{e} \, 3}|$ and 
$|U_{\mathrm{e} \, 4}|$ are related to 
the solar mixing angle $\theta_{\odot}$:
\begin{align}
|U_{\mathrm{e} \, 3}|^2 & =
 \cos^2 \theta_\odot \  ( 1 - \sum_{i=1}^2 
|U_{\mathrm{e} \, i}|^2 ), \nonumber \\
|U_{\mathrm{e} \, 4}|^2 & = \sin^2 \theta_\odot
\  ( 1 - \sum_{i=1}^2 |U_{\mathrm{e} \, i}|^2 ), 
\end{align}
\end{itemize}
where $\theta_\odot$ takes values
in the regions quoted in Table 1.

 In the case of 
the 2+2A scheme 
under discussion, the 
effective neutrino mass
$m_{\nu_e}$, which can be 
determined from the measurement
of the end-point part of 
the $\beta$-spectrum of $^3$H,
is given by
\begin{equation}
m_{\nu_e} 
\simeq \sqrt{m_1^2 + \deltalsnd}
\label{mnu3H2A}
\end{equation}
%
\noindent 
From the results of the 
analysis of the LSND data \cite{LSND} 
it follows that
for $m_1 < 0.1$ eV one has:
\begin{equation}
0.4~{\rm eV} \leq m_{\nu_e} \leq 1.4~{\rm eV}.
\label{mnu3H2Aval}
\end{equation}
%
\noindent If $m_1 \gtap 0.1$ eV, the lower
bound in (\ref{mnu3H2Aval}), in particular, 
will be larger. Consequently, for the 2+2A type of 
neutrino mass spectrum and any value of $m_1$, 
$m_{\nu_e}$ is predicted to lie 
in the range planned to be
probed by the future
Karlsruhe-Mainz-Troitzk  
experiment KATRIN \cite{KATRIN}.
Thus, the realization of the KATRIN project 
will allow to check directly the possibility 
of 2+2A type of neutrino mass spectrum.
A measurement of $m_{\nu_e} \gtap 0.4$ eV and 
and a more accurate knowledge of \deltalsnd 
would permit to determine the value of
$m_1$. This would allow
to determine also the values of $m_{2,3,4}$
in the case of the 2+2A spectrum. 

  Using  equation~(\ref{effectivemass}) and 
neglecting $m_1|U_{\mathrm{e} \, 1}|^2$ 
and $m_2|U_{\mathrm{e} \, 2}|^2$ with respect 
to the terms $\sim m_3, m_4$, we can express
the effective mass \meff in terms of
$m_{\nu_e} \simeq \sqrt{m_1^2 + \deltalsnd}$, $\theta_\odot$ and 
the CP-violating phase difference 
$(\alpha_4 - \alpha_3) = (\alpha_{41} -\alpha_{31})$
\cite{BGGKP99}:
\begin{equation}
\meff \simeq m_{\nu_e} \sqrt{ 1 - \sin^2( 2 \theta_\odot) \ 
\sin^2 \frac{\alpha_{31} - \alpha_{41}}{2} }.
\label{meff22a}
\end{equation}
%
In eq.~(\ref{meff22a})
we have neglected 
the contribution of the two lightest neutrinos since
\begin{equation}
\left| m_1 |U_{\mathrm{e} 1}|^2 +
 m_2 |U_{\mathrm{e} 2}|^2 e^{i \alpha_{21}} \right| \leq
m_1 |U_{\mathrm{e} 1}|^2 +
 m_2 |U_{\mathrm{e} 2}|^2  
\ll m_4
\end{equation}

\noindent For $m_1$ satisfying eq. (\ref{m1}), 
this contribution does not exceed approximately 
$1.5 \times 10^{-3}$ eV.

  The effective Majorana mass \meff, eq. (\ref{meff22a}), 
depends on the value of $\theta_\odot$ and 
therefore the predictions for \meff 
will vary with the solution 
of the solar neutrino  problem.

 Consider first the case of 
$(\alpha_{41} - \alpha_{31}) \equiv \alpha_{43}$ taking
CP-conserving values. 
There are two different possibilities.

{\bf Case A.}  
If the neutrinos $\nu_3$ and  $\nu_4$  have the same CP parities
$\phi_3 = \phi_4$ (i.e. 
$\alpha_{41} = \alpha_{31} = 0, \pm \pi$),
the expression for 
\meff reduces to
\begin{equation}
 \meff \simeq m_{\nu_e} \simeq \sqrt{m_1^2 + \deltalsnd}. 
\end{equation}
\noindent 
In this case \meff coincides
(up to corrections which do not exceed
approximately $10^{-2}$ eV) with the mass 
$m_{\nu_e}$, measured in $^3$H $\beta-$decay
experiments. 
Using the 95\% C.L. results of 
the analysis \cite{LSND} 
one finds that 
for any value of $m_1$, 
\meff should satisfy:
\begin{equation}
\meff \geq 0.4 \eV. 
\label{22app}
\end{equation}
\noindent 
If $m_1$ is sufficiently small so that 
eq. (\ref{m1}) is fulfilled, 
one has also: $\meff \leq 1.4 \eV$.
Comparing the lower bound (\ref{22app}) 
with the upper bounds on \meff given by 
eqs. (\ref{76Ge00}) and (\ref{IGEX00})
one can conclude that an improvement
of these upper bounds by a factor $\sim (3 - 4)$
would essentially rule out the 
possibility of equal CP-parities of
$\nu_3$ and $\nu_4$ for the 2+2A 
neutrino mass spectrum.

{\bf Case B.}
If the  neutrinos $\nu_3$ and  $\nu_4$ 
 have  opposite CP parities
$\phi_3 = -  \phi_4$ (i.e. 
$\alpha_{41} = \alpha_{31} + \pi = 0, \pm \pi$),
  the effective Majorana mass \meff is given by:
\begin{equation}
 \meff \simeq \sqrt{m_1^2 + \deltalsnd} \, | \cos 2 \theta_\odot|. 
\end{equation}

In this case \meff depends  
on the value of $\cos 2 \theta_\odot$ 
and therefore on the particular 
solution of the solar neutrino problem
considered. If the allowed region 
for $\cos 2 \theta_\odot$ includes 
$\cos 2 \theta_\odot =0$, 
for instance, there would be no significant 
lower bound on \meff. 
More specifically, we have:
\begin{enumerate}
\item
For the LMA MSW solution, the allowed values of 
$\cos 2 \theta_\odot$ depend on the value
of $\cos^2 \beta$:
the  results of the analysis performed in
\cite{Gonza4nu} show that
$\cos 2 \theta_\odot > 0$  
both for $\cos^2 \beta= 0.3 $ at 90\% C.L.  
and for $\cos^2 \beta= 0.5 $ at 90\% and 99\% C.L., 
while in the solution region found  
for $\cos^2 \beta= 0.3 $ at  99\% C.L. 
one has $|\cos 2 \theta_\odot| \geq 0$. 
Using the relevant 90\% (99\%) C.L. 
results for $\theta_\odot$ 
one gets for $m_1 < 0.1$ eV:

\begin{align}
 0.09~(0.0) \eV \leq & \meff \leq 0.7 \ (0.8) \eV,  & 
 \qquad \cos^2 \beta= 0.3;\\
 0.13~(0.02) \eV \leq & \meff \leq 0.46 \ (0.67) \eV,  & 
 \qquad \cos^2 \beta= 0.5 .
\end{align}
Note that the 90\% C.L. allowed values of \meff
lie entirely in the region which will be probed in the
currently running and next generation of 
\betabeta-decay experiments \cite{NEMO3,CUORE,EXO,GENIUS}.
For $0.1~{\rm eV} \leq m_1 \ltap 1.0$ eV,
both the non-zero lower bounds 
and the upper bounds increase by factors
1.3 and 1.2, respectively. 
A better determination of the value
of the solar neutrino mixing
angle $\theta_\odot$ would be very 
important for limiting
further the range of possible values of \meff.

\item
For the SMA MSW solution
we have $\cos 2 \theta_\odot \simeq 1$ and
this case of CP-conservation cannot be 
distinguished from the earlier considered one, 
$\phi_3 = \phi_4$, or from the case of violation of the 
CP-symmetry due to $\alpha_{43}$: in all these cases we get
\begin{equation}
 \meff \simeq m_{\nu_e} \simeq \sqrt{m_1^2 + \deltalsnd}
\end{equation}
%
\noindent and, correspondingly, 
the same allowed values of \meff
as in eq.~(\ref{22app}).
Obviously, if the neutrino mass 
spectrum is of the 2+2A type and
the SMA MSW solution 
(or Case A) is realized,
the measurement of  
$\meff \gtap 0.4$ eV, or of 
$m_{\nu_e}\gtap 0.4$ eV,
and of \deltalsnd would allow to
determine the lightest 
neutrino mass $m_1$, 
and, correspondingly, the neutrino 
mass spectrum. 

\item
For the LOW-QVO solution, the 
$90\%$ C.L. and $99\%$ C.L.
results of the analysis
of ref.~\cite{Gonza4nu} do not exclude 
the possibility of having $\meff \cong 0$ eV. 
Noting that for $\cos^2 \beta = 0.5$, 
the LOW-QVO solution of the solar neutrino problem
is excluded at 90\% C.L. and is allowed at 
99\% C.L., we find using the 90\% (99\%) C.L.
allowed values of 
$|\cos 2 \theta_\odot|$
and 95\% C.L. values of \deltalsnd:
\begin{align}
 0.08~(0.0) \eV \leq & \meff \leq 0.28 \ (0.66) \eV,  & 
 \qquad \cos^2 \beta= 0.3;\\
(0.02  \eV ) \leq & \meff \leq  (0.52 \eV)  & 
 \qquad \cos^2 \beta= 0.5 .
\end{align}
\noindent These results 
correspond again to $m_1 < 0.1$ eV.
If $0.1~{\rm eV} \leq m_1 \ltap 1.0$ eV,
both the non-zero lower bounds 
and the upper bounds increase by factors
1.3 and 1.2, respectively.

\end{enumerate}

 In Fig.~\ref{figure:2+2A01} and 
in Fig.~\ref{figure:2+2A02} 
we show the allowed values of \meff
for the LMA and LOW-QVO solutions 
of the solar neutrino problem, respectively,
as a function of $\sqrt{\deltalsnd}$
for the two cases of CP-conservation
and in the case of CP-violation.
The allowed region corresponding to CP-violation
not only covers the two regions
of CP-conservation, but extends also between 
the latter. Thus, there exists a region, 
marked by dark-grey color in each of the two figures, 
which can be spanned {\it only
if the CP-parity is not conserved}, 
the ``just-CP-violation'' region: 
if the neutrino mass spectrum is 
of the 2+2A type,
an experimental point in this region 
would signal 
CP-violation in the lepton sector.
We have already noticed 
that for the SMA solution,
the prediction for \meff
in the two CP-conserving cases and 
the CP-violating one practically 
coincide since $\sin^22\theta_\odot < 10^{-2}$.

  For all solutions of the solar neutrino problem 
we have $\meff \leq 1.4 \eV$ 
if $m_1 < 0.1$ eV. 
This maximal value is already excluded by the most 
stringent upper bounds \cite{76Ge00,IGEX00} on \meff, 
eqs. (\ref{76Ge00}) and (\ref{IGEX00}).
In most of the allowed regions of the 
relevant parameter
space one has $\meff \geq 0.01 \div 0.02 \eV$,
which can be tested in the \betabeta-decay
experiments of the next generation.

  As the predictions for \meff for 
the LMA and the LOW-QVO solutions and 
the 2+2A neutrino mass spectrum
under study depend crucially
on the value of $\cos 2 \theta_\odot$, 
we show in Fig.~\ref{figure:2+2A03}
$\meff$ as function of $\cos 2 \theta_\odot$.
The ``just-CP-violation'' region is marked 
by dark-grey color. The uncertainties 
in the predicted values of \meff 
in the two cases of CP-conservation are 
mainly due to the uncertainty 
in the value of \deltalsnd. 
A better determination of 
\deltalsnd in the upcoming
experiment MiniBooNE would lead to 
a ``contraction'' of the 
CP-conservation regions. 
Once \deltalsnd and 
$\cos 2 \theta_\odot$ are known 
with sufficient accuracy
and if $\cos 2 \theta_\odot$ lies in the
region of the LMA or LOW-QVO solution,
it might be possible
to establish whether the CP-symmetry
is violated and to determine the value 
of one of the Majorana CP-violating phases,
$(\alpha_{41} - \alpha_{31})$ by measuring 
$m_{\nu_e}\gtap 0.4$ eV, and $\meff \neq 0$.
Indeed, through equation~(\ref{meff22a}), 
the value of the 
phase $(\alpha_{41} - \alpha_{31})$  
can be related to the 
experimentally measurable quantities 
$\meff^2$, $m^2_{\nu_e} \simeq (m_1^2 + \deltalsnd)$
and $\theta_\odot$ \cite{BGGKP99}:
\begin{equation}
\sin^2 \frac{\alpha_{41} - \alpha_{31}}{2} \simeq
  \frac{1}{\sin^2 2 \theta_\odot} 
\left( 1 - \frac{\meff^2}{m_1^2 + \deltalsnd} \right) \simeq
  \frac{1}{\sin^2 2 \theta_\odot} 
\left( 1 - \frac{\meff^2}{m_{\nu_e}^2} \right).
\end{equation}

In Fig.~\ref{figure:2+2A04} 
we show the allowed ranges 
of $\sin^2 ((\alpha_{41} - \alpha_{31})/2)$ 
as a function of \meff
in the case $m_1^2 \ll \deltalsnd$,
for the LMA and LOW-QVO
solutions of the solar neutrino problem
(we remind the reader that in the case
of the SMA MSW solution the value of \meff
practically does not depend on whether
CP-invariance holds or not).
For a given \meff, the present uncertainties 
in the value of $\sin^2 (\alpha_{41} - 
\alpha_{31})/2$, are determined by 
the uncertainties in the knowledge of 
\deltalsnd and $\sin^2 2 \theta_\odot$.
As the latter will diminish,
$\sin^2 (\alpha_{41} - 
\alpha_{31})/2$, will be restricted to 
lie in a  smaller region and 
the measurement of \meff
can fix the value of 
$\sin^2 (\alpha_{41} - \alpha_{31})/2$
with some uncertainty. Obviously, 
this would be possible 
if the neutrino mass spectrum is
of the 2+2A type. We would like to emphasize 
that in the case of the 2+2A spectrum,
the Majorana CP-violating phases 
$\alpha_{21}$ and $\alpha_{31}$ cannot be 
constrained by the
measurement of $\meff \gtap 10^{-2}$ eV.
These two phases can be a source of CP-violation
even if one finds that 
$(\alpha_{41} - \alpha_{31}) = 0, \pm \pi$.


\section{\textbf{ The   2+2B  neutrino mass spectrum}}
\label{2+2B}

The 2+2B neutrino mass spectrum 
can be represented graphically as:

\unitlength=1mm
\begin{center}
\begin{picture}(100,40)
\put(20,5){\line(1,0){20}}
\put(20,7){\line(1,0){20}}
\put(20,30){\line(1,0){20}}
\put(20,35){\line(1,0){20}}
\put(30,5){\line(0,1){30}}
\put(45,4.5){\vector(0,1){3}}
\put(45,7.5){\vector(0,-1){3}}
\put(45,30){\vector(0,1){5}}
\put(45,35){\vector(0,-1){5}}
\put(72,5){\vector(0,1){30}}
\put(72,35){\vector(0,-1){30}}
\put(48,6){\deltasol}
\put(48,32){\deltaatm}
\put(75,18){\deltalsnd}
\put(17,4){1}
\put(17,7){2}
\put(17,30){3}
\put(17,34.2){4}
\end{picture}
\end{center}
\noindent 
It can be defined through the inequalities 
$m_1 \simeq m_2 <~(\ll)~ m_3 < m_4$.
This spectrum can also be characterized 
by the following conditions:
\begin{gather}
m_1 <~(\ll)~ \sqrt{\Delta m^2_{41}}, \nonumber \\
\sqrt{\Delta m^2_{21}} \ll \sqrt{\Delta m^2_{43}} 
\ll \sqrt{\Delta m^2_{41}}. 
\label{22Bspectrum}
\end{gather}
%
%
On the basis of (\ref{22Bspectrum}),
one can make the identification:
%
\begin{gather}
\Delta m^2_{21}     \equiv    \Delta  m^2_{\odot}, \ \ 
\Delta m^2_{43}   \equiv    \Delta  m^2_{\mathrm{atm}}, \ \ 
\Delta m^2_{41}    \equiv    \Delta  m^2_{\mathrm{SLB}}.
\label{22bpattern}  
\end{gather}
\noindent
From (\ref{22Bspectrum}) and (\ref{22bpattern}) we get:
\begin{align}
m_2 & = \sqrt{m_1^2 + \deltasol}, \\
m_3 & \simeq m_4 = \sqrt{m_1^2 + \deltalsnd},
\end{align}
where \deltasol takes values in the intervals
given in Table 1 and the values \deltalsnd 
can have are given in eq.~(\ref{mlsnd}).
In the expression for 
$m_3 = \sqrt{m_1^2 + \deltalsnd - \deltaatm}$
we have neglected terms 
$\sim \deltaatm/(m_1^2 + \deltalsnd) \ll 1$
and in this approximation 
\meff does not depend on \deltaatm.
The contribution of the indicated terms 
in \meff, however, does not exceed
$10^{-4}$ eV. This follows from 
eqs. (\ref{mlsnd}) and (9)-(10) and the fact that 
$|U_{\mathrm{e} \, 3}|^2 \leq 1.0 \times 10^{-2}$
(see further). One will have
$m_1 \ll \sqrt{\deltalsnd}$ if
\begin{equation}
m_1 < 0.25 \sqrt{\deltalsnd} \Rightarrow m_1 < 0.1 \eV.
\label{m122B}
\end{equation}

 We find for the elements 
of interest of the lepton 
mixing matrix, $U_{ej}$: 
\begin{itemize}
\item[i)]
$|U_{\mathrm{e} \, 1}|$ and 
$|U_{\mathrm{e} \, 2}|$ are related
to the solar neutrino 
mixing angle $\theta_{\odot}$:
\begin{align}
|U_{\mathrm{e} \, 1}|^2 & = \cos^2 \theta_\odot \ 
 ( 1 - \sum_{i=3}^4 |U_{\mathrm{e} \, i}|^2 ), \\
|U_{\mathrm{e} \, 2}|^2 & = \sin^2 \theta_\odot \ 
 ( 1 - \sum_{i=3}^4 |U_{\mathrm{e} \, i}|^2 ), 
\label{2BUe12}
\end{align}
%
where $\tan^2\theta_\odot$ takes values 
in the regions reported in Table 1;
\item[ii)]
$|U_{\mathrm{e} \, 3}|$ and $|U_{\mathrm{e} \, 4}|$
are constrained by the neutrino oscillation data from 
the short baseline (SBL) experiments LSND and BUGEY
\cite{LSND,BUGEY}:
\begin{equation}
2.0 \times 10^{-4} \leq |U_{\mathrm{e} \, 3}|^2 + 
|U_{\mathrm{e} \, 4}|^2 \leq 1.0 \times 10^{-2}.
\label{Ue342B}
\end{equation}
\end{itemize}

 Using eq.~(\ref{effectivemass}) and 
neglecting  \ $|U_{\mathrm{e} 3}|^2 +|U_{\mathrm{e} 4}|^2$  \ 
with respect to 1 in eqs. (63) - (\ref{2BUe12}),
we obtain for the effective Majorana mass \meff:
\begin{equation}
\begin{split}
\meff \simeq &  | m_1 \cos^2 \theta_\odot 
  +  \sqrt{m_1^2 + \deltasol} 
 \sin^2 \theta_\odot  e^{i \alpha_{21}} \\ 
\mbox{} &  + \ \sqrt{m_1^2 + \deltalsnd} 
( |U_{\mathrm{e} 3}|^2 
e^{i \alpha_{31}} + |U_{\mathrm{e} 4}|^2 e^{i \alpha_{41}})  |
\end{split}
\label{meff22b}
\end{equation}
\noindent where $\alpha_{21}$, $\alpha_{31}$ 
and $\alpha_{41}$ are 
three CP-violating phases.

 We will study next the magnitude of 
the contributions given 
by each of the four terms present 
in the r.h.s. of equation~(\ref{meff22b}).
Using the allowed values of  
$\deltalsnd$ found in ref.~\cite{LSND} at $95\%$ C.L.
as well as the values 
$\tan^2\theta_\odot$ and \deltasol obtained
in ref.~\cite{Gonza4nu} at $90\%$ ($99\%$) C.L. 
for $\cos^2 \beta = 0.3$ [$\cos^2 \beta = 0.5$]
and taking into account the constraint 
(\ref{Ue342B}), one finds if $m_1 < 0.1$ eV: 

\begin{align}
0  \leq |<m>|_1 \equiv  m_1 \cos^2 \theta_\odot  &  \leq \begin{cases}
                      7.5 \ (8.0) 
                      [6.7 \ (7.4)] \times 10^{-2} \eV,  & \text{LMA MSW}; \\
                      0.1  \ (0.1)
                      [ 0.1  \ (0.1)]  \eV,   & \text{SMA MSW}; \\
                      6.0 \ (6.9) 
                      [ \  (6.5)] \times 10^{-2}    \eV,  & \text{LOW-QVO};
                                           \end{cases} \\ 
 |<m>|_2 \equiv \sqrt{m_1^2 + \deltasol}~ \sin^2 \theta_\odot  &  \leq 
                                           \begin{cases}
                      3.9 \ (5.5) 
                      [3.3 \ (3.3)] \times 10^{-2}  \eV,  & \text{LMA MSW}; \\
                      1.0 \ (1.5) 
                      [ 1.0 \ (1.5)] \times 10^{-4} \eV,   & \text{SMA MSW}; \\
                      4.0 \ (7.4) 
                      [ \ (6.9)] \times 10^{-2}   \eV,   & \text{LOW-QVO};
                                           \end{cases} \\ 
 |<m>|_2 \equiv \sqrt{m_1^2 + \deltasol}~ \sin^2 \theta_\odot    & \geq 
                                        \begin{cases}
                      10 \ (7.0) 
                      [1.3 \  (1.2)] \times 10^{-4}  \eV,  & \text{LMA MSW}; \\
                      8.0 \ (4.0) 
                      [ 8.0 \ (4.0)] \times 10^{-7} \eV,   & \text{SMA MSW}; \\
                      13 \ (0.8) 
                      [ \ (0.9)] \times 10^{-5} \eV,  & \text{LOW-QVO};
                                           \end{cases} \\ 
 0 \leq |<m>|_3 \equiv   \sqrt{m_1^2 + \deltalsnd} |U_{\mathrm{e} 3}|^2   &  \leq 5.0 \times 10^{-3} \eV; \\ 
 0 \leq |<m>|_4 \equiv  \sqrt{m_1^2 + \deltalsnd} |U_{\mathrm{e} 4}|^2    &  \leq 5.0 \times 10^{-3} \eV. 
\end{align}
\noindent
 The upper bounds in eqs. (67), (68) and (70 (or (71))
will increase approximately by factors of 10, 3 and 2.2, respectively,
if $m_1 \simeq 1.0$ eV, while the lower bounds in eq. (69)
will be larger by a factor of $\sim 10$.

 This analysis shows that all contributions in \meff, 
eq. (\ref{meff22b}), can be of the same order. 
Thus, in the case 
under study all three Majorana CP-violating phases 
$\alpha_{21}$, $\alpha_{31}$ and $\alpha_{41}$, 
are relevant and must be taken into account in order 
to establish the allowed range for \meff, 
and, in particular, to answer the question 
regarding the possibility 
of cancellations between the different contributions.

  Let us note that $|<m>|_3$ and $|<m>|_4$ 
are not independent since
$ 2.0 \times 10^{-4} \leq |U_{\mathrm{e} 3}|^2 + |U_{\mathrm{e} 4}|^2 
\leq 1.0 \times 10^{-2}. $ \
As a consequence, at least one of 
the two contributions 
is of the order of $10^{-3} \eV$ and
they cannot both be smaller than $10^{-4} \eV$.

  The terms $|<m>|_1$ and  $|<m>|_2$ depend strongly on 
the values of $m_1$ and $\theta_\odot$. 
More specifically we have:
\begin{enumerate}
\item
for $m_1 \leq 10^{-5} \eV$ and the LMA solution 
of the solar neutrino problem, 
$|<m>|_1$ is negligible with respect to 
$|<m>|_2$. Indeed, one has $|<m>|_1 \leq 0.05 \ |<m>|_2$, 
while $|<m>|_2$ can be of the same order as
$|<m>|_3$ and $|<m>|_4$, i.e., $|<m>|_2 \leq 10^{-2} \eV$. 
The expression for \meff reduces to the 
sum of three terms with two 
relative phases $(\alpha_{41} - \alpha_{21})$ 
and $(\alpha_{31} - \alpha_{21})$:
\begin{equation}
\meff \simeq   | \sqrt{\deltasol} 
 \sin^2 \theta_\odot  
  + \sqrt{\deltalsnd} 
(|U_{\mathrm{e} 3}|^2 
e^{i \alpha_{31}} + |U_{\mathrm{e} 4}|^2 e^{i \alpha_{41}})
e^{-i \alpha_{21}} |
\label{meff22b1LMA}
\end{equation}
%
\noindent For ${\rm max}(\deltasol) = 
2.0~(7.0)\times 10^{-4}~{\rm eV^2}$,
we have $\sqrt{\deltasol} \sin^2 \theta_\odot \ltap
0.7~(1.3)\times 10^{-2}~{\rm eV}$.
Since, as can be shown, the contribution of the 
term $\sim \sqrt{\deltalsnd}$ does not exceed 
$5.0\times 10^{-3}~{\rm eV}$, 
for the maximal value of \meff one finds:
$\meff \ltap 1.2~(1.8)\times 10^{-2}~{\rm eV}$.

In the cases of the LOW-QVO and SMA solutions, 
both $|<m>|_1$ and  $|<m>|_2$ are negligible 
with respect to $|<m>|_3$ and $|<m>|_4$ since
$|<m>|_1,|<m>|_2 \leq 10^{-4} \eV$. 
In these cases only one relative CP-violating 
Majorana phase, $(\alpha_{41} - \alpha_{31})$,  
has effectively to be taken into account: 
\begin{equation}
\meff \simeq  |\sqrt{\deltalsnd} 
(|U_{\mathrm{e} 3}|^2 
e^{i \alpha_{31}} + |U_{\mathrm{e} 4}|^2 e^{i \alpha_{41}})|.
\label{meff22b1LOW}
\end{equation}
%
\noindent However, one has 
$\meff \ltap 5.0\times 10^{-3}~{\rm eV}$.
\item
For $10^{-5} \eV \ll m_1 \ll 10^{-1} \eV$ 
all terms in eq. (\ref{meff22b}) can be of
the same order 
and can give nonnegligible contributions 
to the effective Majorana mass \meff. 
The three CP-violating phases $\alpha_{21}$, 
$\alpha_{31}$ and $\alpha_{41}$, have 
to be taken into account. Only in the 
case of the SMA solution, 
$|<m>|_2$ is negligible 
with respect to the other 
three terms.
\item
If $m_1 \gtap 10^{-1} \eV$ and for the 
LMA and LOW-QVO solutions, 
$|<m>|_1$ and $|<m>|_2$ 
give the dominant contributions 
in \meff, the contribution of 
$|<m>|_3$ and $|<m>|_4$ being 
at least by an order 
of magnitude smaller. 
Thus, up to corrections 
$\sim 5.0\times 10^{-3}~{\rm eV}$,
only the CP-violating phase 
$\alpha_{21}$ is effectively relevant for 
the determination of \meff, when 
$\meff \gtap 2\times 10^{-2}$ eV:
\begin{equation}
\meff \gtap 2\times 10^{-2}~ {\rm eV}:~~~~
\meff \simeq m_1| \cos^2 \theta_\odot 
  + \sin^2 \theta_\odot e^{i \alpha_{21}} |.~~~~~~~~~~~ 
\label{meff22b31}
\end{equation}
\noindent
Correspondingly, one finds:
\begin{equation}
\meff \gtap 2\times 10^{-2}~ {\rm eV}:~~~~
m_1| \cos 2\theta_\odot| \leq \meff \leq m_1.~~~~~~~~~~~~~~
\label{meff22b32}
\end{equation}
%
\noindent
In the case of the SMA solution, $|<m>|_2$, $|<m>|_3$ 
and $|<m>|_4$ are negligible in comparison with 
$|<m>|_1$ and we have:
\begin{equation}
\meff \cong m_1\cos^2\theta_{\odot} \cong m_1.
\end{equation}
\end{enumerate}

In the future, with a better determination 
of the values of the relevant 
parameters, the allowed ranges for 
the different contributions will be considerably
restricted and the analysis of the various 
possibilities will be simplified.

 In the case of CP-invariance 
the massive Majorana neutrinos 
can have different CP-parities and 
it is possible to have cancellations 
between the different terms in eq. (\ref{meff22b}).
Cancellations can occur also if 
the CP-invariance does not hold.
Thus, in the case of the 2+2B neutrino mass 
spectrum under discussion
one cannot obtain at present 
non-trivial lower bounds on \meff.
Majorana CP-violating phases
can be present in \meff, but this
might not be recognizable from the 
measurement of $\meff \neq 0$
unless there exists a hierarchy between
the different contributions in \meff. 

 Furthermore, in the case of the 2+2B neutrino 
mass spectrum all the contributions tend to be 
relatively small. Therefore the maximal
possible  values of \meff are of particular 
interest in view of the planned sensitivity of the
the next generation of 
\betabeta-decay experiments.
These maximal values depend, obviously,
on the solution of the solar neutrino problem.
They can be found by assuming 
that CP-invariance holds and 
by taking the CP parities 
of all four massive Majorana 
neutrinos to be equal, and are reported 
in Fig.~\ref{figure:2+2B02} as a function of $m_1$.
For $m_1$ negligible (e.g., a very light sterile neutrino)
and $\cos^2 \beta = 0.3$ [$\cos^2 \beta = 0.5$], we get:
\begin{equation}
\meff \leq   \begin{cases}
           8.9 \  (10)~[8.8 \ (8.8)] \times 10^{-3} \eV, & 
\text{LMA MSW solution at $90 \%$ ($99\%$) C.L.;} \\
           5.0~[5.0]\times 10^{-3} \eV & 
\text{SMA MSW solution}; \\ 
           5.0~[5.0]\times 10^{-3} \eV & 
\text{LOW-QVO solution}. 
           \end{cases}
\label{maxmeff22b}
\end{equation}
%
%
As $m_1$ increases, 
the maximal values of \meff also increase,
reaching for $m_1 \simeq 0.1 \eV$
\begin{equation}
\meff \leq 0.10 \eV
\end{equation}
for all solutions of the solar neutrino problem. 
For $m_1 \gtap 0.1 \eV$ we have 
${\rm max}(\meff) \simeq m_1$.
One can have $\meff \gtap 3.0\times 10^{-2}$ eV
for the 2+2B type of neutrino 
mass spectrum only if $m_1 \gtap 10^{-2}$ eV.
Values of $\meff \gtap 10^{-2}$ eV are in the region
of the planned sensitivity of the  
next generation of \betabeta-decay experiments.
As it follows from Fig.~\ref{figure:2+2B02},
a measured value of $\meff \gtap 0.03$ eV in the case
of the neutrino mass spectrum under discussion
would imply a lower limit on the mass $m_1$,
$\meff \gtap 10^{-2}$ eV;
and it would essentially determine the value of 
$m_1$ if the SMA MSW solution of the solar neutrino
problem turns out to be the correct one.


\section{\textbf{ The 3+1A  mass spectrum}}
\label{2+2B}

The 3+1A mass spectrum  is characterized
by three nearly degenerate neutrinos, $\nu_{2,3,4}$,
having masses sufficiently larger than the fourth one, $\nu_1$:
$m_1 <~(\ll)~ m_2,m_3,m_4$. There exist two possibilities  
which can be presented schematically as follows:
\unitlength=1mm
\begin{center}
\begin{picture}(200,40)
\put(10,5){\line(1,0){20}}
\put(10,28){\line(1,0){20}}
\put(10,33){\line(1,0){20}}
\put(10,35){\line(1,0){20}}
\put(20,5){\line(0,1){30}}
\put(33,27.5){\vector(0,1){5.5}}
\put(33,33){\vector(0,-1){5.5}}
\put(33,33){\vector(0,1){2.5}}
\put(33,35.5){\vector(0,-1){2.5}}
\put(48,5){\vector(0,1){30}}
\put(48,35){\vector(0,-1){30}}
\put(34.5,34){\deltasol}
\put(34.5,28.5){\deltaatm}
\put(49.5,15){\deltalsnd}
\put(7,4){1}
\put(7,27){2}
\put(7,31.5){3}
\put(7,34.6){4}
\put(17,0){3+1A-i case}
\put(70,18){\mbox{or}}
\put(85,5){\line(1,0){20}}
\put(85,28){\line(1,0){20}}
\put(85,30){\line(1,0){20}}
\put(85,35){\line(1,0){20}}
\put(95,5){\line(0,1){30}}
\put(108,27.5){\vector(0,1){2.5}}
\put(108,30){\vector(0,-1){2.5}}
\put(108,30){\vector(0,1){5.5}}
\put(108,35.5){\vector(0,-1){5.5}}
\put(123,5){\vector(0,1){30}}
\put(123,35){\vector(0,-1){30}}
\put(110,28){\deltasol}
\put(110,33){\deltaatm}
\put(125,15){\deltalsnd}
\put(82,4){1}
\put(82,26.5){2}
\put(82,30){3}
\put(82,34.2){4}
\put(92,0){3+1A-ii case}
\end{picture}
\end{center}

\noindent
One has $m_1 \ll m_4$,
if $m_1 < 0.1$ eV.
These patterns of neutrino masses can also be 
characterized by the inequalities:
\begin{align}
\Delta m^2_{43} \ll \Delta m^2_{32} \ll \Delta m^2_{41}, & 
 \qquad \text{3+1A-i case};
\label{31Aispectrum} \\
\Delta m^2_{32} \ll \Delta m^2_{43} \ll \Delta m^2_{41}, & 
 \qquad \text{3+1A-ii case}. 
\label{31Aiispectrum}
\end{align}
\noindent
Using relation~(\ref{31Aispectrum}) in the 3+1A-i case,
one can make the identification:
\begin{equation}
\Delta m^2_{43}  
         \equiv    \Delta  m^2_{\odot}, \ \ 
\Delta m^2_{32}   
               \equiv    \Delta  m^2_{\mathrm{atm}}, \ \ 
\Delta m^2_{41}  
  \equiv  \Delta  m^2_{\mathrm{SBL}}. 
\label{31apattern}  
\end{equation}
%
The corresponding constraints on the 
elements $|U_{\mathrm{e} \, j}|$ of the 
neutrino mixing matrix $U$ read:
\begin{itemize}
\item[i)]
$|U_{\mathrm{e} \, 1}|$ is limited by the data of the
SBL experiments \cite{LSND,BUGEY}: 
$2.0 \times 10^{-4} \leq |U_{\mathrm{e} \, 1}|^2< 1.0 
\times 10^{-2}$;
\item[ii)]
$|U_{\mathrm{e} \, 2}|$ should satisfy the CHOOZ 
limit \cite{Gonza3nu,CHOOZ}: $|U_{\mathrm{e} \, 2}|^2 <0.08$
(99\% C.L.);
\item[iii)]
$|U_{\mathrm{e} \, 3}|$ and $|U_{\mathrm{e} \, 4}|$ 
are related to the solar neutrino mixing angle $\theta_{\odot}$:
\begin{align}
|U_{\mathrm{e} \, 3}|^2 & = \cos^2 \theta_\odot  \ ( 1 - \sum_{i=1}^2 |U_{\mathrm{e} \, i}|^2 ), \\
|U_{\mathrm{e} \, 4}|^2 & = \sin^2 \theta_\odot \ ( 1 - \sum_{i=1}^2 |U_{\mathrm{e} \, i}|^2 ). 
\end{align}
\end{itemize}

  Using 
eq. (\ref{31apattern})
one finds :
\begin{equation}
m_2 = \sqrt{m_1^2+\deltalsnd-\deltasol-\deltaatm},~
m_3 = \sqrt{m_1^2+\deltalsnd-\deltasol},~  
m_4 = \sqrt{m_1^2+\deltalsnd}.
\label{m23431A0}
\end{equation}
\noindent 
Neglecting terms $\sim \deltasol/\deltalsnd$ and
$\sim \deltaatm/\deltalsnd$ in $m_{2,3}$, whose 
contributions in \meff do not exceed 
respectively $10^{-3}$ eV and $1.5\times 10^{-3}$ eV,
we get:
\begin{equation}
m_2 \simeq m_3  \simeq m_4 = \sqrt{m_1^2 + \deltalsnd}.
\label{m23431A1}
\end{equation}

  The 3+1A-ii neutrino mass spectrum 
can be obtained from the 3+1A-i one 
by requiring that $\Delta m^2_{43} \equiv \deltaatm$,
$\Delta m^2_{32} \equiv \deltasol$ 
and by interchanging $|U_{\mathrm{e} 2}|^2$ 
and $|U_{\mathrm{e} 4}|^2$.
However, in the approximation (\ref{m23431A1}) 
we are working (i.e., neglecting terms smaller than
$\sim 10^{-3}$ eV in \meff), \deltaatm and \deltasol 
do not enter into the expression for \meff. 
Since in this approximation $m_2 \simeq m_3 \simeq m_4$,
we have also the freedom to interchange 
$|U_{\mathrm{e} 2}|^2$ and $|U_{\mathrm{e} 4}|^2$,
obtaining the expression for \meff in the case of the 
3+1A-i spectrum.
Consequently, as far as the approximation
(\ref{m23431A1}) holds, i.e., up to corrections
$\sim 10^{-3}$ eV,
the 3+1A-i and the 3+1A-ii patterns
of neutrino masses lead to the  
same predictions for 
the effective Majorana mass \meff. 
In what follows we will consider the 
3+1A-i spectrum only and will refer 
generically to the two spectra as the 
3+1A spectrum.

  In the 3+1A schemes under discussion, 
as like in the 2+2A scheme, the 
neutrino mass $m_{\nu_e}$
measured in the $^3$H $\beta-$decay 
experiments is given, up to correction which do 
not exceed $\sim 10^{-2}$ eV,
by:
\begin{equation}
m_{\nu_e} 
\simeq \sqrt{m_1^2 + \deltalsnd}.
\label{mnu3H2A}
\end{equation}
%
\noindent 
From the results of the 
analysis of the LSND data \cite{LSND} 
it follows that
$m_{\nu_e} \gtap 0.4~{\rm eV}$
for any value of $m_1$~
\footnote{That $m_{\nu_e}$
is not necessarily small in certain
3+1 schemes was noticed also in 
\cite{Vissani01}.}.
Thus, the possibility that 
the neutrino mass spectrum is of 
the 3+1A type can also be tested 
in the KATRIN $^3$H $\beta-$decay 
experiment \cite{KATRIN}.  
The observation made for the case of the
2+2A spectrum,
that a measurement 
of $m_{\nu_e} \gtap 0.4$ eV and 
a more precise knowledge of \deltalsnd 
would permit to determine the value of
$m_1$ and would allow
to fix the values of $m_{2,3,4}$ as well,
is valid also for the 
3+1A spectrum. 

  For  $m_1 < 0.1$ eV we have 
$m_1|U_{\mathrm{e} 1}|^2 < 10^{-3}$ eV and in
the approximation we have adopted this term 
in \meff can be neglected. 
If $m_1 \simeq 1.0$ eV, the 
contribution of the
neglected term in \meff would not 
exceed $10^{-2}$ eV.
Since $|U_{\mathrm{e} 1}|^2 \leq 10^{-2}$,
one can neglect also 
$|U_{\mathrm{e} 1}|^2$ in comparison with
$(1 - |U_{\mathrm{e} 2}|^2) > 0.91$.
Correspondingly, the expression for \meff
reduces to:  
\begin{equation}
\begin{split}
\meff \simeq  \sqrt{m_1^2 + \deltalsnd} |~|U_{\mathrm{e} 2}|^2 +
( 1 - |U_{\mathrm{e} 2}|^2) ( \cos^2 \theta_\odot e^{i (\alpha_3- \alpha_2)}  
+  \sin^2 \theta_\odot  e^{i (\alpha_4 - \alpha_2)} ) | .
\end{split}
\label{31ameff}
\end{equation}
%
\noindent
Thus, in the case of 3+1A neutrino 
mass spectrum under study and up to terms which 
are not larger than $\sim 10^{-3}$ eV and/or 
$m_1|U_{\mathrm{e} 1}|^2$, 
\meff depends on 
$\sqrt{m_1^2 + \deltalsnd}$, $\theta_\odot$ and $|U_{\mathrm{e} 2}|^2$ 
and on two Majorana CP-violating phases
(or phase differences) 
$(\alpha_{3} -\alpha_{2}) = 
(\alpha_{31} -\alpha_{21}) \equiv \alpha_{32}$ 
and $(\alpha_{4} -\alpha_{2}) = (\alpha_{41} -\alpha_{21})
\equiv \alpha_{42}$.

   It follows from eq. (\ref{31ameff}) that the expression for
\meff for the mass spectrum under discussion
is very similar to the expression for \meff
in the case of the three quasi-degenerate neutrinos, 
which has been analyzed recently in detail in ref.~\cite{BPP1}
(see Section 6 in \cite{BPP1}).
More specifically,
\begin{itemize}
\item[i)]
the common neutrino mass $m$, which appears 
as an overall mass scale in \meff in the case of
three quasi-degenerate Majorana neutrinos (see, e.g., \cite{BPP1}), 
corresponds to $\sqrt{m_1^2 + \deltalsnd}$ in eq.  
(\ref{31ameff}). 
Thus, we can make the identification 
$m \equiv \sqrt{m_1^2 + \deltalsnd}$, 
%
where $m$ in the 3-neutrino mixing analysis 
can vary in the interval \cite{BPP1}
 $0.1 \eV \leq m \leq 2.5 \eV$;
\item[ii)]
in both cases the expression for 
\meff contains two 
Majorana CP-violating phases: 
$\alpha_{32} \equiv (\alpha_{31} - \alpha_{21})$ 
and $\alpha_{42} \equiv (\alpha_{41}- \alpha_{21})$ 
are present in the 4-neutrino mixing
expression (\ref{31ameff}),
while $\alpha_{31}$ and $\alpha_{21}$ 
are present in the corresponding 
3-neutrino mixing one. 
As was shown in ref.~\cite{BPP1}, 
it would be possible, in principle, 
to constrain the CP-violating phases $\alpha_{31}$ 
and $\alpha_{21}$ if the values of the 
relevant parameters
entering into the 3-$\nu$ expression for \meff, 
$m$, $|U_{\mathrm{e} 1}|$ and $\theta_\odot$,
as well as the value of \meff, were known
with sufficient accuracy 
(see eq. (82) and Fig. 16 in \cite{BPP1}). 
The Majorana phases $\alpha_{31}$ 
and $\alpha_{21}$ are the only 
possible source
of CP-violation effects in \meff 
in the case of 3-neutrino mixing.
If 4-neutrino mixing takes place and the
neutrino mass spectrum is of the 3+1A type,
it would still be possible, in principle,
to constrain two of the physical Majorana 
CP-violating phases,  
but not the third one, $\alpha_{21}$, 
which can be an additional source of CP-violation.
\end{itemize}

  Further, we have $|U_{\mathrm{e} 2}|^2 < 0.08$ 
according to the (99\% C.L.) results
of ref.~\cite{Gonza3nu}. 
Therefore, the terms containing $|U_{\mathrm{e} 2}|^2$ 
as a factor in eq. (\ref{31ameff})
give non-dominant contribution to \meff
and the results for \meff in the case of the
the 3+1A spectrum are quite similar to those
derived for the 2+2A spectrum. Actually,
the expressions for \meff in the two cases 
coincide in the limit of $|U_{\mathrm{e} 2}|^2 =0$. 
For $|U_{\mathrm{e} 2}|^2 =0$ the difference in 
the predictions for \meff
arises only due to the fact
that, for the 2+2A neutrino mass spectrum
one has to use the results of 
a 4-neutrino mixing analysis of the 
relevant data, while 
if the spectrum is of the 3+1A type,
the results of the 3-neutrino mixing 
analysis of the atmospheric, 
solar and CHOOZ neutrino 
oscillation data should be 
utilized, as is discussed in Section 2.

 We will treat the 3+1A-i case in what follows
taking into account the $|U_{\mathrm{e} 2}|^2 \neq 0$
terms in \meff, i.e., using expression
(\ref{31ameff}) for \meff.
First we will consider the CP-conserving cases.

{\bf Case A.}
If the neutrinos $\nu_3$ and $\nu_4$ 
have the same CP-parities, 
$\phi_3 = \phi_4 = \pm \phi_2$ 
(i.e.  $\alpha_{41} =  \alpha_{31} = 0, \pm \pi,
 \   \alpha_{21} = 0, \pm \pi$),
\meff is given by:
\begin{equation}
\meff = \sqrt{m_1^2 + \deltalsnd} 
\left( 1 - |U_{\mathrm{e} 2}|^2 \pm |U_{\mathrm{e} 2}|^2 \right).
\end{equation}
%
Using the data on \deltalsnd and $|U_{\mathrm{e} 2}|^2$
obtained respectively in ref.~\cite{LSND}
(at $95 \%$~C.L.) and  in ref.~\cite{CHOOZ} (at $90 \%$~C.L.),
we find that for any $m_1$:
\begin{align}
\meff \geq 0.4 \eV, \ \ \ & \ \
 \text{for $\phi_3 = \phi_4 =  \phi_2$;} \\
\meff \geq 0.3 \eV, \ \ \ & \ \ 
\text{for $\phi_3 = \phi_4 = -   \phi_2$}.
\end{align}
%
\noindent
If $m_1^2 \ll \deltalsnd$, we get for 
the maximal allowed values of \meff:
$\meff \leq 1.4$ eV for $\phi_3 = \phi_4 =  \phi_2$,
and $\meff \leq 1.1$ eV if 
$\phi_3 = \phi_4 = -   \phi_2$.
As in the similar case of 2+2A neutrino mass spectrum,
the experimental upper limits \cite{76Ge00,IGEX00},
eqs. (\ref{76Ge00}) and (\ref{IGEX00}),
exclude part (if not all) 
of the allowed regions of 
values (89) and (90) of \meff. An improvement
of these limits by a factor $\sim (3 - 4)$
would essentially rule out the considered 
possibility.

{\bf Case B.}
If the CP-parities of the neutrinos $\nu_{2,3,4}$ satisfy
$\phi_3 = -  \phi_4 = \pm \phi_2$ 
(i.e., if  $\alpha_{41} =  \alpha_{31} + \pi = 0, \pm \pi, \
   \alpha_{21} = 0, \pm \pi)$, one has:
\begin{equation}
\meff = \sqrt{m_1^2 + \deltalsnd} \left|  
( 1 - |U_{\mathrm{e} 2}|^2)\cos 2 \theta_\odot 
\pm |U_{\mathrm{e} 2}|^2 \right|.
\end{equation}
%
The predictions for the effective Majorana mass
\meff depend  on the solution
of the solar neutrino problem:
\begin{enumerate}
\item
In the case of the LMA MSW solution, 
the $90\% ~(99\%)$~ C.L. results 
obtained in ref.~\cite{Gonza3nu}
do not exclude the possibility 
of cancellations between the different terms 
in eq.~(\ref{31ameff}):
\begin{align}
0  \ (0)  \eV \leq \meff \leq 9.8 \ (10) \times 10^{-1} \eV, \ \ \ & \ \
 \text{for $\phi_3 = -  \phi_4 =  \phi_2$;} \\
0  \ (0)   \eV \leq \meff \leq 9.5 \ (10) \times 10^{-1} \eV, \ \ \ & \ \ 
\text{for $\phi_3 = -  \phi_4 = -   \phi_2$}.
\end{align}
%
\noindent 
The maximal values of \meff correspond to
$m_1 < 0.1$ eV. They will be larger by a factor
of $\sim 1.2$ if $m_1 \simeq 1.0$ eV. 
A better determination of the value
of $\theta_\odot$ will be crucial 
for reducing the allowed ranges of 
\meff in this case.
This might lead, in particular, to a 
non-trivial lower bound on \meff.
\item
For the SMA MSW solution 
one has  $\cos 2 \theta_\odot \simeq 1$
and we obtain the following intervals of 
allowed values of \meff corresponding to the 
$90 \%$~C.L. ($99 \%$~C.L.) solution regions
and $m_1^2 \ll \deltalsnd$:
\begin{align}
0.4  \eV \leq \meff \leq 1.4 \eV, \ \ \ & \ \ 
\text{for $\phi_3 = -\phi_4 =  \phi_2$;} \\
0.3  \eV \leq \meff \leq 1.1 \eV, \ \ \ & \ \ 
\text{for $\phi_3 = -\phi_4 = -   \phi_2$}.
\end{align}
This case cannot be distinguished 
from the earlier considered case A,
$\phi_3 =  \phi_4 = \pm \phi_2$.
The comments made at the end of the discussion
of the case A are valid also in the present case.
\item
The values of $\theta_\odot$, corresponding to the
90\% C.L. ( $99 \%$~C.L.)
LOW-QVO solution, allow 
cancellations between the different terms 
in \meff to take place and for $m_1^2 \ll \deltalsnd$
one finds:
\begin{align}
0 \ (0)  \eV \leq \meff \leq 7.0 \ (9.7) \times 10^{-1} \eV, \ \ \ & \ \ 
\text{$\phi_3 = -  \phi_4 =  \phi_2$;} \\
0  \ (0)   \eV \leq \meff \leq 6.7 \ (9.4) \times 10^{-1} \eV, \ \ \ & \ \ 
\text{$\phi_3 = -  \phi_4 = -   \phi_2$}.
\end{align}
%
\noindent As in the cases considered above, the 
upper limits are by a factor of $\sim 1.2$ larger
for $m_1 \simeq 1.0$ eV.
\end{enumerate}

  We see that in all cases of CP-conservation
\meff can take values in the region 
$\meff \gtap 0.01$ eV
which can be tested in the next generation of 
\betabeta-decay experiments.

  We show in Fig.~\ref{figure:3+1A01}  
the regions of allowed values  of \meff 
as a function of  $\sqrt{m_1^2 + \deltalsnd} 
\cong \sqrt{\deltalsnd}$ 
for the LMA and LOW-QVO solutions 
of the solar neutrino problem.
The regions were obtained 
assuming CP-invariance does not hold
as well as for the different 
cases of CP-invariance considered above.
The CP-violation region includes
all the CP-conserving ones: all regions
marked by different grey scales
are allowed if CP-invariance does not hold.
 The region marked by dark-grey color
is the ``just-CP-violating'' region:
it can be spanned by the values of \meff
only if the CP-symmetry is violated.
In a large region of the relevant 
parameter space one can have
values of $\meff \geq 0.1 \eV$, 
which can be tested 
by the current and future
\betabeta-decay experiments.
A positive result obtained in
these experiments and 
a less ambiguous determination of 
\deltalsnd might allow to establish,
provided the neutrino mass spectrum
is of the 3+1A type, 
whether the CP-symmetry is violated, 
and to get information on the relative CP-parities
of the massive Majorana neutrinos $\nu_{2,3,4}$
if the phases  $\alpha_{32}$ and 
$\alpha_{42}$ take their CP-conserving values.
We would like to remind the reader that 
for the spectrum under discussion,
the CP-violating phase $\alpha_{21}$
cannot be constrained by  the data 
on \meff, \deltalsnd and $\tan^2\theta_{\odot}$
and can be a source of CP-violation
in other processes.

 According to eq.~(\ref{31ameff}), \meff should exhibit
a rather strong dependence on $\cos 2 \theta_\odot$.
This dependence is illustrated in 
Fig.~\ref{figure:3+1A02}, where 
we plot $\meff / \sqrt{\deltalsnd}$
as a function of $\cos 2 \theta_\odot$.
The ``just-CP-violating'' region
in shown in Fig.~\ref{figure:3+1A02}
in dark-grey color.

  Equation (\ref{31ameff}) allows to find a relation
between the CP-violating 
phases $(\alpha_3- \alpha_2) \equiv \alpha_{32}$ and 
$(\alpha_4- \alpha_2) \equiv \alpha_{42}$
and the observable quantities \meff,
$|U_{\mathrm{e} 2}|^2$, 
$m_{\nu_e} \simeq \sqrt{m_1^2 + \deltalsnd}$ and $\theta_\odot$:
\begin{equation}
\begin{split}
1 - \frac{\meff^2}{(m_1^2 + \deltalsnd)} 
& = 2 |U_{\mathrm{e} 2}|^2 (1 - |U_{\mathrm{e} 2}|^2)
\left ( 1 - \cos^2  \theta_\odot \cos \alpha_{32} - 
\sin^2  \theta_\odot \cos \alpha_{42} \right ) \\
& + (1 - |U_{\mathrm{e} 2}|^2)^2 \sin^2 2\theta_\odot 
\sin^2 {{(\alpha_{32} - \alpha_{42})}\over{2}} 
\label{31ameffphases}
\end{split}
\end{equation}
 
\noindent A sufficiently accurate 
measurement of \meff and of 
the other observables 
which enter into the above equation 
might allow to obtain significant constraints on 
the phases $\alpha_{32}$ and $\alpha_{42}$
and thus to get information on the 
CP-violation in the lepton sector.
Figures ~\ref{figure:3+1A03} and \ref{figure:3+1A04}
illustrate this possibility: we show in these 
figures the allowed values of 
$\cos \alpha_{32}$ and
$\cos \alpha_{42}$ 
for a set of values of 
$|U_{\mathrm{e} 2}|^2 = 0.0, 0.01, ~{\rm and}~ 0.04, 0.08$, 
respectively. We have used the best fit value for 
$\cos 2 \theta_\odot$ \cite{Gonza3nu} and 
for each value of $|U_{\mathrm{e} 2}|^2$
we consider a set of 
values of \meff given by 
$\meff = \sqrt{0.15 + 0.1 n} \sqrt{\deltalsnd}$ with
$n = 1,2,...,8$.

  The constraint (\ref{31ameffphases}) will simplify 
considerably if the  mixing parameter $|U_{\mathrm{e} 2}|^2$,
which is limited by the CHOOZ data, turned out to be relatively
small, say $|U_{\mathrm{e} 2}|^2 \leq 10^{-2}$.
Let us note that 
the MINOS experiment \cite{MINOS} 
is planned to be sensitive
to values of $|U_{\mathrm{e} 2}|^2 \gtap 5\times 10^{-3}$.
Neglecting terms $\sim |U_{\mathrm{e} 2}|^2$ with 
respect to 1 in
eq. (\ref{31ameffphases}) we get:
\begin{equation}
\sin^2{{(\alpha_{32} - \alpha_{42})}\over{2}} 
\simeq {1\over{\sin^22\theta_{\odot}}} 
\left ( 1 - \frac{\meff^2}{(m_1^2 + \deltalsnd)} \right )~ 
\simeq {1\over{\sin^22\theta_{\odot}}} 
\left ( 1 - \frac{\meff^2}{m_{\nu_e}^2} \right )~.
\label{31ameffphases2} 
\end{equation}

\noindent The measurement of $\meff \neq 0$ 
and of $m_{\nu_e} \gtap 0.4$ eV
in this case 
can provide information 
on one of the Majorana CP-violating
phases, $(\alpha_{32} - \alpha_{42}) = \alpha_{34}$.

   Equation (\ref{31ameffphases}) takes 
a simpler form in the case of the SMA MSW solution of 
the solar neutrino problem as well.
For the SMA MSW solution one has \cite{Gonza3nu} 
$\sin^2\theta_{\odot} \ltap 2.5\times 10^{-3}~{\rm eV^2}$ 
and neglecting terms $\sim \sin^2\theta_{\odot}$,
eq. (\ref{31ameffphases}) reduces to:  
\begin{equation}
4|U_{\mathrm{e} 2}|^2 (1 - |U_{\mathrm{e} 2}|^2)~ 
\sin^2{{\alpha_{32}}\over{2}} \simeq 
\left ( 1 - \frac{\meff^2}{(m_1^2 + \deltalsnd)} \right )~
\simeq 
\left ( 1 - \frac{\meff^2}{m_{\nu_e}^2} \right )~.
\label{31ameffphases3} 
\end{equation}
\noindent Note that, as in the previous case, only one 
Majorana CP-violating phase enters into 
eq. (\ref{31ameffphases3}) and
could be constrained by using data on  
\meff, $m_{\nu_e} \simeq \sqrt{m_1^2 + \deltalsnd}$ and 
$|U_{\mathrm{e} 2}|^2$. For $m_1$
not known, eq. (\ref{31ameffphases3}) 
allows to obtain a correlated constraint
on $m_1$ and $\alpha_{32}$.

 If the SMA MSW solution is the 
correct solution of the 
solar neutrino problem and 
$|U_{\mathrm{e} 2}|^2$ is rather small,
e.g., $|U_{\mathrm{e} 2}|^2 \ltap 5\times 10^{-3}$,
the measurement of $\meff \gtap 10^{-2}$ eV will hardly provide
any information on the violation of the 
CP-symmetry in the lepton sector, since 
under the above circumstances we would have 
for the 3+1A neutrino mass spectrum 
under discussion:
\begin{equation}
\meff \simeq \sqrt{m_1^2 + \deltalsnd}~(1 + O(10^{-2}))~
\simeq m_{\nu_e}~.
\label{31ameffnophases} 
\end{equation}

\noindent However, as eq. (\ref{31ameffnophases}) indicates, 
such a measurement can give information on the mass of the 
lightest neutrino $m_1$. Let us remind the reader that 
similar result holds for the SMA MSW solution
(or in the CP-conserving case A) if the neutrino
mass spectrum is of the 2+2A type.


\section{\textbf{ The 3+1B  mass spectrum}}
\label{3+1B}

 The 3+1B neutrino mass spectrum 
corresponds to three nearly-degenerate
neutrinos $\nu_{1,2,3}$ with masses 
sufficiently smaller
than the mass of the fourth one,
$m_{1} \simeq m_{2} \simeq m_{3} <~(\ll)~m_{4}$,
as is indicated graphically below:

\unitlength=1mm
\begin{center}
\begin{picture}(100,40)
\put(20,5){\line(1,0){20}}
\put(20,8){\line(1,0){20}}
\put(20,13){\line(1,0){20}}
\put(20,35){\line(1,0){20}}
\put(30,5){\line(0,1){30}}
\put(45,4.5){\vector(0,1){3.5}}
\put(45,8){\vector(0,-1){3.5}}
\put(45,8){\vector(0,1){5}}
\put(45,13){\vector(0,-1){5}}
\put(72,5){\vector(0,1){30}}
\put(72,35){\vector(0,-1){30}}
\put(48,4){\deltasol}
\put(48,9.5){\deltaatm}
\put(75,18){\deltalsnd}
\put(17,4){1}
\put(17,8){2}
\put(17,12.3){3}
\put(17,34.2){4}
\end{picture}
\end{center}

\noindent
This neutrino mass spectrum can also be characterized by:
\begin{equation}
\Delta m^2_{21}   \ll \Delta m^2_{32} \ll \Delta m^2_{41}
\label{31Bspectrum}
\end{equation}
\noindent
Equation ~(\ref{31Bspectrum}) allows one 
to relate the neutrino mass-squared differences
$\Delta m^2_{jk}$ to those determined 
from the solar, atmospheric and LSND 
neutrino oscillation data:
\begin{equation}
\Delta m^2_{21}   
            \equiv    \Delta  m^2_{\odot}, \ \ 
\Delta m^2_{32}  
            \equiv    \Delta  m^2_{\mathrm{atm}}, \  \
\Delta m^2_{41}  
            \equiv    \Delta  m^2_{\mathrm{SBL}}.
\label{31bpattern}  
\end{equation}
\noindent The three neutrinos
$\nu_{1,2,3}$ can have a hierarchical mass spectrum,
a spectrum with partial hierarchy or 
a quasi-degenerate spectrum (see, e.g., \cite{BPP1}).

 Using ~(\ref{31bpattern}) 
we can express the masses $m_{2,3,4}$ as functions 
of $m_1$ and the measured neutrino mass-squared
differences:
\begin{align}
m_2 = & \sqrt{m_1^2 + \deltasol}, \nonumber  \\
m_3 = & \sqrt{m_1^2 + \deltasol + \deltaatm}, \nonumber  \\ 
m_4 = & \sqrt{m_1^2 + \deltalsnd}.
\label{31bmj}
\end{align}
%
\noindent We will require also that, 
as in the case of the 3-neutrino mass spectrum 
with hierarchy or partial mass hierarchy,
$m_1$ satisfies \cite{BPP1}: $m_1 \leq 0.2 \eV$. 
For the spectrum with partial mass hierarchy 
one has \cite{BPP1}
$0.02~{\rm eV} \ltap m_1 \leq 0.2 \eV$,
while for the hierarchical spectrum
$m_1 \ll 0.02$ eV. In the latter case
the contribution of 
$m_1$ to \meff is negligible. 
Given the allowed values of 
\deltalsnd, eq. (7), and eq. (\ref{31bmj}),
the inequality $m_1 \leq 0.2 \eV$
implies $m_1^2 \ll m_4^2$. For 
$m_1 > 0.2 \eV$ one gets essentially 
a quasi-degenerate $\nu_{1,2,3}$ mass spectrum.
We will discuss the predictions for \meff
in this latter case at the end of 
the Section. 

 For the elements of interest  
of the neutrino mixing matrix, 
$|U_{\mathrm{e} \, j}|$,
one finds:
\begin{itemize}
\item[i)]
$|U_{\mathrm{e} \, 1}|$ and $|U_{\mathrm{e} \, 2}|$ 
are related to the solar mixing angle $\theta_{\odot}$:
\begin{align}
|U_{\mathrm{e} \, 1}|^2  =  & \cos^2 \theta_\odot \  
( 1 -  |U_{\mathrm{e} \, 3}|^2 ), \nonumber  \\
|U_{\mathrm{e} \, 2}|^2  =  &  \sin^2 \theta_\odot  \ 
( 1 -  |U_{\mathrm{e} \, 3}|^2 );
\label{31b} 
\end{align}
%
\item[ii)]
$|U_{\mathrm{e} \, 3}|^2$ should satisfy the CHOOZ  upper limit 
which at $99 \%$~C.L. reads \cite{CHOOZ,Gonza3nu}:
$|U_{\mathrm{e} \, 3}|^2 < 0.08$;
\item[iii)]
$|U_{\mathrm{e} \, 4}|^2$ is constrained 
by the data from the SBL 
experiments \cite{LSND,BUGEY} and 
we have $ 2 \times 10^{-4} < |U_{\mathrm{e} \, 4}|^2< 1 \times 10^{-2}$.
\end{itemize}

\noindent

 Using eqs. (\ref{effectivemass}) and (\ref{31b}), 
one can express the effective Majorana  mass 
\meff in terms of the 
quantities measured in 
the neutrino oscillation experiments, 
$m_1$ and the Majorana 
CP-violating phases $\alpha_{21}$, 
$\alpha_{31}$ and $\alpha_{41}$:
\begin{equation}
\begin{split}
\meff \simeq &  |m_1  (1 - |U_{\mathrm{e} \, 3}|^2)~\cos^2 \theta_\odot  + 
\sqrt{m_1^2 + \deltasol} (1 - |U_{\mathrm{e} \, 3}|^2)
\sin^2 \theta_\odot e^{i \, \alpha_{21}} \\
 & +  \sqrt{m_1^2 + \deltasol + \deltaatm}|U_{\mathrm{e} \, 3}|^2 
e^{i \, \alpha_{31}} +  \sqrt{m_1^2 + 
\deltalsnd} |U_{\mathrm{e} \, 4}|^2 e^{i \, \alpha_{41}} |.
\end{split}
\label{31bmeff}
\end{equation}
%
%
Since \meff depends on three CP-violating phases 
in this case, the general analysis 
of the allowed values of \meff is rather complicated. 
It can be simplified if we consider first 
the absolute value of the sum 
of the first three contributions in 
eq.~(\ref{31bmeff}), $|<m>|_{3-\nu}$:
\begin{equation}
\begin{split}
|<m>|_{3-\nu} \equiv | m_1 (1 - |U_{\mathrm{e} \, 3}|^2) \cos^2 \theta_\odot
+  \sqrt{m_1^2 + \deltasol}~(1 - |U_{\mathrm{e} \, 3}|^2)~
\sin^2 \theta_\odot e^{i \, \alpha_{21}} \\
 + \sqrt{m_1^2 + \deltasol + \deltaatm} 
|U_{\mathrm{e} \, 3}|^2 e^{i \, \alpha_{31}}|.
\label{meff3nu}
\end{split}
\end{equation}
The term  $|<m>|_{3-\nu}$ of interest  
coincides with  the effective Majorana mass 
for the neutrino mass spectrum with
hierarchy or with partial mass hierarchy
in the 3-neutrino mixing case, which have 
been studied in detail recently in ref.~\cite{BPP1} 
(see Sections 4 and 7 in  \cite{BPP1}).
It was found in \cite{BPP1} that  
$|<m>|_{3-\nu}$ is constrained to lie in 
the intervals
\begin{align}
0 \leq |<m>|_{3-\nu} \leq &  \ 7.4 \ (14) \times 10^{-3} \eV,  
\qquad \mbox{LMA MSW solution}; \nonumber  \\
0 \leq |<m>|_{3-\nu} \leq &  \ 1.9 \ (3.6) \times 10^{-3} \eV,  
\qquad \mbox{SMA MSW solution}; \nonumber  \\
 0 \leq |<m>|_{3-\nu} \leq &  \ 3.5 \ (3.6) \times 10^{-3} \eV,  
\qquad \mbox{LOW-QVO solution},   
\label{31b3meffh}
\end{align}
for negligible $m_1$
(hierarchical $3-\nu$ spectrum), and
\begin{align}
8.0 \times 10^{-4 } \ ( 0) \leq |<m>|_{3-\nu} \leq & \  0.2  \ (0.2)  \eV,  
\qquad \mbox{LMA MSW solution}; \nonumber  \\
1.6 \ (1.5) \times 10^{-2} \leq |<m>|_{3-\nu} \leq & \  0.2  \ (0.2) \eV,  
\qquad \mbox{SMA MSW solution}; \nonumber  \\
 0 \ (0 ) \leq |<m>|_{3-\nu} \leq & \  0.2 \ (0.2)  \eV,  
\qquad \mbox{LOW-QVO solution},   
\label{31b3meffph}
\end{align}
for the spectrum with partial mass hierarchy,
i.e., for $0.02~{\rm eV} \ltap m_1 \leq 0.2~{\rm eV}$. 
The values (the values in brackets) in 
eqs. (\ref{31b3meffh}) and (\ref{31b3meffph})
correspond to 90\% (99\%) C.L. solution regions \cite{BPP1}.
The upper bounds in eq. (\ref{31b3meffph})
correspond to the maximal assumed value of $m_1$:
${\rm max}(|<m>|_{3-\nu}) = {\rm max}(m_1)$.
The fourth term in eq.~(\ref{31bmeff}), $|<m>|_{4}$, is constrained  
by the SBL data ~\cite{LSND} (at $95 \%$~C.L.):
\begin{equation}
3.0 \times 10^{-4} \eV \leq |<m>|_4 \equiv  \sqrt{m_1^2 + \deltalsnd}  
|U_{\mathrm{e} \, 4}|^2 \leq 5.0 \times 10^{-3} \eV.
\label{31bmeff4th}
\end{equation}
%
It follows from (\ref{31b3meffh}) and (\ref{31bmeff4th})
that for the 3+1B spectrum one can have
$\meff \gtap 3\times 10^{-2}$ eV only if
$m_1 \gtap 2\times 10^{-2}$ eV and that 
a measured value of 
$\meff \gtap 4\times 10^{-2}$ eV 
would imply:  
\begin{equation}
\meff \gtap 4\times 10^{-2}~{\rm eV}:~~~~~~~~~
\meff \simeq m_1 (1 - |U_{\mathrm{e} \, 3}|^2) 
\left ( 1 - \sin^2 2\theta_\odot~\sin^2{{\alpha_{21}}\over{2}} \right ).
~~~~~
\label{31bmeffbig}
\end{equation}
%
\noindent For 
$\sin^2 2\theta_\odot \sin^2(\alpha_{21}/2)$
sufficiently smaller than 1
(e.g., $\sin^2 2\theta_\odot \sin^2(\alpha_{21}/2) \ltap 0.5$)
the above relation offers the possibility 
to determine the mass of the lightest neutrino $m_1$.
The indicated condition is realized, e.g., 
for the SMA MSW solution of 
the solar neutrino problem 
($\sin^2 2\theta_\odot \ll 1$) and/or 
for ${\alpha_{21}} = 2\pi k$, $k=0,1,...$.

\begin{figure}
\begin{center}
\epsfig{file=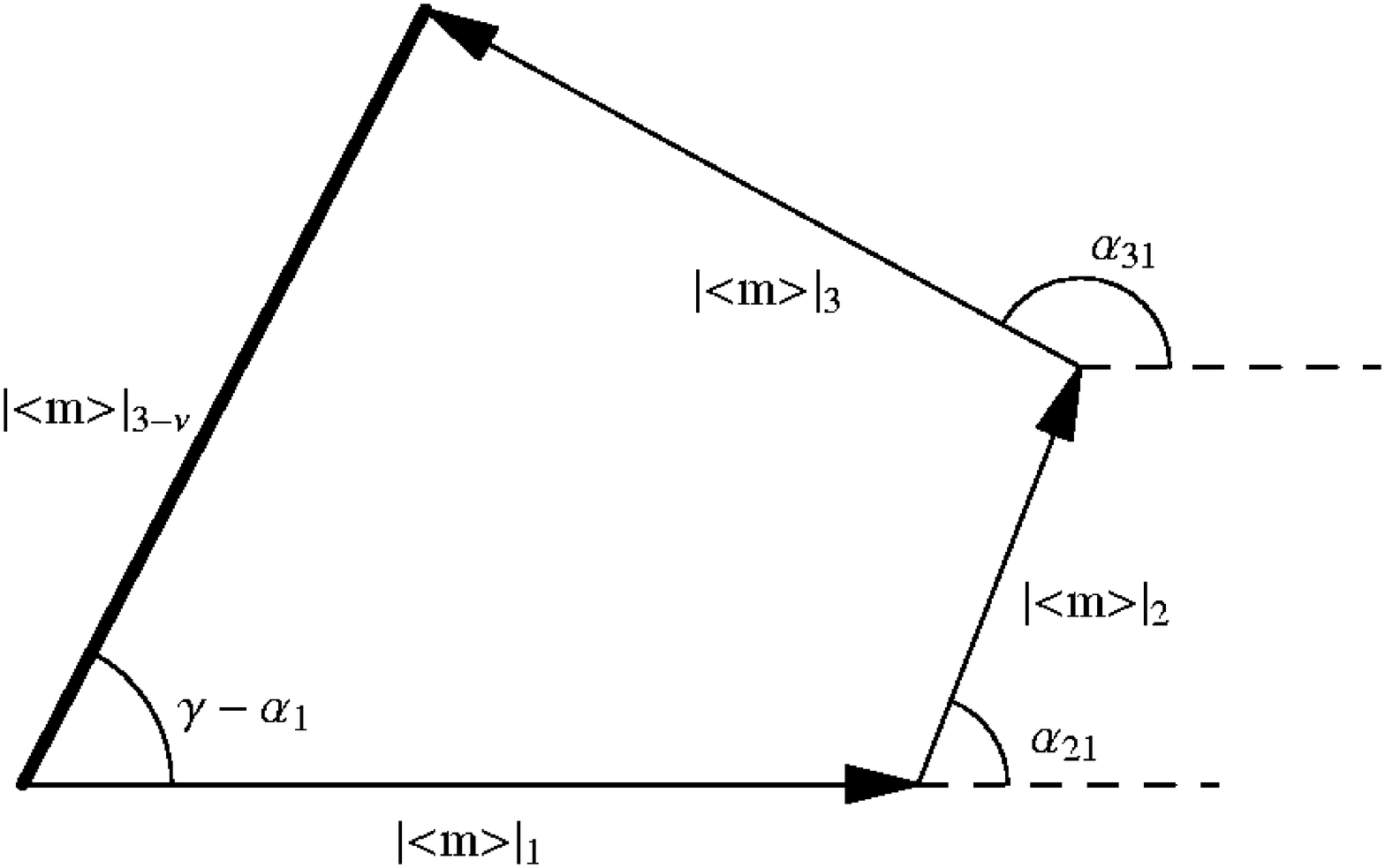, height=6cm
}
\end{center}
\caption{The contribution $\meff_{3-\nu}$ to the effective Majorana mass 
represented as the absolute 
value of the vector
sum of the three contributions $\meff_1$, $\meff_2$ 
and $\meff_3$ in eqs. (\ref{meff3nu}), 
each of which is  
expressed as a vector in the complex plane
(see text for details).
The CP-violating phases $\alpha_{21}$ and $\alpha_{31}$ 
shown in the figure can vary between 0 and $2 \pi$.
 The effective CP-violating phase $(\gamma - \alpha_1)$ is also 
shown in the figure.  
}
\label{figure:3+1B}
\end{figure}

\begin{figure}
\begin{center}
\epsfig{file=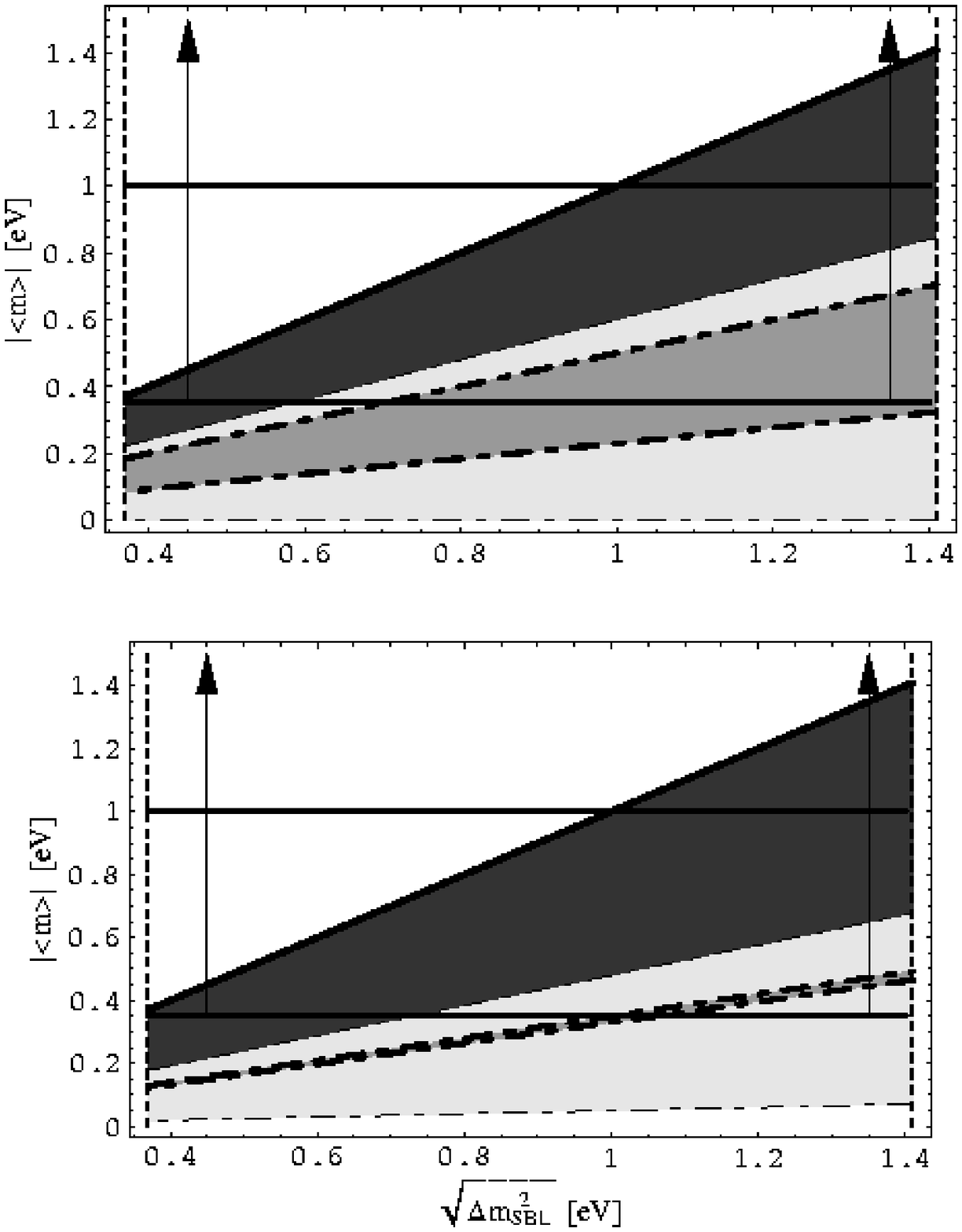, height=15cm
}
\end{center}
\caption[22A01]
{The effective Majorana mass 
\meff as a function of $\sqrt{\deltalsnd}$ 
for the 2+2A  neutrino mass spectrum, eq. (\ref{22Aspectrum}). 
The allowed regions (in grey) 
correspond to the LMA solution 
of ref.~\cite{Gonza4nu} 
for $\cos^2 \beta = 0.3$ (upper panel) 
and  $\cos^2 \beta = 0.5$ (lower panel).
In the case of CP-invariance and for the
$90\%~ (99\%)$ C.L. LMA
solution region, \meff can have values 
{\it i}) for $\phi_3 = \phi_4$ - 
on the upper doubly-thick solid line
 (upper doubly-thick solid line)
 and 
{\it ii}) for $\phi_3 = - \phi_4$ - in the medium grey  
(light grey and medium grey)  
region limited
by the  thick (thin) 
dash-dotted lines.  If CP is not conserved,
 \meff can lie in any of the 
regions marked by different grey scales.     
The dark-grey region 
corresponds to ``just-CP-violation'':
\meff can have value in this region 
{\it only if the CP-parity is not conserved}. 
The two horizontal 
(thick) lines show the upper limits 
\cite{76Ge00}, quoted in eq. (\ref{76Ge00}).}
\label{figure:2+2A01}
\end{figure}

\begin{figure}
\begin{center}
\epsfig{file=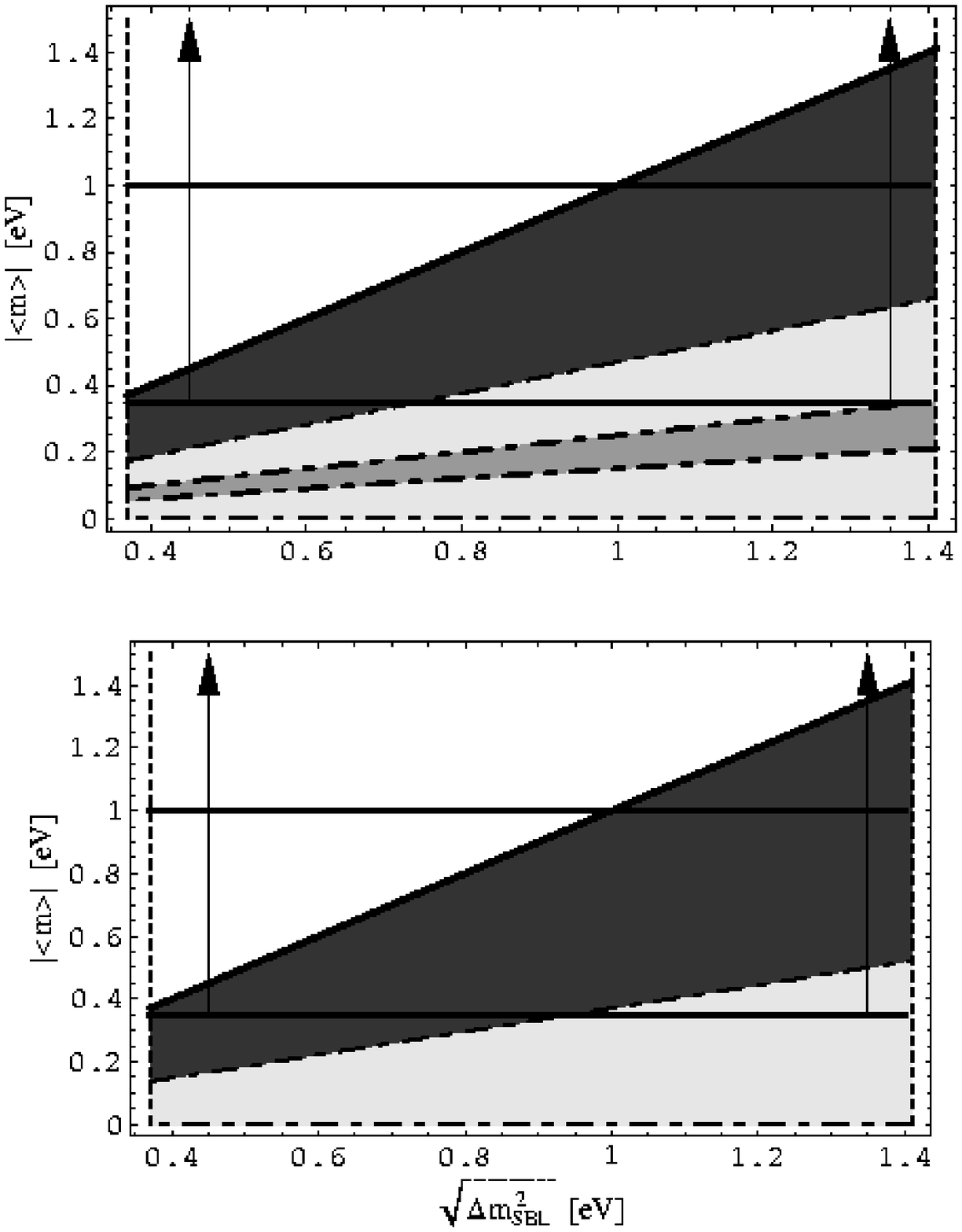, height=15cm
}
\end{center}
\caption[22A02]{
The same as in Fig. (\ref{figure:2+2A01})
for the $90\%$ ($99\%$) C.L. LOW-QVO solution
of ref. \cite{Gonza4nu}
for $\cos^2 \beta = 0.3$ (upper panel) 
and  $\cos^2 \beta = 0.5$ (lower panel). 
The doubly-thick
(doubly-thick) solid lines
and the medium grey  
(light grey and medium grey) lower region,
bounded by the thick (thin) dash-dotted lines,
correspond to the two cases of CP-invariance, 
$\phi_3 =  \phi_4$  and $\phi_3 = -  \phi_4$,
respectively. If CP is not conserved, 
\meff can lie in any of the 
regions marked by different grey scales.      
The ``just-CP-violation'' region is shown in dark-grey color:
\meff can have value in this region 
{\it only if the CP-symmetry is violated}.
}
\label{figure:2+2A02}
\end{figure}

\begin{figure}
\begin{center}
\epsfig{file=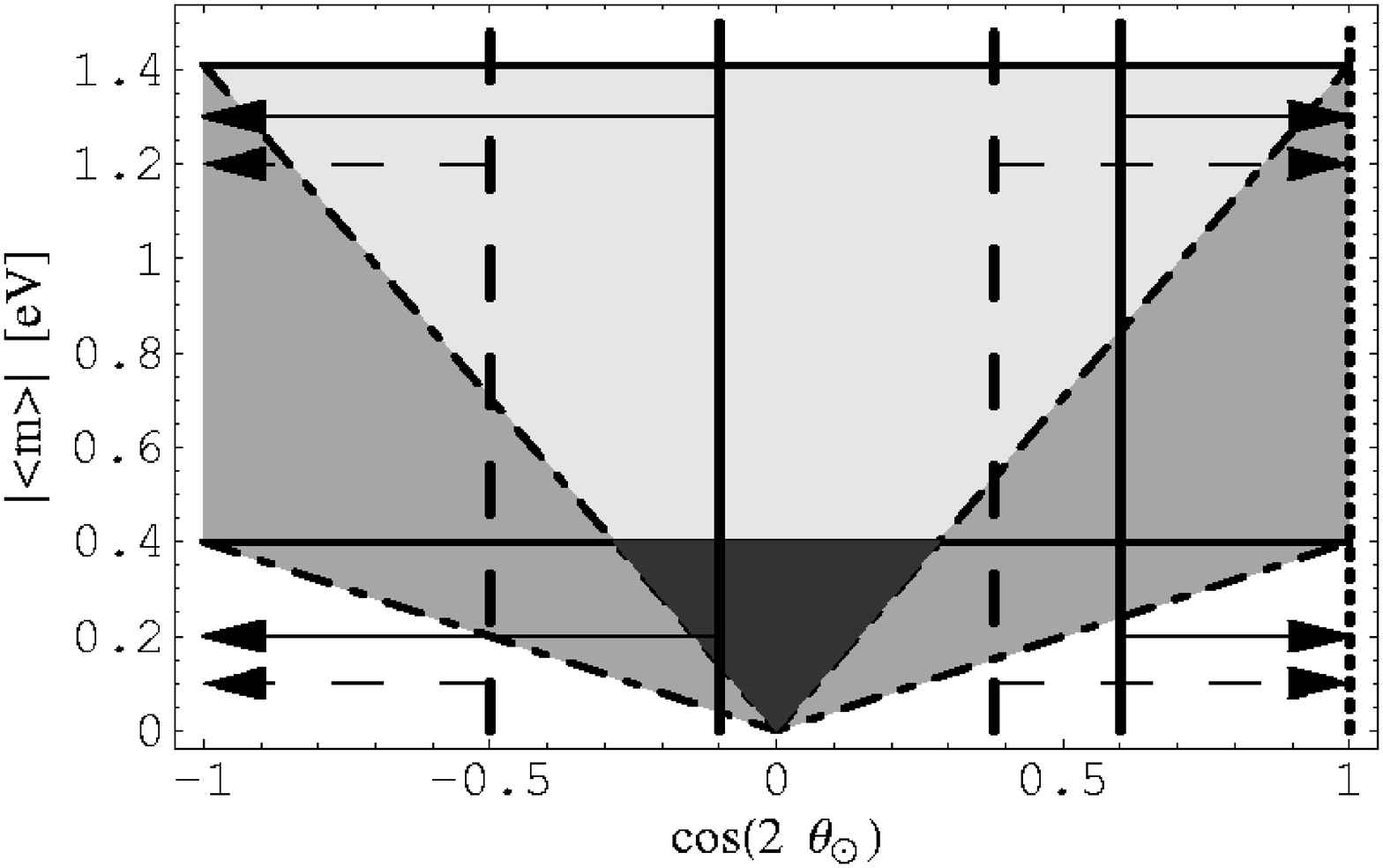, height=6.5cm
}
\end{center}
\caption{The dependence of 
\meff on $\cos 2 \theta_\odot$ 
for the 2+2A neutrino mass spectrum, 
eq. (\ref{22Aspectrum}). 
The region between the two thick horizontal 
solid lines (in light  grey and medium grey colors)
and the two triangular regions between the thick 
dash-dotted lines (in medium grey color),
correspond to the two CP-
conserving cases, $\phi_3 = \phi_4$ 
and $\phi_3 = - \phi_4$, respectively.
The ``just-CP-violation'' region is denoted 
by dark-grey color. 
The regions between each of the two pairs 
of vertical lines of a given type
- doubly thick solid and doubly thick dashed,
correspond to the intervals of values of 
$\cos 2 \theta_\odot$ 
for the LMA solution and the LOW-QVO solution
 derived (at $99\%$~C.L.) 
in ref.~\cite{Gonza4nu} for $\cos^2 \beta =0.3$
while the vertical dotted line
 corresponds to the SMA solution.
}
\label{figure:2+2A03}
\end{figure}

\begin{figure}
\begin{center}
\epsfig{file=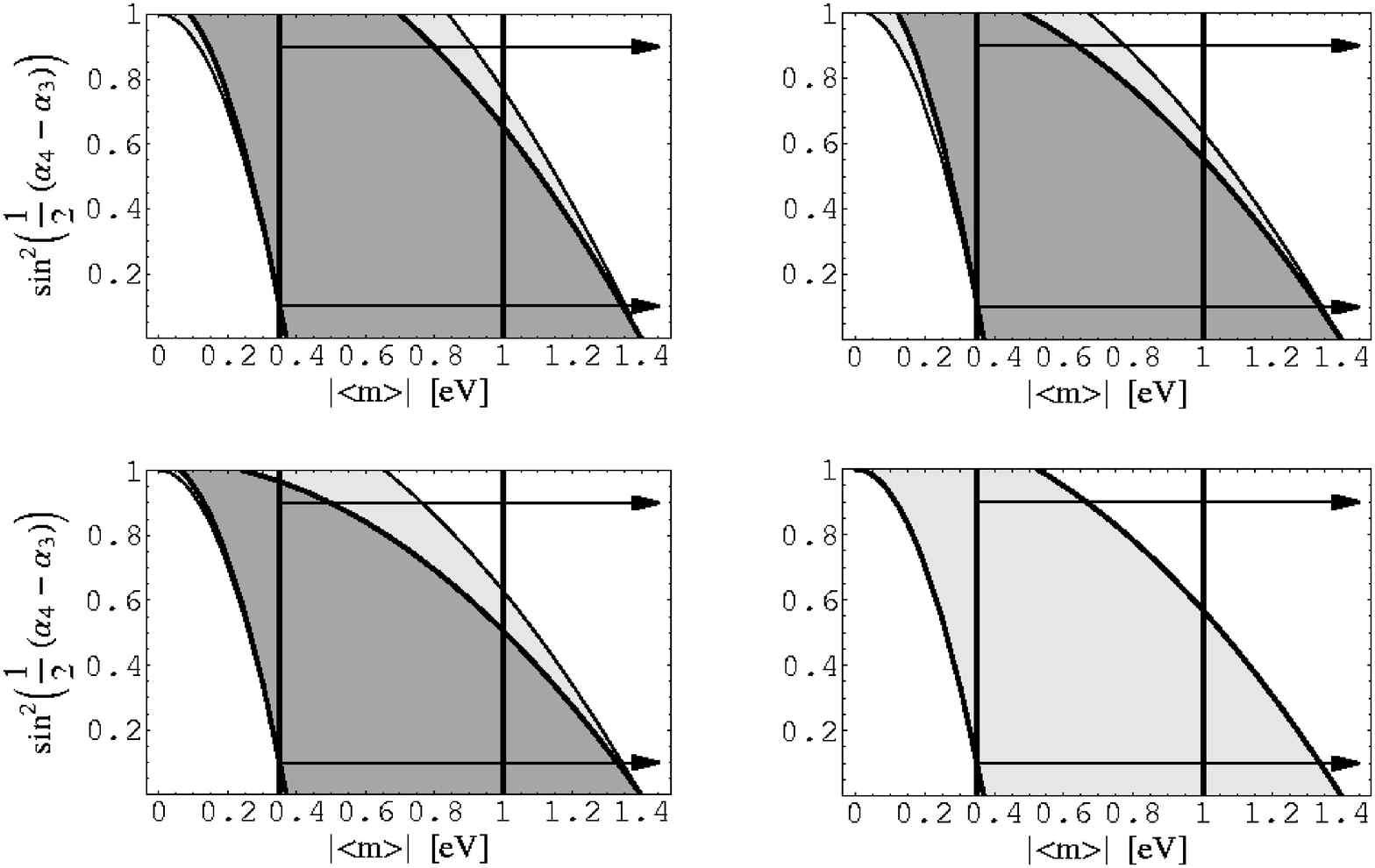, height=12cm, width=17cm
}
\end{center}
\caption[22A04]{
The CP-violation factor 
$\sin^2 (\alpha_4 - \alpha_3)/2$ 
as a function of \meff
in the case of 2+2A  mass  spectrum, 
eq. (\ref{22Aspectrum}),  
for  the LMA  (upper  panels) and 
LOW-QVO (lower panels) solutions of ref.~\cite{Gonza4nu} 
and for $\cos^2 \beta = 0.3$ (upper and lower left panels)
and  $\cos^2 \beta = 0.5$ (upper and lower right panels). 
The results correspond to the 
 $90\%$ C.L. (medium grey region 
with thick contours) and $99\%$ C.L.
(light grey  and  medium grey region limited by 
one ordinary solid line and one thick solid line)
 solution regions. 
A value of $\sin^2 (\alpha_4 - \alpha_3)/2 \neq 0,1$, 
would signal CP-violation.
The two vertical 
(thick) lines show the upper limits 
\cite{76Ge00}, quoted in eq. (\ref{76Ge00}).
}
\label{figure:2+2A04}
\end{figure}

    The effective Majorana mass \meff
can be re-written in terms of $|<m>|_{3-\nu}$ and 
$|<m>|_4$ as follows:
\begin{equation}
\meff = \left| |<m>|_{3-\nu} e^{i (\gamma - \alpha_1)} + 
|<m>|_4  e^{i \alpha_{41}} \right|
\end{equation}
%
%
where $\gamma $ is the overall CP-violating 
phase of the sum of the first three terms.
Let us note that $\gamma$ cannot be simply related 
to the CP-violating phases $\alpha_{21}$ and $\alpha_{31}$
entering into the expression for the
sum of the first three terms in eq. (\ref{31bmeff}).
The physical meaning of the phase 
$\gamma$ becomes more clear from Fig.~\ref{figure:3+1B}, where
we have represented graphically the term $|<m>|_{3-\nu}$ 
in the complex plane. 
%
%
%
As is shown in Fig.~\ref{figure:3+1B}, 
the phase $(\gamma - \alpha_1)$ is determined not only by 
$\alpha_{21}$ and $\alpha_{31}$,
but depends also on $\meff_1$, $\meff_2$ 
and $\meff_3$ - the absolute values of the three
terms whose sum is $<m>_{3-\nu}$.
If CP is conserved, $\gamma$ and $\alpha_4$
can be either 0 or $\pm \pi$.
The phase $(\gamma - \alpha_1)$
determines in the case of CP-conservation 
the sign of the contribution of the term $<m>_{3-\nu}$,
while  $(\gamma - \alpha_1 - \alpha_{41})$
fixes the relative sign of 
$<m>_{3-\nu}$ and $<m>_4$.
According to the different CP-conserving 
values the phases
$(\gamma - \alpha_1)$ and $\alpha_{41}$
can assume, we have two possibilities.

{\bf Case A.}
If $\gamma- \alpha_1 =\alpha_{41} = 0, \pm \pi$, 
\meff is bounded to lie in the intervals
\begin{align}
3.0 \ (3.0)  \times 10^{-4} \eV  \leq \meff  \leq &  \ 1.2 \ (1.9) 
\times 10^{-2} \eV,  \qquad \mbox{LMA MSW solution}; \nonumber  \\
3.0 \ (3.0)  \times 10^{-4} \eV  \leq  \meff  \leq &  \ 6.9 \ (8.6) \times 10^{-3} \eV,  \qquad \mbox{SMA MSW solution}; \nonumber  \\
3.0 \ (3.0)  \times 10^{-4} \eV   \leq  \meff  \leq &  \ 8.5 \ (8.6) \times 10^{-3} \eV,  \qquad \mbox{LOW-QVO solution},   
\end{align}
if  $m_1$ is negligible 
(hierarchical 3-$\nu$ mass spectrum). For  
$0.02~{\rm eV} \ltap m_1 \leq 0.2~{\rm eV}$ 
(3-$\nu$ mass spectrum with partial 
hierarchy \cite{BPP1}) one gets
\begin{equation}
3.0 \ (3.0)  \times 10^{-4} \eV  \leq \meff  \leq   \ 0.2 \ (0.2)  \eV   
\label{31bApmh}
\end{equation}
for all solutions of the solar 
neutrino problem, where 
${\rm max}(\meff) \simeq m_1$.
Equation (\ref{31bApmh}) was obtained 
by using the $90 \%$~C.L. ($95 \%$~C.L.) allowed values of 
$\theta_\odot$,  \deltasol 
and $|U_{\mathrm{e} 3}|^2$ from ref.~\cite{Gonza3nu} 
and the allowed values of 
\deltalsnd and $|U_{\mathrm{e} 4}|^2$
at $95 \%$~C.L. from ref.~\cite{LSND}.

{\bf Case B.} If $(\gamma - \alpha_1)$ and $\alpha_{41}$ 
satisfy $(\gamma - \alpha_1) = 
\alpha_{41} + \pi = 0, \pm \pi$, 
there can be  cancellations 
between $|<m>|_{3-\nu}$ and $\meff_4$ and 
there is no significant lower bound on \meff.
We get for negligible  $m_1$,
\begin{align}
0 \  (0)\leq \meff \leq &  \ 7.0 \ (14) \times 10^{-3} \eV,  
\qquad \mbox{LMA MSW solution}; \nonumber  \\
0 \  (0) \leq \meff  \leq &  \ 1.5 \ (3.2) \times 10^{-3} \eV,  
\qquad \mbox{SMA MSW solution}; \nonumber  \\
 0  \  (0)\leq \meff  \leq &  \ 3.1 \ (3.2) \times 10^{-3} \eV,  
\qquad \mbox{LOW-QVO solution},   
\end{align}
\noindent while for
$0.02~{\rm eV} \ltap m_1 \leq 0.2~{\rm eV}$ 
one finds
\begin{equation}
0 \  (0) \eV  \leq \meff  \leq   \ 0.2 \ (0.2)  \eV   
\end{equation}
for all solutions of the solar neutrino 
problem, with ${\rm max}(\meff) \simeq m_1$.
This result was obtained 
for the allowed ranges of values of 
$\theta_\odot$, \deltasol, 
$|U_{\mathrm{e} 3}|^2$ 
\deltalsnd and $|U_{\mathrm{e} 4}|^2$,
which were used to get eq. (\ref{31bApmh})
and which are described after eq. (\ref{31bApmh}).
%
In Fig.~\ref{figure:31b01} we show
the allowed values of \meff 
for the different solutions
of the solar neutrino problem 
as a function of $m_1$.
As the figure demonstrates, 
there exists an upper bound on \meff, but 
not a non-trivial lower one: $\meff \cong 0$ eV 
is not excluded. However, 
one has $\meff \geq 0.01 \eV$ in 
a large region of the relevant 
parameter space, which can be tested 
in the \betabeta-decay
experiments of the next generation. 
Note also that for the neutrino mass spectrum
under discussion a value of 
$\meff \gtap 0.03$ eV is possible 
{\it only if $m_1 \gtap 0.02$} eV.

  If the CP-symmetry is violated, 
it would be rather difficult 
to obtain information on the Majorana
CP-violating phases 
$\alpha_{21}$, $\alpha_{31}$ and $\alpha_{41}$  
even if it were possible to measure 
all the other parameters entering 
into the expression for \meff, eq.~(\ref{31bmeff}),
and \meff with sufficient
accuracy. The measured values of 
these parameters and of \meff
would define a hypersurface 
in the 3-D parameter space of the three
CP-violating phases 
$\alpha_{21}$, $\alpha_{31}$ and $\alpha_{41}$, 
on which the phases would be
constrained to lie. If such hypersurface does 
not contain any of the points corresponding 
to the cases of CP-conservation,
i.e., $\alpha_{21}, \alpha_{31},\alpha_{41}=0, \pm \pi$,
one could conclude that the CP-symmetry
is violated in the lepton sector.
Otherwise, it would not be possible to draw 
any conclusions concerning the CP-symmetry violation.

 There are several physically interesting cases, however,
in which the expression for \meff simplifies somewhat
and includes effectively only two or just one 
Majorana CP-violating phases:
i) $m_1$ is negligible, ii) SMA MSW solution 
of the solar neutrino problem, 
iii) $|U_{\mathrm{e} \, 3}|^2$, which is limited 
by the CHOOZ data, is rather small, e.g.,
$|U_{\mathrm{e} \, 3}|^2 \ltap (0.5 - 1.0)\times 10^{-2}$, 
and iv) a combination of any two of the 
above three possibilities.

  If $m_1$ is negligible,  
the first term in equation~(\ref{31bmeff}) would 
``drop'' and only two relative CP-violating 
phases would be relevant.
Once \deltasol, \deltaatm, \deltalsnd and 
the relevant neutrino mixing 
angles are determined experimentally 
with a sufficient accuracy,
the measurement of $\meff \neq 0$ would allow
to obtain constraints 
on the phase differences 
$\alpha_{32}$ and $\alpha_{42}$ 
using the relation:
\begin{equation}
\begin{split}
\meff^2 = & (\sqrt{\deltasol}(1 - |U_{\mathrm{e} \, 3}|^2)
\sin^2\theta_{\odot})^2 +
 ( \sqrt{\deltaatm}|U_{\mathrm{e} \, 3}|^2 )^2 +
( \sqrt{\deltalsnd}|U_{\mathrm{e} \, 4}|^2 )^2 \\ 
 &  + 2\sqrt{\deltasol} \sqrt{\deltaatm}~
(1 - |U_{\mathrm{e} \, 3}|^2)~
|U_{\mathrm{e} \, 3}|^2~\sin^2\theta_{\odot}~\cos \alpha_{32} \\
& + 2 \sqrt{\deltasol}~\sqrt{\deltalsnd}~
(1 - |U_{\mathrm{e} \, 3}|^2)~|U_{\mathrm{e} \, 4}|^2
~\sin^2\theta_{\odot}~\cos \alpha_{42} \\
&  +  2 \sqrt{\deltaatm} \sqrt{\deltalsnd}~|U_{\mathrm{e} \, 3}|^2 
|U_{\mathrm{e} \, 4}|^2~ \cos (\alpha_{42} - \alpha_{32})~.
\label{31bphasec}
\end{split}
\end{equation}
%
\noindent
Figs.~\ref{figure:3+1B02} and ~\ref{figure:3+1B03} 
illustrate this possibility.

  If the mass $m_1$ is negligible,
one would be able to obtain 
information on the violation of
CP-symmetry in the lepton sector utilizing 
eq. (\ref{31bphasec})
even if the mixing matrix element 
$|U_{\mathrm{e} \, 3}|^2$
will be shown to be exceedingly small, so that
the terms $\sim |U_{\mathrm{e} \, 3}|^2$ 
give negligible
contribution in \meff. In this case 
the data on \deltasol, $\sin^2\theta_{\odot}$, 
\deltalsnd and 
$|U_{\mathrm{e} \, 4}|^2$ together with a measured value of
$\meff \neq 0$ eV could provide information
on the Majorana CP-violating phase $\alpha_{42}$:
\begin{equation}
4\sqrt{\deltasol} \sqrt{\deltalsnd}~
|U_{\mathrm{e} \, 4}|^2~\sin^2\theta_{\odot}~
\sin^2{{\alpha_{42}}\over{2}}
\simeq (\sqrt{\deltasol}~\sin^2\theta_{\odot} +
\sqrt{\deltalsnd}~|U_{\mathrm{e} \, 4}|^2)^2 - \meff^2~.   
\label{31bphasec1}
\end{equation}

  For the SMA MSW solution of 
the solar neutrino problem we have both
$\sin^2\theta_{\odot} \ltap 2.5\times 10^{-3}$ and
$\deltasol \ll \deltaatm$. Neglecting terms 
$\sim \sin^2\theta_{\odot}$ and $\sim \deltasol/ \deltaatm$,
we get from eq. (\ref{31bmeff}):
\begin{equation}
\begin{split}
\meff^2 \simeq & (m_1  (1 - |U_{\mathrm{e} \, 3}|^2))^2 
 +  \deltaatm|U_{\mathrm{e} \, 3}|^4 
+ \deltalsnd |U_{\mathrm{e} \, 4}|^4 \\
& + 2m_1\sqrt{m_1^2 + \deltaatm}|U_{\mathrm{e} \, 3}|^2  
(1 - |U_{\mathrm{e} \, 3}|^2)\cos\alpha_{31} \\
& + 2m_1\sqrt{m_1^2 + \deltalsnd}|U_{\mathrm{e} \, 4}|^2  
(1 - |U_{\mathrm{e} \, 3}|^2)\cos\alpha_{41} \\
& + 2\sqrt{m_1^2 + \deltaatm}
\sqrt{m_1^2 + \deltalsnd}|U_{\mathrm{e} \, 3}|^2|U_{\mathrm{e} \, 4}|^2
\cos(\alpha_{31} - \alpha_{41})
\end{split}
\label{31bmeffSMAph}
\end{equation}

\noindent where we took into account the
upper limits $|U_{\mathrm{e} \, 3}|^2 < 0.08$ and
$|U_{\mathrm{e} \, 4}|^2 < 0.01$.
If the element 
$|U_{\mathrm{e} \, 3}|$  of the neutrino
mixing matrix is rather small,
say $|U_{\mathrm{e} \, 3}|^2 \ltap (0.5 - 1.0)\times 10^{-2}$,
we would have a further simplification
of the expression for \meff,
as it follows from eq. (\ref{31bmeff}):
\begin{equation}
\meff \simeq  |m_1 + 
\sqrt{m_1^2 + \deltalsnd}~|U_{\mathrm{e} \, 4}|^2 e^{i \, \alpha_{41}} |.
\label{31bmeffSMAUe3}
\end{equation}
\noindent For $m_1 \gg \sqrt{\deltalsnd}~|U_{\mathrm{e} \, 4}|^2$,
i.e., for $m_1 \gtap  0.04$ eV (see eq. (\ref{31bmeff4th})),
eq. (\ref{31bmeffSMAUe3}) reduces to
$\meff \simeq m_1$, while if
$m_1 \ll \sqrt{\deltalsnd}~|U_{\mathrm{e} \, 4}|^2$,
$\meff \simeq \sqrt{\deltalsnd}~|U_{\mathrm{e} \, 4}|^2 \ltap 
5\times 10^{-3}$ eV. For the ``intermediate''
values of $m_1 \sim \sqrt{\deltalsnd}~|U_{\mathrm{e} \, 4}|^2$,
the following relation would be valid:
\begin{equation}
4m_1\sqrt{\deltalsnd}~|U_{\mathrm{e} \, 4}|^2~  
\sin^2{{\alpha_{41}}\over{2}} = 
(m_1 + \sqrt{\deltalsnd} |U_{\mathrm{e} \, 4}|^2)^2 - \meff^2
\label{31bmeffSMAUe3ph}
\end{equation}
%
\noindent 
In deriving this relation
we have neglected 
the terms which are either much smaller 
than $m_1^2$, or are smaller
than $\sim 10^{-5}~{\rm eV^2}$.
Thus, in the case of the SMA MSW solution 
of the solar neutrino problem,
sufficiently small value of 
$|U_{\mathrm{e} \, 3}|^2$ 
and for the 3+1B neutrino 
mass spectrum under discussion, 
the measurement of \deltalsnd, 
$|U_{\mathrm{e} \, 4}|^2$ and \meff might allow to 
obtain combined constraints on the mass of the
lightest neutrino, $m_1$ and on the CP-violating 
phase $\alpha_{41}$.

 The relation (\ref{31bphasec}), 
valid for negligible $m_1$,
simplifies also in
the case of the SMA MSW solution 
of the solar neutrino problem. 
For this solution 
the terms $\sim \sqrt{\deltasol}\sin^2\theta_{\odot}$ 
in eq. (\ref{31bphasec}) can be neglected. 
In this approximation only one Majorana 
CP-violating phase ``survives'' in eq. (\ref{31bphasec}) 
and we get:
\begin{equation}
\begin{split}
& 4\sqrt{\deltaatm} \sqrt{\deltalsnd}~|U_{\mathrm{e} \, 3}|^2~ 
|U_{\mathrm{e} \, 4}|^2~ \sin^2{{(\alpha_{42} - \alpha_{32})}\over{2}}\\ 
& \simeq (\sqrt{\deltaatm}~|U_{\mathrm{e} \, 3}|^2  +
\sqrt{\deltalsnd}~|U_{\mathrm{e} \, 4}|^2)^2 - \meff^2~.   
\label{31bphasec2}
\end{split}
\end{equation}
%

 We have considered so far 
the cases of 3+1B neutrino mass spectrum,
with  $m_1 \ll 0.02$ eV 
and $0.02~{\rm eV} \ltap m_1 \leq 0.02$ eV,
corresponding to mass spectra of
$\nu_{1,2,3}$ of
hierarchical type 
and with partial hierarchy.
For $m_1 \gtap 0.3$ eV,
\meff is equal, 
up to corrections which do not exceed
$\sim 5\times 10^{-3}$ eV,
to the effective Majorana mass 
$|<m>|_{3-\nu}$
in the case of mixing of 
3 quasi-degenerate
Majorana neutrinos.
Detailed predictions for
$|<m>|_{3-\nu}$ for the latter case 
were derived in ref.~\cite{BPP1} 
(Section 6).
Up to the corrections 
indicated above,
the predictions obtained in \cite{BPP1}
are valid also
for the spectrum of the 3+1B type
and we are not going to 
reproduce them here.
It is interesting to note
that for $m_1 \gtap 0.3$ eV,
the masses of the three lighter neutrinos
coincide up to corrections smaller than 10\%:
$m_{1} \simeq m_{2} \simeq m_{3}$.
As it is not difficult to convince oneself,
under the above conditions 
we have $m_{\nu_e} \simeq m_1$,
where $m_{\nu_e}$ is the mass measured 
in $^3$H $\beta-$decay experiments.
Therefore, the KATRIN  experiment
\cite{KATRIN} can provide unique information 
on $m_1$ if the 3+1B scheme is realized.
The measurement of $m_1 \gtap 0.4$ eV
will allow to determine, in particular,
the values of the masses of the 
neutrinos $\nu_{2,3,4}$, i.e.,
the neutrino mass spectrum.


\section{\textbf{ The 3+1C mass spectrum}}
\label{3+1C}

 The 3+1C neutrino mass spectrum 
differs from the 3+1B spectrum 
by the role played by $\Delta m^2_{21}$ and
$\Delta m^2_{32}$. We have again
three nearly-degenerate
neutrinos $\nu_{1,2,3}$ with masses sufficiently 
smaller than the mass of the fourth one,
$m_{1} \simeq m_{2} \simeq m_{3} <~(\ll)~ m_{4}$, 
as is shown graphically below:

\unitlength=1mm
\begin{center}
\begin{picture}(100,40)
\put(20,5){\line(1,0){20}}
\put(20,10){\line(1,0){20}}
\put(20,13){\line(1,0){20}}
\put(20,35){\line(1,0){20}}
\put(30,5){\line(0,1){30}}
\put(45,4.5){\vector(0,1){5.5}}
\put(45,10){\vector(0,-1){5.5}}
\put(45,10){\vector(0,1){3.5}}
\put(45,13.5){\vector(0,-1){3.5}}
\put(72,5){\vector(0,1){30}}
\put(72,35){\vector(0,-1){30}}
\put(48,11){\deltasol}
\put(48,6){\deltaatm}
\put(75,18){\deltalsnd}
\put(17,4){1}
\put(17,9){2}
\put(17,12.3){3}
\put(17,34.2){4}
\end{picture}
\end{center}
%
\noindent
This mass pattern 
corresponds to
\begin{equation}
\Delta m^2_{32}  \ll \Delta m^2_{21} \ll \Delta m^2_{41}.
\label{31Cspectrum}
\end{equation}
Equation ~(\ref{31Cspectrum}) leads to
the identification:
\begin{equation}
\Delta m^2_{32}  
              \equiv    \Delta  m^2_{\odot},\ \ 
\Delta m^2_{31}  
              \equiv    \Delta  m^2_{\mathrm{atm}}, \ \ 
\Delta m^2_{41}  
              \equiv   \Delta  m^2_{\mathrm{LSND}}.
\label{31cpattern}  
\end{equation}
The masses of the three lighter neutrinos, $m_{1,2,3}$,
can form a spectrum of the inverted hierarchy, 
partial inverted hierarchy, or of quasi-degenerate 
type (see, e.g., \cite{BPP1}).
In the first case one 
has $m_1 \ll m_{2,3} \simeq \sqrt{\deltaatm}$,
which implies $m_1 \ll 0.02$ eV; 
in the second one has
$0.02~{\rm eV} \ltap m_1 \leq 0.2$ eV,
while the third corresponds to
$m_1 > 0.2$ eV \cite{BPP1}.
We will comment on the predictions
for \meff when $m_1 > 0.2$ eV
after discussing in detail the
other two possibilities. 

 In accordance with eq. (\ref{31cpattern}) we have:
\begin{itemize}
\item[i)]
$|U_{\mathrm{e} \, 1}|^2$ is constrained by the  
CHOOZ limit, which at 
$99 \%$~C.L. reads \cite{CHOOZ,Gonza3nu}
$|U_{\mathrm{e} \, 1}|^2 <0.08$;
\item[ii)]
$|U_{\mathrm{e} \, 2}|$ and $|U_{\mathrm{e} \, 3}|$ 
are related to the solar neutrino 
mixing angle $\theta_{\odot}$,
\begin{align}
|U_{\mathrm{e} \, 2}|^2  =  & \cos^2 \theta_\odot  
\ ( 1 -  |U_{\mathrm{e} \, 1}|^2 ), \nonumber  \\
|U_{\mathrm{e} \, 3}|^2  =  &  \sin^2 \theta_\odot 
\  ( 1 -  |U_{\mathrm{e} \, 1}|^2 ); 
\end{align}
%
%
\item[iii)]
$|U_{\mathrm{e} \, 4}|^2$ is bounded from above and from below
by SBL data \cite{LSND,BUGEY}: 
$ 2 \times 10^{-4} < |U_{\mathrm{e} \, 4}|^2
< 1 \times 10^{-2}$.
\end{itemize}

\noindent

Using 
eq.~(\ref{31cpattern}), one obtains for 
the neutrino masses $m_{2,3,4}$:
\begin{align}
m_2 = & \sqrt{m_1^2 - \deltasol + \deltaatm}; \nonumber  \\
m_3 = & \sqrt{m_1^2 + \deltaatm}; \nonumber  \\ 
m_4 = & \sqrt{m_1^2 + \deltalsnd}.
\end{align}
%
For $\deltasol \ltap 2\times 10^{-4}~{\rm eV^2}$ we have
$\deltasol/\deltaatm \ll 1$. Correspondingly, 
the \deltasol correction in \meff, which is
proportional to
$ 0.5\deltasol/\sqrt{m_1^2 + \deltaatm}$,
does not exceed approximately $1.1\times 10^{-3}$ eV
and we will neglect it. In this approximation
\meff does not depend on \deltasol. If values 
of $\deltasol \simeq 7\times 10^{-4}~{\rm eV^2}$ 
are allowed by the solar neutrino and CHOOZ data
\cite{Gonza3nu},  \deltasol 
in the expression for 
$m_2$ can be  non-negligible for
values of $\deltaatm \simeq 
(1.5 - 3.0)\times 10^{-3}~{\rm eV^2}$ and 
$m_1 \ltap \deltaatm$.
For $m_1 \simeq 0$ eV, 
$\deltaatm = 1.5\times 10^{-3}~{\rm eV^2}$
and $\deltasol = 2\times 10^{-4};~ 
7\times 10^{-4}~{\rm eV^2}$, for instance,
we have $m_2 \simeq 3.6\times 10^{-2};~ 
2.8\times 10^{-2}~{\rm eV}$.
For $m_1^2 \gtap \deltaatm$, the 
relative contribution
of \deltasol in $m_2$ is even smaller.
Thus, taking the effect due to 
the presence of $\deltasol$ 
in the expression for 
$m_2$ into account leads to a 
change by not more than $\sim$30\%  
in the prediction for the contribution
of the term $\sim m_2$ in \meff.

 In the case of the neutrino mass spectrum 
of the 3+1C type under discussion 
the effective Majorana mass  \meff is given by:
\begin{equation}
\begin{split}
\meff \simeq &  |m_1 |U_{\mathrm{e} \, 1}|^2  + 
\sqrt{m_1^2 - \deltasol + \deltaatm}~(1 - |U_{\mathrm{e} \, 1}|^2)
~\cos^2 \theta_\odot e^{i \, \alpha_{21}} \\
& +
\sqrt{m_1^2 + \deltaatm}(1 - |U_{\mathrm{e} \, 1}|^2)
~\sin^2 \theta_\odot e^{i \, \alpha_{31}}   
 +  \sqrt{m_1^2 + \deltalsnd} |U_{\mathrm{e} \, 4}|^2 e^{i \, \alpha_{41}} |.
\end{split}
\label{meff31cc}
\end{equation}
%
In the analysis which follows we 
will neglect  \deltasol in the expression for $m_2$.
In this approximation the expression for \meff 
simplifies somewhat:
\begin{equation}
\begin{split}
\meff \simeq &  | m_1 |U_{\mathrm{e} \, 1}|^2 + 
 \sqrt{m_1^2 + \deltaatm}(1 - |U_{\mathrm{e} \, 1}|^2)
(\cos^2 \theta_\odot e^{i \, \alpha_{21}}  + 
\sin^2 \theta_\odot e^{i \, \alpha_{31}})  \\
 &  +  \sqrt{m_1^2 + \deltalsnd} e^{i \, \alpha_{41}} 
|U_{\mathrm{e} \, 4}|^2|.
\end{split}
\label{meff31c}
\end{equation}

  The different terms in the r.h.s of 
eq. (\ref{meff31c}) for \meff can, in general,
be of the same order. Thus, all
three CP-violating phases 
$\alpha_{21}$, $\alpha_{31}$ and $\alpha_{41}$ 
have to be taken into account in the analysis of the
allowed ranges of values of \meff. 
This analysis simplifies considerably, however, 
due to the fact that the absolute value 
of the sum of the first three contributions 
in \meff due to $\nu_{1,2,3}$, $|<m>|_{3-\nu}$, 
is equal to the absolute value of the 
effective Majorana mass
in the case of 3-neutrino mixing
and neutrino mass spectrum of inverted hierarchy
or of partial inverted hierarchy type.
Detailed predictions for $|<m>|_{3-\nu}$
have been obtained recently 
in ref.~\cite{BPP1}.  

 Indeed, one can rewrite  \meff as follows:
\begin{equation}
\meff = | |<m>|_{3-\nu} e^{i( \gamma- \alpha_1)} + 
 \sqrt{ \deltalsnd} |U_{\mathrm{e} \, 4}|^2 e^{i \, \alpha_{41}} |,
\label{meff31c2}
\end{equation}
%
%
where $\gamma$ is  the overall phase 
of the sum of the first three terms 
in eq. (\ref{meff31c}).
The CP-violating phase $\gamma$ 
is related to the CP-violating phases
$\alpha_{21}$ and  $\alpha_{31}$
in a complicated way which depends  
on $\meff_1 =  m_1 |U_{\mathrm{e} \, 1}|^2$, 
$\meff_2 = \sqrt{m_1^2 + \deltaatm} 
(1 - |U_{\mathrm{e} \, 1}|^2)\cos^2 \theta_\odot$,
and $\meff_3 = \sqrt{m_1^2 + \deltaatm}
(1 - |U_{\mathrm{e} \, 1}|^2) \sin^2 \theta_\odot$.  
Representing $<m>_{3-\nu}$ as a vector 
in the complex plane,
which is a sum of three vectors,
$<m>_1$, $<m>_2$ and
$<m>_3$, as is 
sketched in Fig.~\ref{figure:3+1B},
it becomes clear that
the phase $(\gamma -\alpha_1)$ is 
equal to the angle formed by the 
$<m>_1$ (i.e., the $x$-axis) 
and $<m>_{3-\nu}$. If, e.g., 
the first term $\meff_1$ is negligible,
one finds:
\begin{equation}
\gamma - \alpha_1 = \alpha_{21} + 
\arctan \frac{\sin \alpha_{32}}{\arctan^2 \theta_\odot 
+ \cos \alpha_{32}}.
\end{equation}
In the case of CP-conservation, 
the phase $(\gamma -\alpha_1)$ 
takes into account 
the overall sign of the term 
$<m>_{3-\nu}$.

  The allowed intervals of values of $|<m>|_{3-\nu}$ 
found at $90 \%$~C.L. ($99 \%$~C.L) 
in ref.~\cite{BPP1} read:
\begin{align}
0 \  (0) \eV \leq |<m>|_{3-\nu} \leq & 6.8 \ (  8.1)  \times 10^{-2} \eV,  
\qquad m_1 \ll 0.02\mbox{~eV}; 
         \\
0  \  (0) \eV \leq |<m>|_{3-\nu} \leq & 2.1 \ (2.2) \times 10^{-1} \eV,  
\qquad 0.02~{\rm eV} \ltap m_1 \leq 0.2
\mbox{~eV}. 
\label{31c3meffv}
\end{align}
%
The upper bounds in eqs. (132) - (\ref{31c3meffv}), as can 
be shown \cite{BPP1},
do not depend on $\theta_{\odot}$ and thus on the solution 
of the solar neutrino problem. The lower bounds in  
(124) and (\ref{31c3meffv}), however, depend 
on $\cos2\theta_{\odot}$
and vary with the solution of 
the solar neutrino problem.
The values reported correspond to the LMA MSW and 
to the LOW-QVO solutions, for which $\cos2\theta_{\odot} = 0$
is possible. For the SMA MSW solution the lower 
bounds on $|<m>|_{3-\nu}$ read \cite{BPP1}:
$3.7~(3.3)  \times 10^{-2}$ eV  
for the spectrum with inverted hierarchy
(or $m_1 \ll 0.02$ eV),
and $2.8~(1.8)  \times 10^{-2}$ eV 
for the spectrum with partial inverted hierarchy
($0.02~{\rm eV} \ltap m_1 \leq 0.2$ eV).

The  fourth term 
$\sim |U_{\mathrm{e} \, 4}|^2$ in eq. (\ref{meff31c})
can take the following values:
\begin{equation}
3.0   \times 10^{-4} \eV \leq |<m>|_4 \equiv  
\sqrt{m_1^2 + \deltalsnd}|U_{\mathrm{e} \, 4}|^2 
\leq 5.0 \times 10^{-3} \eV,
\end{equation}
%
where we have used the allowed values of 
$|U_{\mathrm{e} \, 4}|^2$ 
and \deltalsnd, found at $95 \%$~C.L. 
in ref.~\cite{LSND}.

If CP-parity is conserved there exist
two possibilities.

{\bf Case A.}
If $\gamma - \alpha_1 = \alpha_{41} = 0, \pm \pi$,
the two contributions in eq. (\ref{meff31c2})
sum up and we have:
\begin{align}
3.0 \times 10^{-4} \eV \leq \meff  \leq &  
6.8 \ (  8.1) \times 10^{-2} \eV,  \qquad 
m_1 \ll 0.02 \mbox{~eV~(negligible $m_1$)}; \\
3.0 \times 10^{-4} \eV \leq \meff \leq & 2.1 \ (2.2) \times 10^{-1} \eV, 
\qquad \mbox{for $0.02~\eV \ltap m_1 \leq 0.2 \eV$}.
\end{align}
%
\noindent The lower limits in these 
intervals of allowed values
are due to the term $|<m>|_4$, while the 
maximal values are the same as
those in eqs (124) and (\ref{31c3meffv}) 
and correspond to
the LMA and LOW-QVO solutions. For the SMA MSW solution 
they are given in the text after eq. (\ref{31c3meffv}).

{\bf Case B.}
If $\gamma - \alpha_1 = \alpha_{41} + \pi = 0, \pm \pi$,
there can be  cancellations 
between the two terms $|<m>|_{3-\nu}$ and $\meff_4$.
Correspondingly, there is no 
non-trivial  lower bound on the 
predicted values of \meff.
The maximal allowed values of \meff are identical  
to those found in the case A:
\begin{align}
0 \leq \meff  \leq &  6.8 \ (  8.1) \times 10^{-2} \eV,  
\qquad m_1 \ll 0.02 \mbox{~eV~(negligible $m_1$)}; \\
0  \leq \meff \leq &  2.1 \ (2.2) \times 10^{-1} \eV,  
\qquad \mbox{for $0.02~\eV \ltap m_1 \leq 0.2 \eV$}.
\end{align}
%

 In Fig.~\ref{figure:3+1C01} 
we show the dependence of \meff 
on $\cos 2 \theta_\odot$ for the two limiting 
cases of values 
of $m_1$, i.e. for $m_1 = 0$ and $m_1 = 0.2 \eV$. 
As Fig.~\ref{figure:3+1C01} illustrates,
in a large region of the parameter space 
one can have values of $\meff \geq 0.01 \eV$, 
which can be tested in the next generation of the 
\betabeta-decay experiments.

  In the case of CP-violation, 
three CP-violating phases 
play a role in the determination of \meff 
and therefore it is rather difficult 
to obtain significant constraints on their values. 
However, such constraints might be obtained
if the term $m_1|U_{\mathrm{e} 1}|^2$  
gives a negligible contribution in \meff
and/or in the case of the SMA MSW solution of the
solar neutrino problem.
If, for instance, $|U_{\mathrm{e} 1}|^2$ 
turns out to be relatively small,
say $|U_{\mathrm{e} 1}|^2 \ltap 0.01$,
for $m_1 \leq 0.2$ eV
one could neglect the term 
$m_1|U_{\mathrm{e} 1}|^2$ in eq. (\ref{meff31c}).
Under this condition
(i.e., up to corrections $\sim 2\times 10^{-3}$ eV)
there will be just
three terms contributing to \meff and
two relevant CP-violating phase differences,
$\alpha_{32}$ and $\alpha_{42}$.
If the other parameters in eq. ~(\ref{meff31c})
are known with sufficient accuracy, a 
measurement of $\meff \neq 0$ would allow 
to constraint the values of  
$\alpha_{32}$ and $\alpha_{42}$, 
which have to satisfy:
\begin{equation}
\begin{split}
& \meff^2 -~(m_1^2 + \deltaatm)~(1 - {1\over{2}}\sin^22\theta_\odot)
~-~\deltalsnd~ |U_{\mathrm{e} 4}|^4  \\
& = {1\over{2}}(m_1^2 + \deltaatm)~\sin^2 2\theta_\odot~\cos \alpha_{32} \\
& +~2\sqrt{(m_1^2 + \deltaatm)  
(m_1^2 + \deltalsnd)}~|U_{\mathrm{e} 4}|^2~ 
(\cos^2 \theta_\odot~\cos \alpha_{42} +
\sin^2 \theta_\odot~\cos (\alpha_{42} - \alpha_{32})). 
\end{split}
\label{cp31c}
\end{equation}
 
To illustrate the correlation between the 
values of  $(\alpha_4 - \alpha_2)$ 
and $(\alpha_3 - \alpha_2)$ 
eq. (\ref{cp31c}) would imply,
we show in Fig. ~\ref{figure:3+1C02} 
the phase $(\alpha_4 - \alpha_2)$ versus
$(\alpha_3 - \alpha_2)$ 
for a number of fixed values of \meff and
for two different sets of values of the  
other parameters entering into eq. (\ref{cp31c}). 

 Let us note that in the case of the SMA MSW solution
of the solar neutrino problem the terms $\sim \sin^2\theta_{\odot}$  
can be neglected in eq. (\ref{cp31c}). Taking this into account one
finds the relation:  
\begin{equation}
4\sqrt{(m_1^2 + \deltaatm)  
(m_1^2 + \deltalsnd)}~|U_{\mathrm{e} 4}|^2~ 
\sin^2 {{\alpha_{42}}\over{2}} 
\simeq (\sqrt{m_1^2 + \deltaatm}  
+ \sqrt{m_1^2 + \deltalsnd})^2 - \meff^2~.
\label{cp31cSMA}
\end{equation}
%
\noindent This relation can be used, in principle, 
to obtain combined constraints on $m_1$ and
$\alpha_{42}$.

 We have assumed so far 
that either $m_1 \ll 0.02$ eV 
or $0.02~{\rm eV} \ltap m_1 \leq 0.2$ eV,
i.e., that the mass spectrum of
the three lighter neutrinos
$\nu_{1,2,3}$ is either of
the inverted hierarchy or of
partial inverted hierarchy type.
If $m_1 \gtap 0.3$ eV, 
the mass spectrum of
$\nu_{1,2,3}$ will be 
of quasi-degenerate type. 
In this case, 
up to corrections which are smaller
than $\sim 5\times 10^{-3}$ eV,
the spectra 3+1B and 3+1C are 
indistinguishable
in what regards the predictions
for \meff
\footnote{The two expressions for $\meff
\simeq |<m>|_{3-\nu}$
differ only by the two Majorana
CP-violating phases
present in \meff and multiplying
terms in \meff, which, apart 
from the phase factors, are identical
for the two spectra:
these phases are, e.g., $\alpha_{21}$
and $\alpha_{31}$ if the spectrum is
of 3+1B type, and
$(\alpha_{31} - \alpha_{21})$ 
and $(-\alpha_{31})$
for the 3+1C spectrum. 
Let us note that at present 
there does not exist 
any experimental information
on the phases  
$\alpha_{21}$ and $\alpha_{31}$.}. 
Correspondingly, the comments 
and conclusions made for the 3+1B spectrum
at the end of Section 7 
concerning \meff
in the case when $m_1 \gtap 0.3$ eV
and the fact that in this case
$m_{\nu_e} \simeq m_1$,
are valid also for the 3+1C spectrum.


\vspace{0.3cm} 
\section{\textbf{Conclusions.}}
\label{Conclusions}

  Assuming mixing of four  
massive Majorana neutrinos,
we have investigated in detail
the implications of the 
solar, atmospheric, LSND 
and other neutrino oscillation data
for the predictions of 
the possible values of
effective Majorana mass, \meff,
which controls the \betabeta-decay
rate. Four-neutrino mixing  
allows to explain the existing evidences 
and indications for neutrino oscillations,
obtained in the solar, atmospheric and
LSND experiments,
which require three different 
neutrino mass-squared differences,
\deltasol, \deltaatm and \deltalsnd.
This is the minimal scheme 
which describes
all available neutrino oscillation data.
The fourth 
neutrino involved in the mixing 
must be \cite{BGG99,4numodels} 
a sterile neutrino, $\nu_s$ \cite{Pont67}.
The sterile neutrino can be accommodated, e.g., 
in extensions of the Standard Theory, 
which include $SU(2)_L$ singlet 
right-handed (RH) neutrino field(s) 
(see, e.g., \cite{4numodels}).
We have considered all different 
types of neutrino mass spectrum,
which are
compatible with the neutrino 
oscillation data and with
the data from $^3$H $\beta-$decay
direct neutrino mass measurements 
\cite{BGG4nu98,Gosw4nu97,Barger4nu98,
Barger31,PeresS31,Giunti31,Grimus31}: 
2+2A, 2+2B, 3+1A, 3+1B and 3+1C.
The results 
of the neutrino oscillation fits of the
latest solar and atmospheric neutrino data
\cite{SKatm00,SKYSuz00,Fogli4nu,Gonza4nu,Fogli00,Gonza3nu},
and of the data from the LSND \cite{LSND}, 
KARMEN \cite{KARMEN} and
BUGEY \cite{BUGEY} experiments, 
have been used in our analysis.
The constraints on the neutrino 
oscillation parameters
obtained in the CHOOZ \cite{CHOOZ} and
Palo Verde \cite{PaloV} 
experiments have also been utilized.

  All through the study we have assumed that 
the \betabeta-decay is 
induced {\it only} 
by the (V-A) charged current weak
interaction via the
exchange of virtual massive Majorana neutrinos,
transforming the initial nucleus into
the final state nucleus and two free
electrons. In the case of mixing 
involving four massive Majorana neutrinos,
the \betabeta-decay effective Majorana mass
\meff depends, in general,
on the lightest neutrino mass $m_1$,
on \deltasol, \deltaatm, \deltalsnd,
on the mixing angle $\theta_{\odot}$
characterizing the transitions
of the solar neutrinos, and 
on three physical (Majorana) CP-violating 
phases, $\alpha_{21}$, $\alpha_{31}$ and
$\alpha_{41}$.
In the 2+2A and 2+2B schemes,
\meff depends indirectly also on the
parameter $\cos^2\beta \leq 1$, 
which determines the fraction
of the solar $\nu_e$ and of the
atmospheric $\nu_{\mu}$ ($\bar{\nu}_{\mu}$), 
which can oscillate into the
sterile neutrino $\nu_s$ 
\cite{Fogli4nu,Gonza4nu}.
However, for most of the
neutrino mass spectra of interest
the expression for \meff,
to a good approximation, takes rather 
simple forms, depending only on 
few of the indicated 
physical quantities.

  For each of the five types of 
neutrino mass spectrum considered we 
have derived detailed predictions for
the values of \meff for the three solutions 
of the solar neutrino problem,
favored by the current solar neutrino data:
the LMA MSW, the SMA MSW and the LOW-QVO one.
In each of these cases we have identified 
the ``just-CP-violation'' region
of values of \meff whenever it existed:
a value of \meff in this
region would unambiguously 
signal the presence of
CP-violation in the lepton sector.
Analyzing the case of CP-conservation, we 
have derived predictions for \meff corresponding 
to all possible sets of values of the
relative CP-parities of 
the massive Majorana neutrinos. 
We have investigated
the possibility of cancellation 
between the different terms contributing to
\meff. The cases when such a cancellation
is impossible and there exists
a non-trivial lower bound
on \meff  were identified and the
corresponding lower bounds
were given.

  We have also analyzed in detail 
the question of whether a measurement of 
$\meff \gtap 0.01$ eV in the next generation of
\betabeta-decay experiments \cite{NEMO3,CUORE,EXO,GENIUS} 
combined with the data from the solar, atmospheric,
reactor and accelerator neutrino 
oscillation experiments and from the 
future $^3$H $\beta-$decay experiment KATRIN
\cite{KATRIN} would allow,
and under what conditions,
i) to determine the absolute values of the 
neutrino masses and thus the 
neutrino mass spectrum, and
ii) to establish the existence of 
CP-violation in the lepton sector. 
We have pointed out, in particular, that 
in certain cases of 4-neutrino 
mass spectra (2+2A, 3+1A) 
there exists a direct 
relation between the effective Majorana
mass \meff and the neutrino mass 
measured in $^3$H $\beta-$decay,
$m_{\nu_e}$, i.e., 
\meff can be proportional to
$m_{\nu_e}$ and we can have
even $\meff \simeq m_{\nu_e}$,
and that the measurement of 
$\meff \gtap 0.01$ eV 
\cite{CUORE,EXO,GENIUS} and of
$m_{\nu_e} \gtap 0.4$ eV \cite{KATRIN}
will give in these cases 
the unique possibility to determine the 
absolute values of all four 
neutrino masses and 
to obtain information on
CP-violation in the lepton sector.

   More specifically, we have found that 
if the neutrino mass 
spectrum is of the 2+2A type 
(with $m_4^2 - m_3^2 = \deltasol$,
see Section 4), or
of 3+1A type (with 
$m_2 \simeq m_3 \simeq m_4 > m_1$, 
see Section 6), one has $m_{\nu_e}
\simeq \sqrt{m_1^2 + \deltalsnd} \gtap 0.4$ eV,
where the lower bound is determined by
the minimal value of \deltalsnd, 
allowed by the LSND data with KARMEN and 
BUGEY limits taken into account. 
Thus, we find that for any value of $m_1$, 
the neutrino mass 
$m_{\nu_e}$ is predicted to lie 
in the range planned to be
probed by the future
Karlsruhe-Mainz-Troitzk  
experiment KATRIN \cite{KATRIN}.
Therefore the realization of the KATRIN project 
will allow to test directly the possibility 
of 2+2A and 3+1A types of neutrino mass spectrum.
A measurement of $m_{\nu_e} \gtap 0.4$ eV and 
a more accurate knowledge of \deltalsnd 
would permit to determine the value of
$m_1$. This would allow
to determine also the values of $m_{2,3,4}$. 

 In the 2+2A scheme, 
up to corrections $\sim 10^{-3}$ eV
only the two heavier 
Majorana neutrinos $\nu_{3,4}$
make contributions to \meff.
The effective Majorana mass
\meff depends on 
one CP-violating phase, 
$\alpha_{34} = \alpha_{31} - \alpha_{41}$. 
For the SMA solution of 
the solar neutrino problem
one has independently
of whether CP is conserved or not,
$\meff \simeq m_{\nu_e}
\simeq \sqrt{m_1^2 + \deltalsnd} \gtap 0.4$ eV.
The same result is valid for the
LMA and the LOW-QVO solutions
in the case of CP-conservation
if $\phi_3 = \phi_4$, 
where $i\phi_{3,4}$ are the CP-parities 
of the Majorana neutrinos $\nu_{3,4}$.
For $\phi_3 = - \phi_4$ we find 
$\meff \simeq \sqrt{m_1^2 + \deltalsnd}|\cos 2\theta_{\odot}| 
\simeq m_{\nu_e}|\cos 2\theta_{\odot}|$
and in most of the allowed 
region of the relevant parameter space
one has $\meff \gtap 0.01$ eV.
The measurement of 
$\meff \gtap 0.01$ eV,
$m_{\nu_e} \gtap 0.4$ eV,
and a more precise determination
of $\cos 2\theta_{\odot}$
would allow in the case of the
LMA MSW or LOW-QVO solutions
of the solar neutrino problem
and neutrino mass spectrum of
2+2A type to establish whether
CP-symmetry is violated in the lepton
sector, and will permit to
determine the relative CP-parity of 
the neutrinos $\nu_{3,4}$
if one of the CP-conservation 
cases indicated above will 
be found to be realized.
There exists, in particular,
a ``just-CP-violation'' region
of values of \meff:
a measured value of \meff in this
region would imply that 
the CP-symmetry is 
violated in the lepton sector.
Some of our results for the 
2+2A spectrum are illustrated 
in Figs. (\ref{figure:3+1A01}) -
(\ref{figure:3+1A04}).

 In the case of neutrino mass spectrum of the
2+2B type ($m_4^2 - m_3^2 = \deltaatm$),
discussed in Section 5,
{\it one can have $\meff \gtap 3.0\times 10^{-2}$ eV
only if $m_1 \gtap 10^{-2}$ eV}:
for $m_1 \leq 10^{-3}$ eV, for instance,
we find $\meff \ltap 2.0\times 10^{-2}$ eV
for the LMA MSW solution, and 
$\meff \ltap 6.0 \times 10^{-3}$ eV
for the SMA MSW and LOW-QVO solutions.
As $m_1$ increases, 
the maximal allowed 
values of \meff also increase,
and for $m_1 \simeq 0.10 \eV$ we get
${\rm max}(\meff) \simeq 0.10$ eV
for all solutions of the solar neutrino problem. 
A measured value of $\meff \gtap 0.03$ eV in the case
of the 2+2B neutrino mass spectrum 
would imply a lower limit on the mass $m_1$,
$m_1 \gtap 10^{-2}$ eV;
and it would essentially determine the value of 
$m_1$ if the SMA MSW solution of the solar neutrino
problem turns out to be the correct one 
since in this case $\meff \simeq m_1$.
These results are illustrated 
in Fig.~\ref{figure:2+2B02}.
For $m_1 \ltap 10^{-2}$ eV,
we have  $\meff \ltap 3.0\times 10^{-2}$ eV,
three CP-violating phases 
enter into the expression for \meff
and the CP-violation analysis 
in this case would be rather
complicated and hardly conclusive.
If $m_1 \gtap 0.10$ eV and  
it is found that 
$\meff \gtap 2\times 10^{-2}$ eV, however,
up to corrections 
$\sim 5.0\times 10^{-3}~{\rm eV}$
only one CP-violating phase, 
$\alpha_{21}$, is relevant for 
the determination of \meff:
$\meff \simeq m_1| \cos^2 \theta_\odot 
  + e^{i \alpha_{21}} \sin^2 \theta_\odot |$.
Correspondingly, one has
$m_1| \cos 2\theta_\odot| \ltap \meff \ltap m_1$;
in the case of the SMA solution
$\sin^2 \theta_\odot \ltap 2\times 10^{-3}$ and  
$\meff \simeq m_1$.

  As we have already indicated above, the 3+1A
scheme can be tested in 
the future $^3$H $\beta-$decay experiment KATRIN
\cite{KATRIN}. In this scheme
$\meff \sim m_{\nu_e}
\simeq \sqrt{m_1^2 + \deltalsnd} \gtap 0.4$ eV.
In a large region of the relevant parameter
space one has $\meff \gtap 0.01$ eV,
and even $\meff \gtap 0.10$ eV,
which can be tested in the current and
future \betabeta-decay experiments.
Up to corrections $\sim 10^{-2} m_1 \leq 10^{-2}$ eV, 
\meff depends on $m_{\nu_e}
\simeq \sqrt{m_1^2 + \deltalsnd}$,
$\theta_{\odot}$,
on the element $|U_{\mathrm{e} 2}|$
of the neutrino mixing matrix, 
which is constrained
by the CHOOZ and Palo Verde results,
and on two CP-violating phases,
$\alpha_{32}$ and $\alpha_{42}$. 
There exists ``just-CP-violating'' region
of values of \meff:
this region can be spanned by 
the values of \meff
only if the CP-symmetry is violated.
A sufficiently accurate 
measurement of $\meff \gtap 0.01$ eV  
and of $m_{\nu_e} \gtap 0.4$ eV,
as well as of \deltalsnd, $\theta_{\odot}$ 
and $|U_{\mathrm{e} 2}| \neq 0$, will allow 
to get information on the 
CP-violation in the lepton sector,
or on the relative CP-parities
of the massive Majorana neutrinos 
if the relevant CP-violating phases  
take their CP-conserving values.
The CP-violation analysis  
simplifies considerably
if $|U_{\mathrm{e} 2}|^2$ 
is negligible as well as in the case of the
SMA MSW solution of the solar neutrino problem.
If both $|U_{\mathrm{e} 2}|^2 \ltap 5\times 10^{-3}$
and the SMA MSW solution is realized,
the measurement of $\meff \gtap 10^{-2}$ eV 
will hardly provide
any information on the violation of the 
CP-symmetry in the lepton sector since
$\meff \simeq \sqrt{m_1^2 + \deltalsnd}~(1 + O(10^{-2}))~
\simeq m_{\nu_e}$.
Our results for the 3+1A spectrum
are illustrated in 
Figs.~\ref{figure:3+1A01} -
\ref{figure:3+1A04}.

 For the 3+1B neutrino mass spectrum 
($m_{1} \simeq m_{2} \simeq m_{3} <~(\ll)~m_{4}$,
$\deltasol = \Delta m^2_{21}$, 
$\deltaatm = \Delta m^2_{32}$,
see Section 7), the effective Majorana mass
\meff coincides, up to corrections
$\sim 5\times 10^{-3}$ eV
with  the effective Majorana mass 
in the 3-neutrino mixing case 
$|<m>|_{3-\nu}$
and 3-neutrino mass spectrum with
hierarchy, or with partial mass hierarchy,
or of quasi-degenerate type.
Detailed predictions for 
$|<m>|_{3-\nu}$ in the indicated
cases, most of which are valid 
for the 3+1B spectrum as well,
have been derived
recently in ref.~\cite{BPP1} 
(Sections 4, 6 and 7).
We have found, in particular, that 
{\it one can have
$\meff \gtap 3\times 10^{-2}$ eV only if
$m_1 \gtap 2\times 10^{-2}$ eV}. 
A measured value of 
$\meff \gtap 4\times 10^{-2}$ eV 
would imply for the 3+1B spectrum:  
$\meff \simeq m_1 (1 - |U_{\mathrm{e} \, 3}|^2) 
( 1 - \sin^2 2\theta_\odot~\sin^2(\alpha_{21}/2))$,
where $|U_{\mathrm{e} \, 3}|^2 < 0.08$ is constrained by the
CHOOZ and Palo Verde data.
For the SMA MSW solution of 
the solar neutrino problem 
($\sin^2 2\theta_\odot \ll 1$) 
and/or for ${\alpha_{21}} \simeq 2\pi k$, $k=0,1,...$,
the above relation offers the possibility 
to determine the mass of the lightest neutrino $m_1$.
Information on $m_1$ can be obtained in the 
future $^3$H $\beta-$decay experiment KATRIN
\cite{KATRIN}: in the 3+1B scheme one has
$m_{\nu_e} \simeq m_1$.
A measurement of $m_{\nu_e} \gtap 0.4$ eV
would allow to determine $m_{2,3,4}$.
Combined with a measurement
of $\meff \gtap 4\times 10^{-2}$ eV
in the case of the LMA or LOW-QVO solution
and a better determination 
of $|U_{\mathrm{e} \, 3}|^2$ 
would allow to obtain 
information on the
CP-violation in the lepton sector.
For $\meff \ltap 3\times 10^{-2}$ eV
the CP-violation analysis is 
rather complicated since it would
involve three CP-violating phases.
It simplifies, however, in
several physically interesting cases,
when the expression for 
\meff includes effectively only two or just one 
CP-violating phases:
i) $m_1$ is negligible, ii) SMA MSW solution 
of the solar neutrino problem, 
iii) $|U_{\mathrm{e} \, 3}|^2$ 
is rather small, e.g.,
$|U_{\mathrm{e} \, 3}|^2 \ltap (0.5 - 1.0)\times 10^{-2}$, 
and iv) a combination of any two of the 
above three possibilities.
Figures \ref{figure:31b01} - \ref{figure:3+1B03} 
illustrate some of the results for the 3+1B
neutrino mass spectrum.

 The 3+1C neutrino mass spectrum (Section 8)
differs from the 3+1B spectrum 
by the role played by $\Delta m^2_{21}$ and
$\Delta m^2_{32}$:
now $\deltasol = \Delta m^2_{32}$ and
$\deltaatm = \Delta m^2_{31} 
\simeq \Delta m^2_{21}$.
The effective Majorana mass
\meff coincides, up to corrections
$\sim 5\times 10^{-3}$ eV
with  the effective Majorana mass 
in the 3-neutrino mixing case, 
$|<m>|_{3-\nu}$,
and 3-neutrino mass spectrum with
inverted hierarchy, or with partial 
inverted hierarchy, or 
of quasi-degenerate type.
Detailed predictions for 
$|<m>|_{3-\nu}$ in the indicated
cases have been obtained in ref.~\cite{BPP1} 
(Sections 5, 6 and 7).
Most of them are valid 
for the 3+1C spectrum. 
As for the other types of neutrino 
mass spectrum considered,
one has $\meff \gtap 0.01$ eV
in a large region of the relevant parameter
space. For the  SMA MSW solution there
exist lower 
bounds on $\meff \simeq |<m>|_{3-\nu}$  \cite{BPP1}:
$|<m>|_{3-\nu} \gtap 3.7~(3.3)  \times 10^{-2}$ eV  
for the spectrum with inverted hierarchy
($m_1 \ll 0.02$ eV),
and $|<m>|_{3-\nu} \gtap 2.8~(1.8)  \times 10^{-2}$ eV 
for the spectrum with partial inverted hierarchy
($0.02~{\rm eV} \ltap m_1 \leq 0.2$ eV).
There are no non-trivial lower bounds 
on $\meff \simeq |<m>|_{3-\nu}$ in the case of
the LMA and LOW-QVO solutions.
The maximal allowed values of  
$\meff \simeq |<m>|_{3-\nu}$
for the 3-neutrino spectrum with inverted hierarchy
($m_1 \ll 0.02$ eV),
and for the spectrum with partial inverted hierarchy
($0.02~{\rm eV} \ltap m_1 \leq 0.2$ eV)
read, respectively,
$6.8~(8.1)  \times 10^{-2}$ eV
and $2.1~(2.2)  \times 10^{-1}$ eV
and are the same for all
solutions of the solar neutrino problem.
In the case of CP-violation, 
three CP-violating phases 
play a role in the determination of \meff 
and therefore it is rather difficult 
to obtain significant constraints on their values
from the measurement of $\meff \gtap 0.01$ eV.
However, such constraints might be obtained
if the term $m_1|U_{\mathrm{e} 1}|^2$,
where $|U_{\mathrm{e} 1}|^2 < 0.08$ 
is limited by the CHOOZ data,  
gives a negligible contribution in \meff
and/or in the case of the SMA MSW solution of the
solar neutrino problem.
Our results for the 3+1C spectrum 
are shown in Figs. \ref{figure:3+1C01} -
\ref{figure:3+1C02}.

 The 3+1B and 3+1C neutrino mass spectra 
are essentially 
indistinguishable in what regards 
the predictions for \meff 
if the lightest neutrino mass
satisfies $m_1 > 0.2$ eV, i.e.,
if the three lighter neutrinos
$\nu_{1,2,3}$ are 
quasi-degenerate in mass.
Up to corrections
$\sim 5\times 10^{-3}$ eV,
\meff is equal to 
the effective Majorana mass 
$|<m>|_{3-\nu}$
in the case of mixing of 3 quasi-degenerate
Majorana neutrinos. 
Predictions for
$|<m>|_{3-\nu}$ in the
indicated case 
were given  in ref.~\cite{BPP1} 
(see Section 6 in  \cite{BPP1}).
For $m_1 \gtap 0.3$ eV
we have (up to corrections smaller than 10\%)
$m_{1} \simeq m_{2} \simeq m_{3}$
and, correspondingly, 
$m_{\nu_e} \simeq m_1$. Thus, the 
future $^3$H $\beta-$decay experiment KATRIN
\cite{KATRIN} can provide information 
on $m_1$. This can allow to determine
the neutrino mass spectrum.

 Finally, in Figs. 
(\ref{figure:gl01}), (\ref{figure:gl02}) and
(\ref{figure:gl03})
we show the possible magnitude of
\meff as a function of $m_1$
for $m_1$ varying continuously
from $10^{-5}$ eV to 1.0 eV
for the five types of 
neutrino mass spectra
2+2A,B and 3+1A,B,C 
compatible with the data 
and respectively for the 
LMA MSW, LOW-QVO  
and SMA MSW solutions of the
solar neutrino problem
\footnote{For the case of 
mixing of three massive
Majorana neutrinos
similar plots were
proposed also in 
refs. \cite{Pols00,KPS00}.}.

 To conclude, as in the earlier 
study \cite{BPP1}, we have found that  
the observation of the
\betabeta-decay with a rate
corresponding to 
$\meff \gtap 0.02~$eV,
which is in the range of sensitivity of the 
future \betabeta-decay experiments, 
can provide unique information 
on the neutrino mass spectrum.
Combined with information on the lightest
neutrino mass, 
which could be provided, e.g.,
by the $^3$H $\beta-$decay experiment 
KATRIN \cite{KATRIN}, 
it can give also information 
on the CP-violation in the lepton sector,
and if CP-invariance holds - on the 
relative CP-parities
of the massive Majorana neutrinos.

\vspace{0.5cm}
\leftline{\bf Acknowledgments.} 
We would like to acknowledge
useful discussions with L. Wolfenstein.
S.M.B. would like to thank the Elementary Particle
Theory Sector at SISSA for kind hospitality and support.
The work of S.T.P. was supported in part by the EEC grant 
ERBFMRXCT960090
and by the  Italian MURST under the program ``Fisica
Astroparticellare.''




\begin{figure}
\begin{center}
\epsfig{file=g4nu2+2B01.epsi, height=6.5cm
}
\end{center}
\caption[22B02]{
The effective Majorana mass 
\meff, allowed by the data from the solar 
and  CHOOZ experiments, as a function of $m_1$ 
for the 2+2B neutrino mass spectrum, eq. (\ref{22Bspectrum}). 
The values of \meff  are obtained 
for the allowed ranges of \deltalsnd and 
$|U_{\mathrm{e} 3}|^2 + |U_{\mathrm{e} 4}|^2$
derived in ref.~\cite{LSND} at $95 \%$~C.L.
and for \deltasol, $\sin^2 \theta_\odot$ from
{\it i}) the $90\%$ C.L. (medium-grey   
and dark-grey regions 
limited by the two doubly-thick 
solid lines and the axes)
and the  $99\%$ C.L. (all grey regions)
LMA MSW solution region,
{\it ii}) the $90\%$ C.L. 
(medium-grey   
and dark-grey regions 
limited by the two doubly-thick 
dash-dotted lines and the axes)
and the $99\%$ C.L. (grey regions below the upper 
doubly-thick dash-dotted line)
LOW-QVO  solution regions, 
and {\it iii}) the $99\%$ C.L.  SMA solution region
(dark-grey region limited 
by the doubly-thick dash-dotted  
and dashed lines and the axes).
The results shown 
correspond to
$\cos^2 \beta = 0.3$. 
The two horizontal lines
show the upper limits
\cite{76Ge00}, given in eq. (\ref{76Ge00}).
}
\label{figure:2+2B02}        
\end{figure}

\begin{figure}
\begin{center}
\epsfig{file=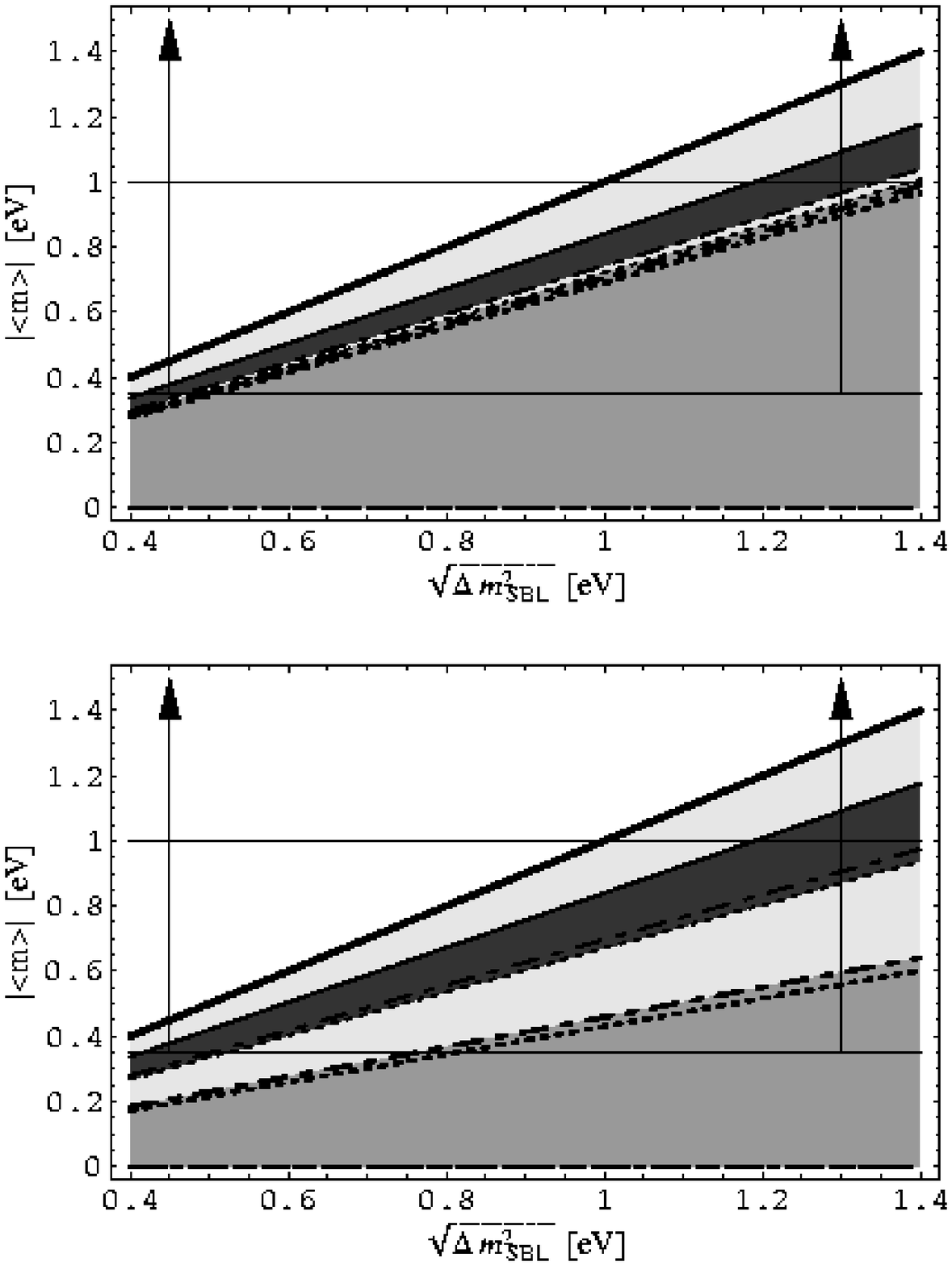, height=15cm
}
\end{center}
\caption[31A01]{
\meff as a function of $\sqrt{\deltalsnd}$
 for the 3+1A neutrino mass spectrum, 
eq. (\ref{31Aispectrum}), 
and the LMA solution (upper panel)
and LOW-QVO solution (lower panel)
 of the $\nu_{\odot}-$problem
found in  ref. \cite{Gonza3nu} at 90\%~C.L. (99\% C.L.). 
The regions allowed in the cases of 
CP-invariance correspond to
{\it i}) $\phi_3= \phi_4 = \phi_2$ - the
thick solid (non-horizontal) line $\meff = \sqrt{\deltalsnd}$,
{\it ii}) $\phi_3= \phi_4 = - \phi_2$ - 
the light grey   triangular region 
between the thick and the normal solid lines 
(light grey   triangular region 
between the thick and the normal solid lines),  
{\it iii}) $\phi_3= - \phi_4 = \pm \phi_2$ (two cases) - 
the medium grey  
triangular regions between the two
 thick dashed-dotted lines
  and between the  two  thick
dotted lines (light-grey and medium-grey region
 between the thin dashed-dotted line and the horizontal axes 
  and between the thin
dotted line and the horizontal axes). 
The ``just-CP-violation'' region is denoted by 
dark-grey color. The two horizontal 
(thick) lines show the upper limits 
\cite{76Ge00}, quoted in eq. (\ref{76Ge00}).
}
\label{figure:3+1A01}
\end{figure}

\begin{figure}
\begin{center}
\epsfig{file=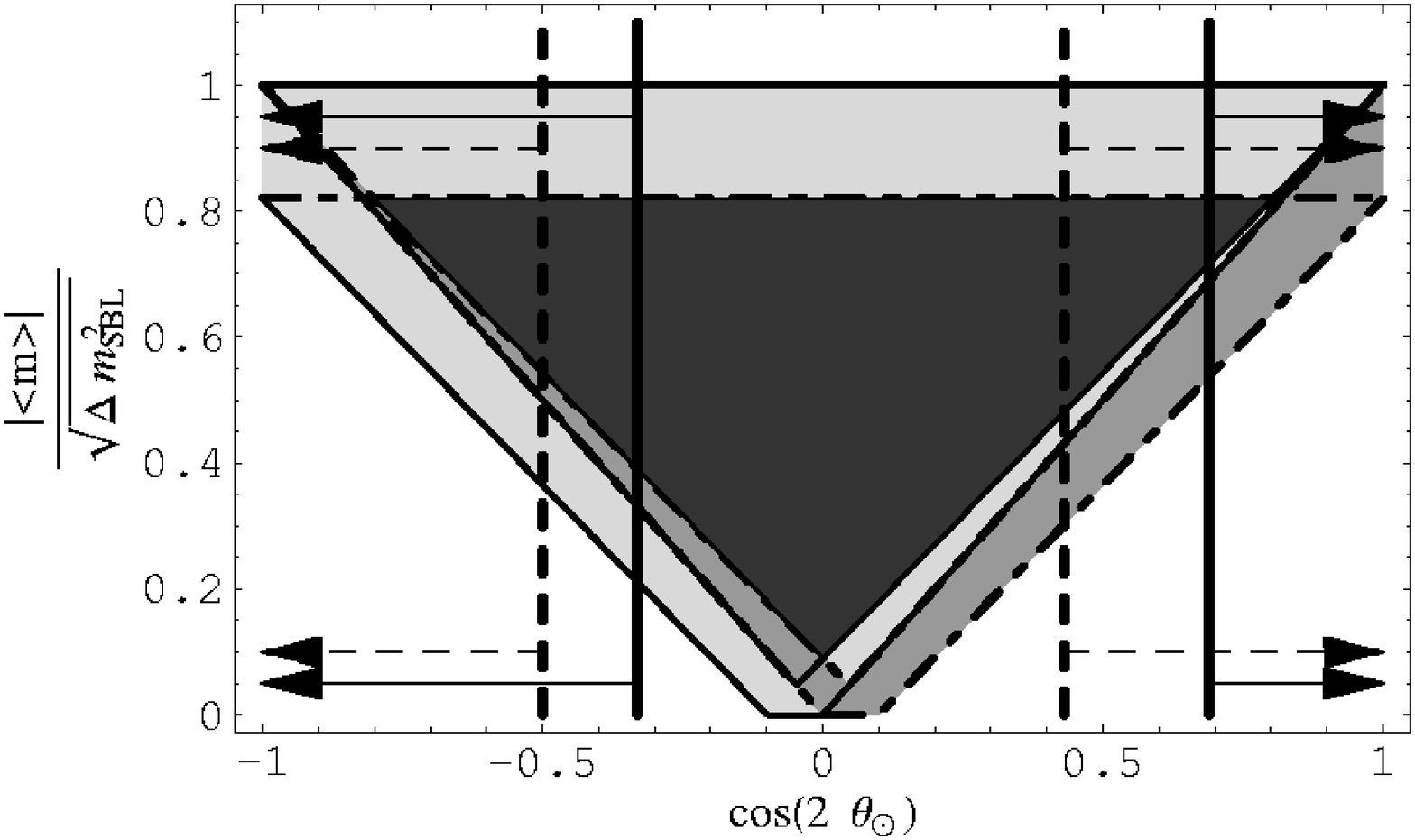, height=6.5cm
}
\end{center}
\caption[31A02]{
The dependence of $\meff/\sqrt{\deltalsnd}$ 
on $\cos 2 \theta_\odot$ 
for the 3+1A 
neutrino mass spectrum, 
eqs. (\ref{31Aispectrum}).
If CP-invariance holds, the values of 
$\meff/\sqrt{\deltalsnd}$ lie:
i) for $\phi_3 = \phi_4 = \phi_2$ -
on the line $\meff / \sqrt{\deltalsnd}  = 1$, 
ii) for $\phi_3 = \phi_4 = - \phi_2$ - 
in the region between the   horizontal thick solid and 
dash-dotted lines (in light  grey and 
medium grey colors),
iii) for $\phi_3 =- \phi_4 = + \phi_2$ - in 
the light  grey polygon with solid-line contours
and iv) for $\phi_3 = -  \phi_4 = - \phi_2$ -
in the medium grey 
polygon with the dash-dotted-line contours.
The ``just-CP-violation'' region is denoted 
by dark-grey color. The 
values of $\cos 2 \theta_\odot$ 
between the doubly thick solid and dashed lines
correspond to the 
99\%~C.L.
LMA and LOW-QVO solution regions
of refs. \cite{Gonza3nu}, respectively. 
}
\label{figure:3+1A02}
\end{figure}

\begin{figure}
\begin{center}
\epsfig{file=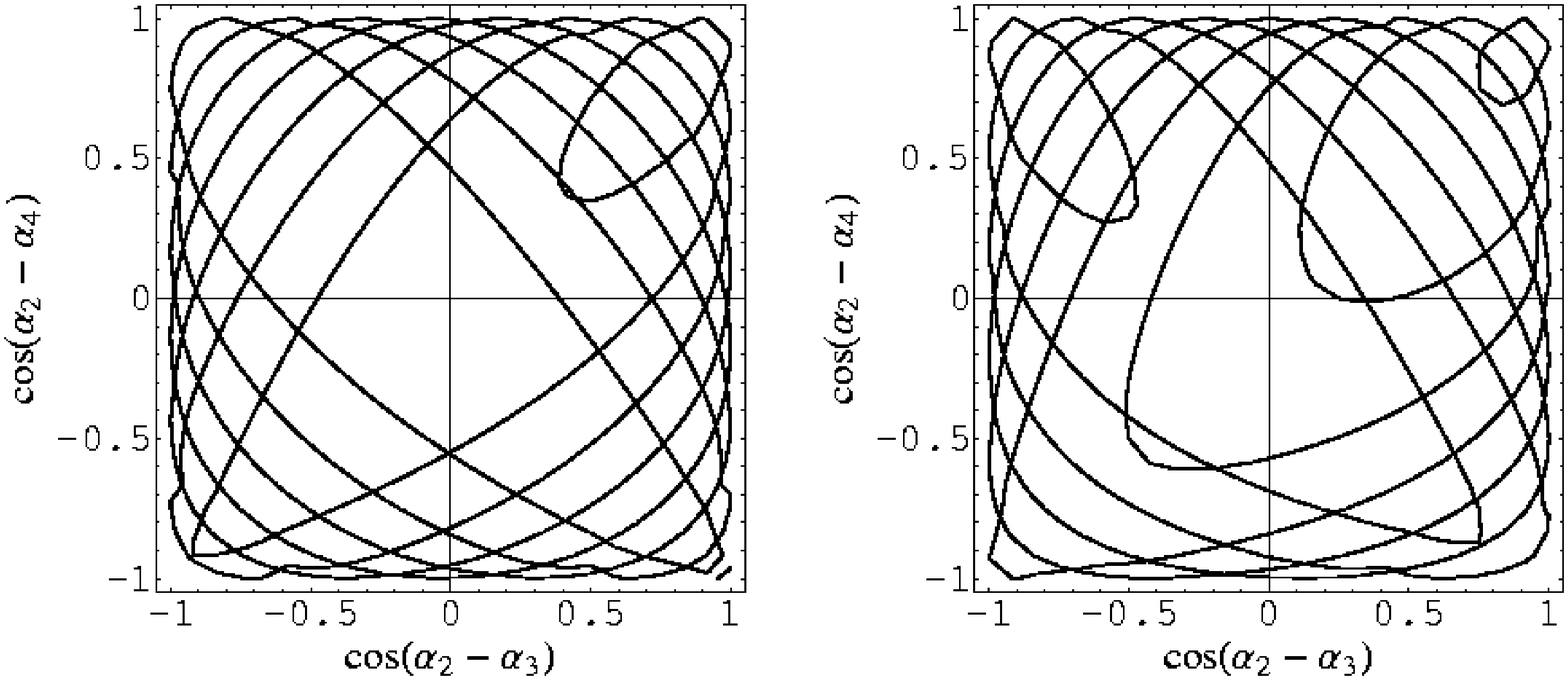, height=6cm
}
\end{center}
\caption{
The interdependence 
of the two CP-violating 
phases, $\alpha_{42}$ and $\alpha_{32}$, for a given value of the ratio
$\meff/\sqrt{\deltalsnd}$ 
in the case of the 3+1A neutrino mass spectrum.
The figures are obtained for 
$ \meff = \sqrt{0.15 + 0.10 n} \sqrt{\deltalsnd} $ with 
$n= 0,1 \ldots 8$  (with increasing \meff from left to right) 
and $|U_{\mathrm{e} 2}|^2 = 0.04$ (left-hand plot)
and $|U_{\mathrm{e} 2}|^2 = 0.08$ (right-hand plot)
 and the best fit values of the $\theta_\odot$ parameter
from \cite{Gonza3nu}
(left-hand plot). 
The values of $\cos \alpha_{32,42} = 0,\pm 1$, 
correspond to CP-invariance.
}
\label{figure:3+1A03}
\end{figure}

\begin{figure}
\begin{center}
\epsfig{file=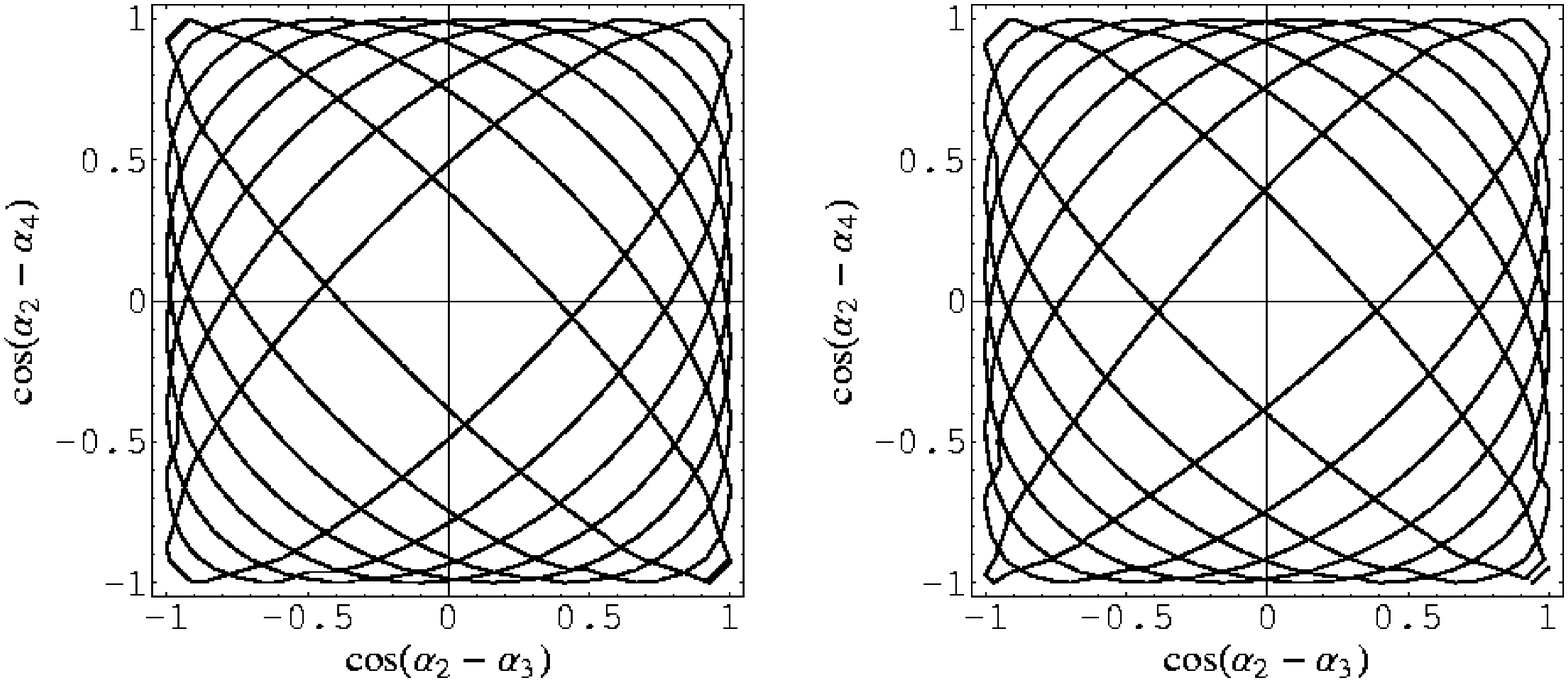, height=6.5cm
}
\end{center}
\caption{
The same as in Fig.~\ref{figure:3+1A03} 
for $|U_{\mathrm{e} 2}|^2 = 0$ (left-hand plot)
and $|U_{\mathrm{e} 2}|^2 = 0.01$ (right-hand plot).
 }
\label{figure:3+1A04}
\end{figure}

\begin{figure}
\begin{center}
\epsfig{file=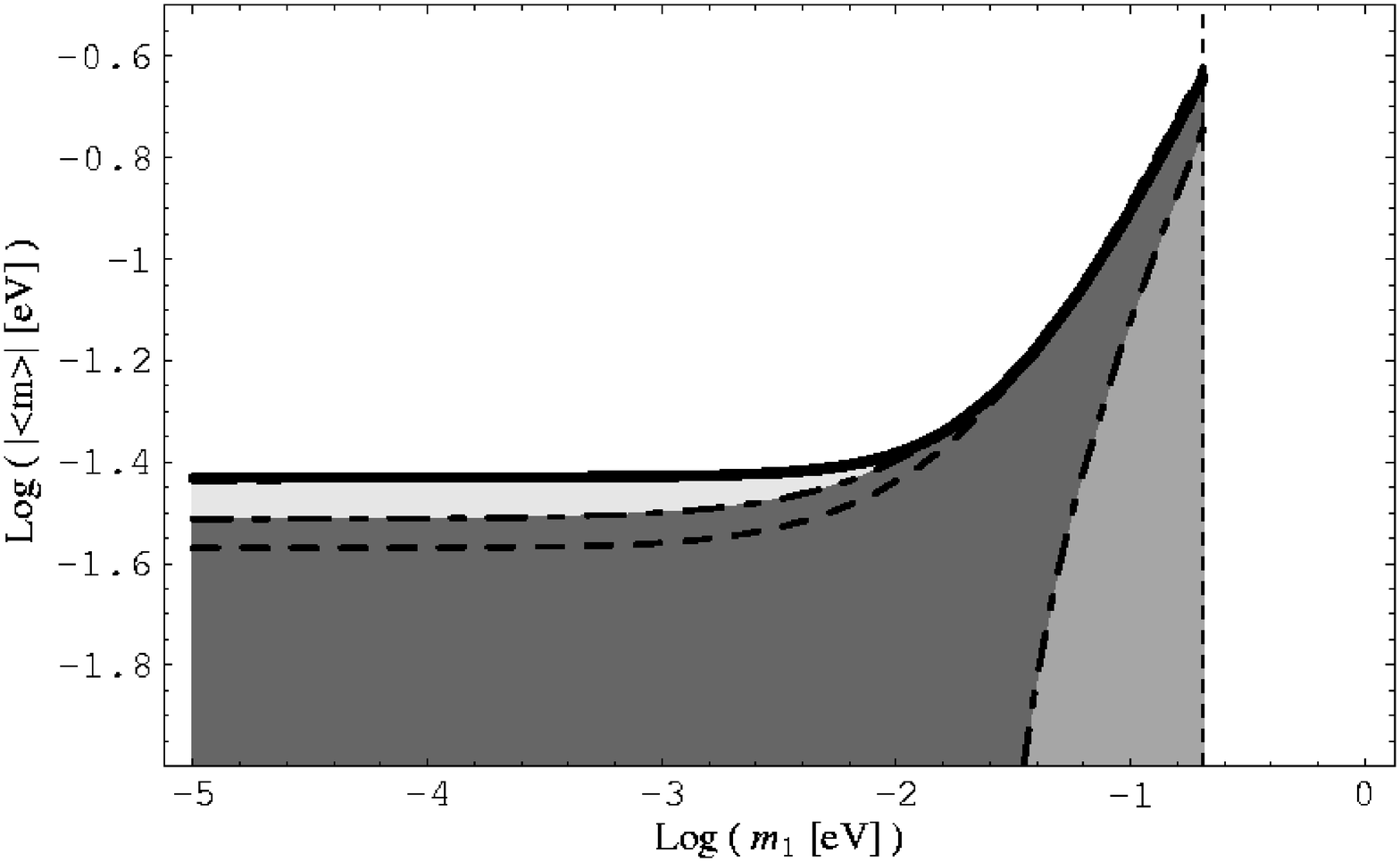, height=6.5cm
}
\end{center}
\caption[31B01]{The effective Majorana mass 
\meff, allowed by the data from the solar 
and  CHOOZ experiments, as a function 
of $m_1$ 
in the 3+1B mass spectrum, eq. (\ref{31Bspectrum}). 
The values of \meff  are obtained 
for the allowed ranges of  \deltalsnd and 
$ |U_{\mathrm{e} 4}|^2$
found in ref.~\cite{LSND} at $95 \%$~C.L.,
for \deltaatm, $ |U_{\mathrm{e} 3}|^2$ 
and
 \deltasol, $\sin^2 \theta_\odot$ from 
{\it i}) the $90\%$ C.L. and $99\%$ C.L.
LMA MSW solution regions,
(all grey regions),
{\it ii}) the $90\%$ C.L. and $99\%$ C.L.
 LOW-QVO  solution regions (medium   
and dark grey regions under 
the thick dashed line)
and {\it iii}) the $99\%$ C.L. SMA MSW solution region
(dark grey region among the dash-dotted  lines and the axes),
derived in ref.~\cite{Gonza3nu}. 
 }
\label{figure:31b01}
\end{figure}

\begin{figure}
\begin{center}
\epsfig{file=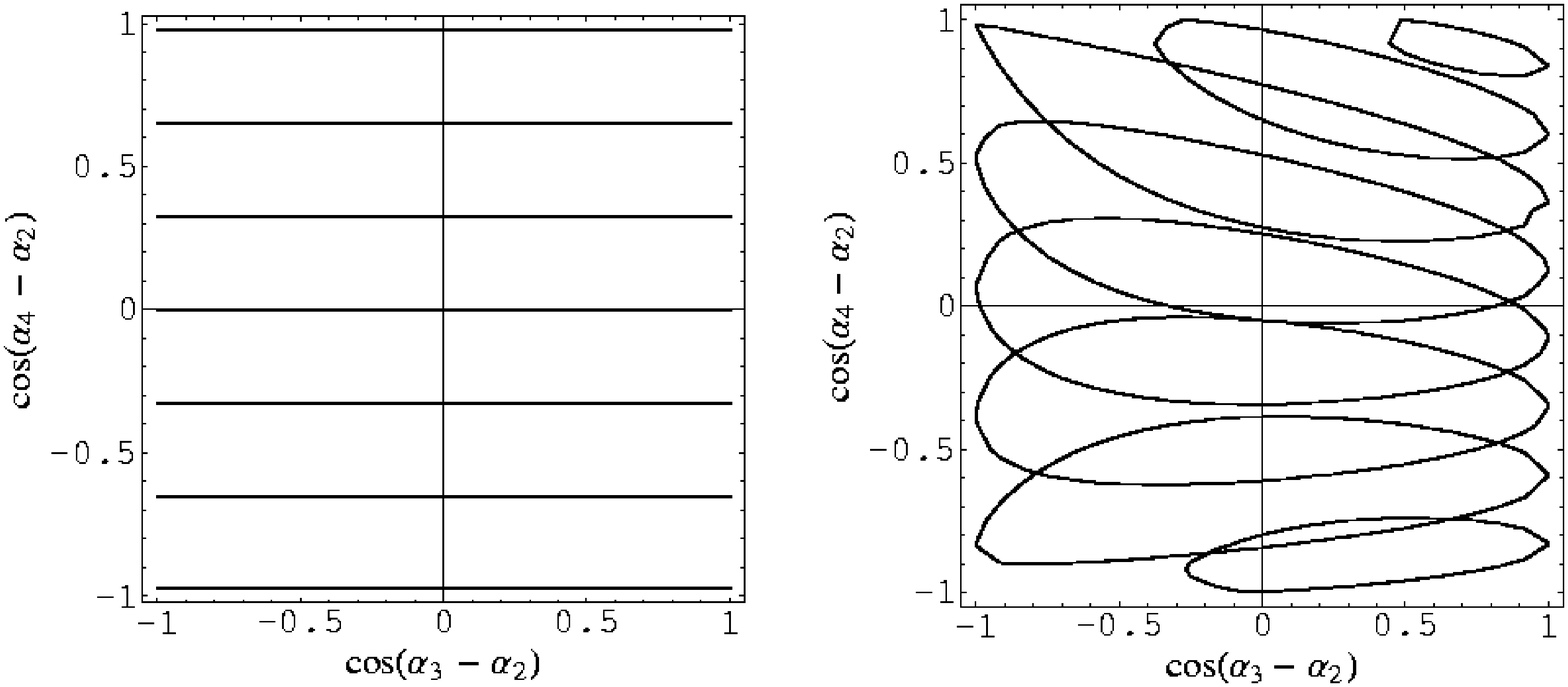, height=6.5cm
}
\end{center}
\caption[31B02]{
The interdependence 
of the two CP-violating 
phases, $\alpha_{42}$ and $\alpha_{32}$, for a given value of 
$\meff$ 
in the case of the 3+1B neutrino mass spectrum.
The figures are obtained
for $|U_{\mathrm{e} 3}|^2 = 0$ and 
$ \meff = \sqrt{0.12 + 0.05 n} \times 10^{-2} \eV $ with 
$n= 0,1 \ldots 6$  (with increasing \meff from left to right)
(left-hand plot)
and  for $|U_{\mathrm{e} 3}|^2 \sqrt{\deltaatm} 
= 7 \times 10^{-4} \eV$ and 
$ \meff = \sqrt{7 + 5 n}  \times 10^{-3} \eV $ with 
$n= 0,1 \ldots 8$  (with increasing \meff from left to right) 
 (right-hand plot)
and for the best fit value of 
$ |U_{\mathrm{e} 2}|^2 \sqrt{\deltasol} 
= 1.6 \times 10^{-3} \eV$ of  ref.~\cite{Gonza3nu},
 for the maximum allowed value of
$ |U_{\mathrm{e} 4}|^2 \sqrt{\deltalsnd} 
= 5.0 \times 10^{-3} \eV$ found  in ref.~\cite{LSND}. 
The values of $\cos \alpha_{32,42} = 0,\pm 1$, 
correspond to CP-invariance.
}
\label{figure:3+1B02}
\end{figure}

\begin{figure}
\begin{center}
\epsfig{file=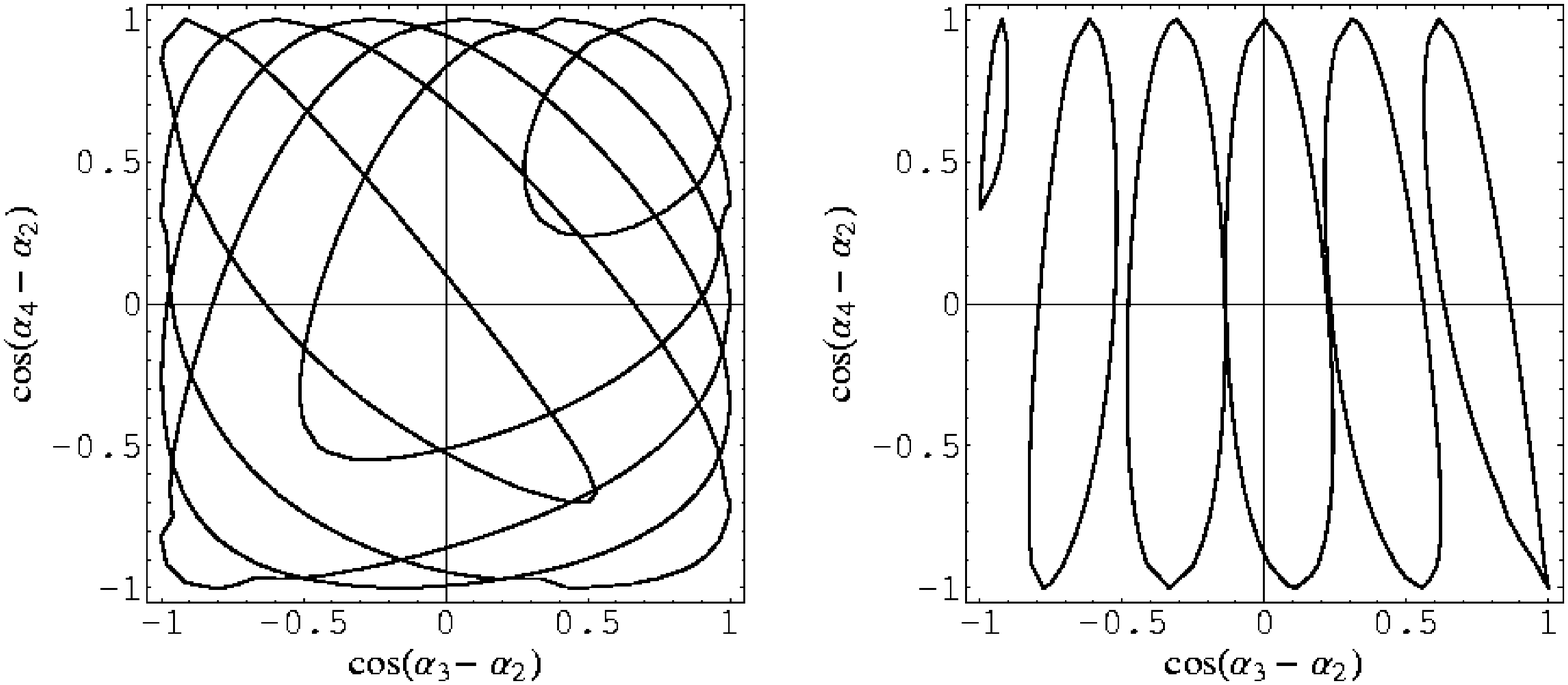, height=6.5cm
}
\end{center}
\caption[31B03]{
The same as Fig.~(\ref{figure:3+1B02}).
The figures are obtained
for
{\it i})  the maximum allowed value of
$|U_{\mathrm{e} 4}|^2 \sqrt{\deltalsnd} 
= 5.0 \times 10^{-3} \eV$ from  ref.~\cite{LSND} and 
$ \meff = \sqrt{0.1 + 0.3 n} \times 10^{-2} \eV $ with 
$n= 0,1 \ldots 5$  (with increasing \meff from left to right)
(left-hand plot)
and  
{\it ii}) the minimum allowed value of
$|U_{\mathrm{e} 4}|^2 \sqrt{\deltalsnd} 
= 3.0 \times 10^{-4} \eV$ found in ref.~\cite{LSND} and 
$ \meff = \sqrt{0.30 + 0.08 n}  \eV $ with 
$n= 0,1 \ldots 5$  (with increasing \meff from left to right)
 (right-hand plot)
and 
{\it iii}) for the best fit value of 
$ |U_{\mathrm{e} 2}|^2 \sqrt{\deltasol} 
= 1.6 \times 10^{-3} \eV$ derived in  ref.~\cite{Gonza3nu},
 for the maximum allowed value of
$ |U_{\mathrm{e} 3}|^2 \sqrt{\deltaatm} 
= 7.0 \times 10^{-3} \eV$ from  ref.~\cite{Gonza3nu}. 
The values of $\cos \alpha_{32,42} = 0,\pm 1$, 
correspond to CP-invariance.
 }
\label{figure:3+1B03}
\end{figure}

\begin{figure}
\begin{center}
\epsfig{file=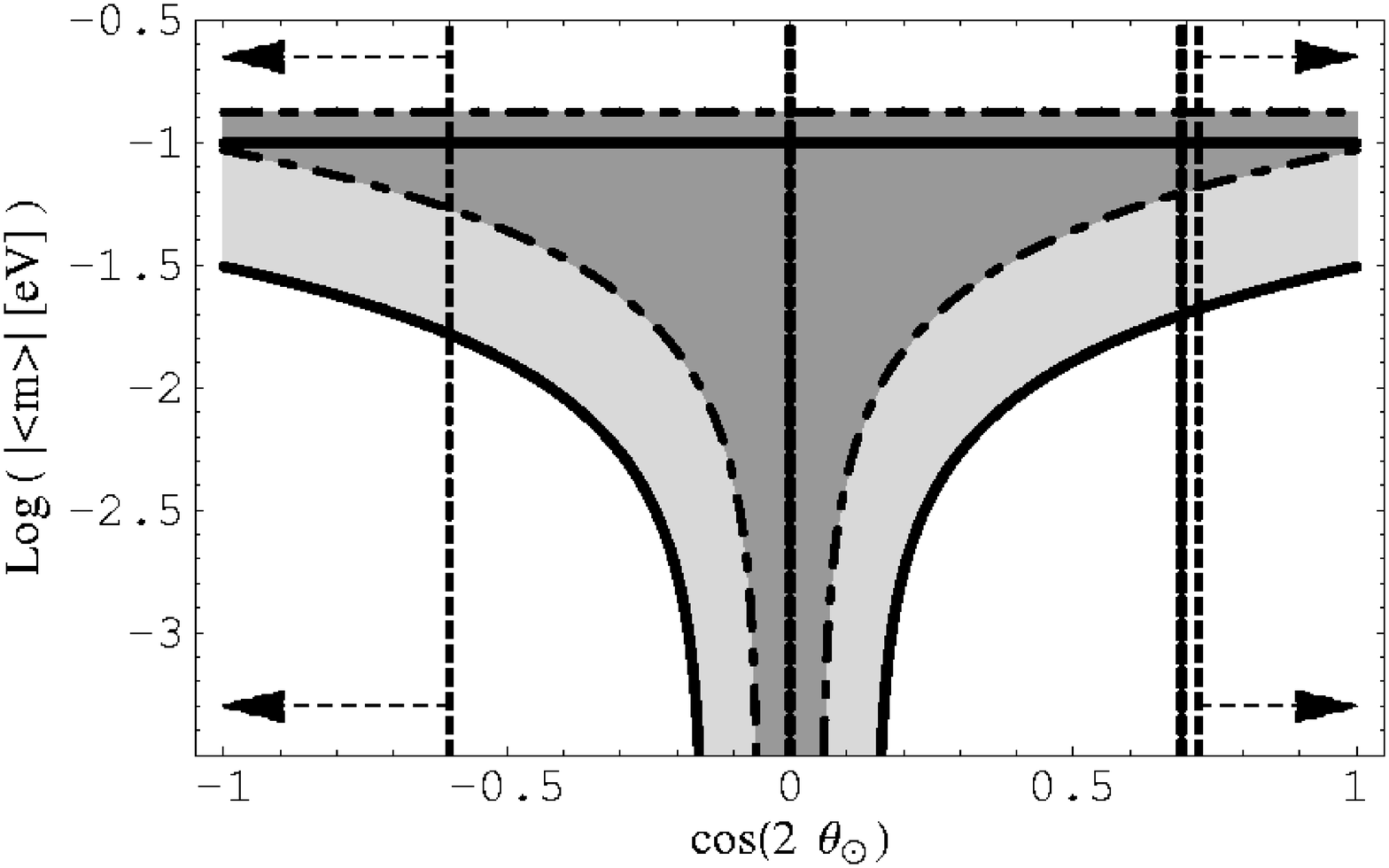, height=6cm
}
\end{center}
\caption[31C01]{
The dependence of $\meff$ 
on $\cos 2 \theta_\odot$ 
in the case of the  3+1C neutrino  mass spectrum, 
eq. (\ref{31Cspectrum}), 
for the 99\%~C.L. allowed values of 
the solar and atmospheric neutrino oscillation parameters
from  ref.~\cite{Gonza3nu} and for \deltalsnd and 
$ |U_{\mathrm{e} 4}|^2$
from  ref.~\cite{LSND} at $95 \%$~C.L.,
and  $m_1 = 0.2 \eV$ (medium grey region 
with thick dash-dotted contours) and 
$m_1 = 0 \eV$ (medium grey and light grey  region 
with thick solid contours).
The  values of $\cos 2 \theta_\odot$ 
between the doubly thick dashed (the 
 thick dashed) lines
correspond to the 
90\%~C.L. (99\%~C.L.) LMA solution regions 
in ref.~\cite{Gonza3nu}. 
}
\label{figure:3+1C01}
\end{figure}

\begin{figure}
\begin{center}
\epsfig{file=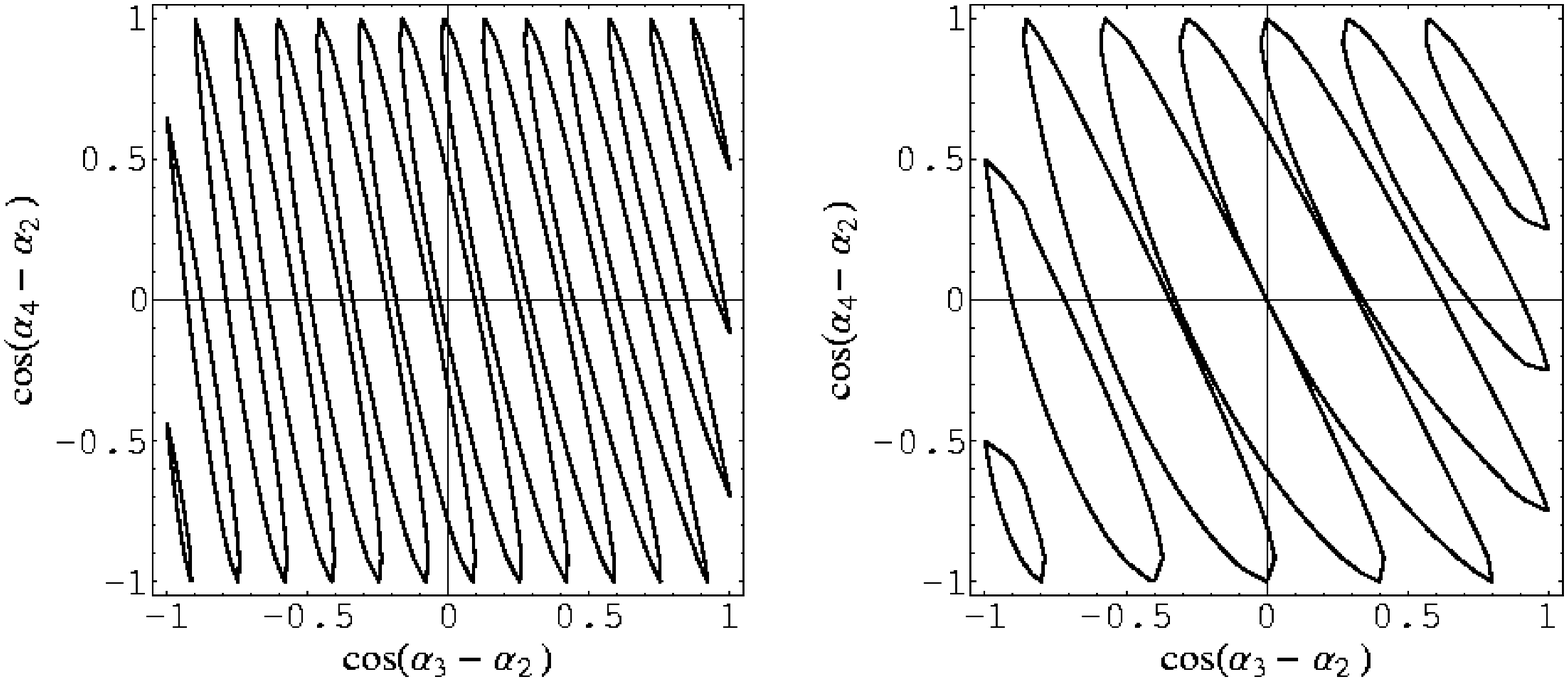, height=6cm
}
\end{center}
\caption[31C02]{
The interdependence 
of the two CP-violating 
phases, $\alpha_{42}$ and $\alpha_{32}$,
 for a given value of  $\meff$ 
in the case of the 3+1C neutrino mass spectrum.
The figures are obtained
for 
{\it i}) the maximum allowed value 
of $\deltaatm = 8.0 \times 10^{-3} \eV$
and $ \meff = \sqrt{0.10 + 0.05 n} \times 10^{-1} \eV $ with 
$n= 0,1 \ldots 15$  (with increasing \meff from left to right)
(left-hand plot)
and  
{\it ii}) the minimum allowed value of 
$\deltaatm = 1.6 \times 10^{-3} \eV$
and $ \meff = \sqrt{3 + 2 n} \times 10^{-2} \eV $ with 
$n= 0,1 \ldots 7$  (with increasing \meff from left to right)
 (right-hand plot)
and for $m_1 =0$, $ |U_{\mathrm{e} 1}|^2 \leq 0.08$, 
 the best fit value of 
$ |U_{\mathrm{e} 2}|^2$ found in ref.~\cite{Gonza3nu} and 
  the maximum allowed value of
$ |U_{\mathrm{e} 4}|^2 \sqrt{\deltalsnd} 
= 5.0 \times 10^{-3} \eV$ from  ref.~\cite{LSND}. 
The values of $\cos \alpha_{32,42} = 0,\pm 1$, 
correspond to CP-invariance.
}
\label{figure:3+1C02}
\end{figure}

\begin{figure}
\begin{center}
\epsfig{file=g4nugloballma.epsi, height=8cm
}
\end{center}
\caption[31C02]{
The dependence of \meff on $m_1$
for the LMA solution of the solar-$\nu$ problem.
For the  2+2 neutrino mass spectra,
the figure is obtained  
using the allowed values of 
 \deltalsnd and $\theta_{\mathrm{SBL}}$
 at $95\%$~C.L. derived in ref.~\cite{LSND}, 
of \deltaatm found at $90\%$~C.L. in  
ref.~\cite{Fogli4nu},
 of \deltasol and $\theta_\odot$ 
obtained for $\cos^2\beta = 0.3$
at $90\%$~C.L. in ref.~\cite{Gonza4nu},
and the 90\% C.L. limit on $\theta_{\mathrm{CHOOZ}}$ 
from  ref.~\cite{CHOOZ}. 
For the 3+1 neutrino mass spectra, 
the allowed regions of  \meff are derived using 
the $95\%$~C.L. allowed ranges of  \deltalsnd 
and $\theta_{\mathrm{SBL}}$
from  ref.~\cite{LSND} and the allowed values 
of \deltaatm, \deltasol, $\theta_\odot$ and 
$\theta_{\mathrm{CHOOZ}}$  at $90\%$~C.L. 
from  ref.~\cite{Gonza3nu}.
The allowed regions  for  \meff correspond 
{\it i}) for the 2+2A  neutrino mass spectrum,
eq. (\ref{22Aspectrum}) - to
light-grey, medium-grey and dark-grey  regions
between the two doubly-thick dash-dotted lines;
{\it ii}) for the 2+2B neutrino mass spectrum,
eq. (\ref{22Bspectrum}) -
to  the dark-grey region 
between the  two thick dash-dotted lines;
{\it iii}) for the 3+1A neutrino mass spectrum,
eq. (\ref{31Aispectrum}) - to the grey regions 
below the upper doubly-thick dash-dotted line;
{\it iv}) for the 3+1B neutrino mass spectrum,
eq. (\ref{31Bspectrum}) - to
 the medium-grey and dark-grey region 
below the  thick solid  line;
{\it v}) for the 3+1C neutrino mass spectrum,
eq. (\ref{31Cspectrum}) -
to the medium-grey and dark-grey region 
below  the  doubly-thick dashed line.
The two horizontal 
lines show the upper limits 
\cite{76Ge00}, quoted in eq. (\ref{76Ge00}).
}
\label{figure:gl01}
\end{figure}

\begin{figure}
\begin{center}
\epsfig{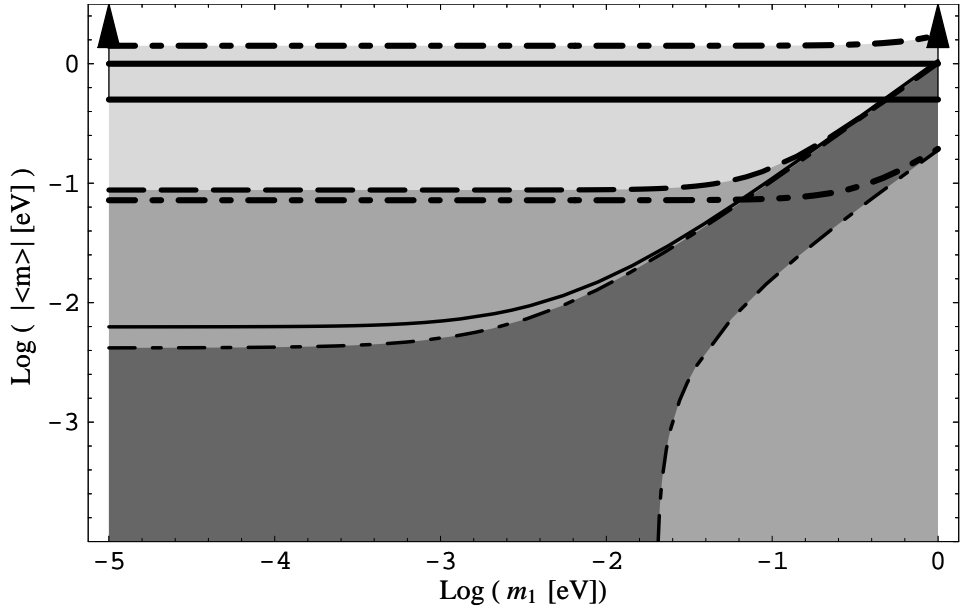}
\end{center}
\caption[31C02]{The same as in Fig.~\ref{figure:gl01}
but for the  LOW-QVO solution of the solar-$\nu$ problem. 
The allowed values of \meff correspond 
{\it i}) for the 2+2A neutrino mass spectrum,
eq. (\ref{22Aspectrum}) -
to  the grey regions 
between the  doubly-thick dash-dotted lines;
{\it ii}) for the 2+2B neutrino mass spectrum,
eq. (\ref{22Bspectrum}) -
to  the dark-grey region 
between the  thick dash-dotted lines;
{\it iii}) for the 3+1A neutrino mass spectrum,
eq. (\ref{31Aispectrum}) - to  
the grey regions 
below the upper doubly-thick dash-dotted line;
{\it iv}) for the 3+1B neutrino mass spectrum,
eq. (\ref{31Bspectrum}) -
to the medium-grey and dark-grey region 
below the  thick solid  line;
{\it iv}) for the 3+1C neutrino mass spectrum,
eq. (\ref{31Cspectrum}) - to
 the medium-grey and dark-grey region 
below the  doubly-thick dashed line.
}
\label{figure:gl02}
\end{figure}

\begin{figure}
\begin{center}
\epsfig{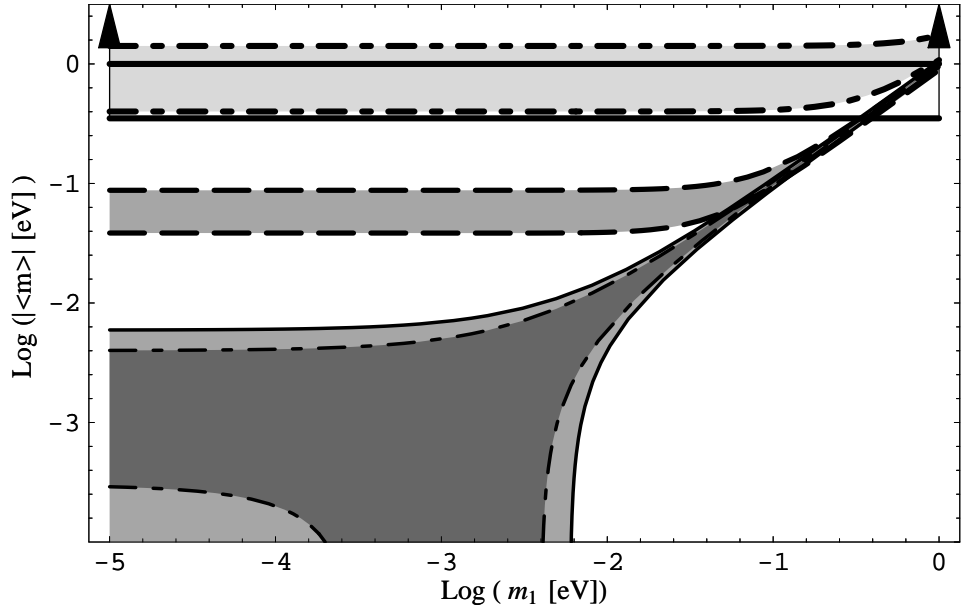}
\end{center}
\caption[31C02]{
The same as in Fig.~\ref{figure:gl01}
but for the  SMA MSW solution of the solar neutrino
problem.  The allowed values of \meff correspond  
{\it i}) for the 2+2A and 3+1A neutrino mass spectra,
eqs. (\ref{22Aspectrum}) and (\ref{31Aispectrum}) -
to  the light-grey  region 
between the two  doubly-thick dash-dotted lines;
{\it ii}) for the 2+2B neutrino mass spectrum,
eq. (\ref{22Bspectrum}) -
to  the dark grey  region 
between the  thick dash-dotted lines and the axes;
{\it iii}) for the 3+1B neutrino mass spectrum,
eq. (\ref{31Bspectrum}) - to
 the medium-grey and dark-grey region 
between  the two  thick solid  lines and the axes;
{\it iv}) for the 3+1C neutrino mass spectrum,
eq. (\ref{31Cspectrum}) - to
 the medium-grey region 
between  the two doubly-thick dashed lines.
}
\label{figure:gl03}
\end{figure}

\end{document}